\documentclass[12pt,preprint]{aastex}
\usepackage{graphicx}
\usepackage[latin1]{inputenc}
\usepackage{txfonts}
\usepackage{lscape}
\usepackage{natbib}

\newcommand{\lre}{$\log r_{\rm e}$}
\newcommand{\re}{$r_{\rm e}$}

\newcommand{\sn}{$n$}
\newcommand{\mie}{$<\! \mu\! >_{\rm e}$}
\newcommand{\ls}{$\log \sigma_0$}

\newcommand{\papdata}{Paper I}
\newcommand{\nablas}{$\nabla_\star$}
\newcommand{\nablat}{$\nabla_t$}
\newcommand{\nablaz}{$\nabla_Z$}
\shorttitle{Stellar population gradients in ETGs}
\shortauthors{La Barbera et al.}

\begin{document}

\title{SPIDER - IV. Optical and NIR color gradients in Early-type galaxies: New Insights into Correlations  with Galaxy Properties}
\author{
La Barbera, F.\altaffilmark{1},
de Carvalho, R.R.\altaffilmark{2}, 
de la Rosa, I.G.\altaffilmark{3,4}, 
Gal, R.R.\altaffilmark{5}, 
Swindle, R.\altaffilmark{5}, 
Lopes, P.A.A.\altaffilmark{6}}

\affil{$^{(1)}$INAF -- Osservatorio Astronomico di Capodimonte, Napoli, Italy }
\affil{$^{(2)}$Instituto Nacional de Pesquisas Espaciais/MCT, S. J. dos Campos, Brazil}
\affil{$^{(3)}$Instituto de Astrofisica de Canarias (IAC), E-38200 La Laguna, Tenerife, Spain}
\affil{$^{(4)}$Depto. de Astrofisica, Universidad de La Laguna (ULL), E-38206 La Laguna, Tenerife, Spain}
\affil{$^{(5)}$University of Hawai'i, Institute for Astronomy, 2680 Woodlawn Drive, Honolulu, HI 96822, USA}
\affil{$^{(6)}$Observat\'orio do Valongo/UFRJ, Rio de Janeiro, Brazil
}


\begin{abstract}
  We  present an  analysis of  stellar population  gradients  in 4,546
  Early-Type Galaxies (ETGs) with  photometry in $grizYHJK$ along with
  optical  spectroscopy. { ETGs were  selected  as bulge-dominated
    systems, displaying passive spectra within the SDSS fibers}. A new
  approach is described which  utilizes color information to constrain
  age  and   metallicity  gradients.   Defining   an  effective  color
  gradient, $\nabla_{\star}$, which  incorporates all of the available
  color  indices,  we  investigate  how $\nabla_{\star}$  varies  with
  galaxy mass proxies,  i.e.  velocity dispersion, stellar ($M_\star$)
  and dynamical  ($M_{dyn}$) masses, as well as  age, metallicity, and
  $[\alpha/{Fe}]$.  ETGs  with M$_{\rm dyn}$ larger than  $ 8.5 \times
  10^{10}  \, M_\odot$  have increasing  age gradients  and decreasing
  metallicity gradients  $wrt$ mass, metallicity,  and enhancement. We
  find that velocity dispersion and $[\alpha/Fe]$ are the main drivers
  of these  correlations.  ETGs with  $ 2.5 \times 10^{10}  \, M_\odot
  \le M_{dyn} \le 8.5 \times  10^{10} \, M_\odot$, show no correlation
  of age, metallicity, and  color gradients $wrt$ mass, although color
  gradients  still correlate with  stellar population  parameters, and
  these  correlations are  independent of  each other.   In  both mass
  regimes,  the striking anti-correlation  between color  gradient and
  $\alpha$-enhancement is significant  at { $\sim$ 5~$\sigma$}, and
  results  from  the fact  that  metallicity  gradient decreases  with
  $[\alpha/Fe]$.  This anti-correlation may reflect the fact that star
  formation and metallicity enrichment  are regulated by the interplay
  between the  energy input from  supernovae, and the  temperature and
  pressure  of  the hot  X-ray  gas in  ETGs.   For  all mass  ranges,
  positive   age   gradients   are   associated  with   old   galaxies
  ($>5-7$~Gyr).   For galaxies  younger than  $\sim 5$~Gyr,  mostly at
  low-mass,  the age  gradient tends  to be  anti-correlated  with the
  $Age$ parameter, with more positive gradients at younger ages.
\end{abstract}

\section{Introduction}

The study of  the stellar population in galaxies  presents some unique
challenges.   First,  the  method   used  to  characterize  a  stellar
population is crucial as it provides constraints on theories of galaxy
formation (e.g.   De Lucia et al.~2006).  Second,  age and metallicity
dominated  indicators should  minimize the  age-metallicity degeneracy
described in the 3/2  Worthey law (Worthey, Trager \& Faber 1995).  Finally, a galaxy's
environment  may  be the  dominant  influence  on  its star  formation
history,  further   reducing  the  adequacy  of   the  Single  Stellar
Population  (SSP)  assumption.   Monolithic  collapse  models  predict
early-type  systems with  a central  region that  is more  metal rich,
slightly younger, and less $\alpha/F_e$ enhanced than the outer parts.
Mergers may wash out any possible gradient present in the progenitors,
although  dissipation  can trigger  star  formation  in  the cores  of
galaxies, creating age gradients. It is important to emphasize that we
still   lack   firm  quantitative   estimates   for   many  of   these
effects~\citep{White:80}. Considering the  impact of such predictions,
we need large and well characterized samples to tackle this issue. The
nearby universe  ($z<0.1$) provides a  starting point, as we  now have
high quality data for large numbers of galaxies.

The  study of  color  gradients  is motivated  by  the possibility  of
discriminating  between  different  models  of galaxy  formation,  and
points to the real need for systematic and accurate measurements for a
large  and  well controlled  sample  of  ETGs.   These systems  become
gradually  bluer  in  the   outskirts  (e.g.  Peletier  et  al.  1990;
Goudfrooij et  al.  1994; Michard 1999;  Wu et al.  2005; Cantiello et
al. 2005;  Suh et al.  2010). Although  the age-metallicity degeneracy
makes   the  interpretation   of  the   color   gradient  troublesome,
metallicity gradient seems to be  main cause (e.g. Saglia et al. 2000;
\citealt{TKA00, TaO00}; La Barbera  et al. 2003).  Small, positive age
gradients      are     also      consistent      with     observations
(e.g.~\citealt{SMG00}),  and have  been detected  in  the low-redshift
population   of   ETGs~\citep{LdC:09,   Clemens:09}.  Positive   color
gradients in  high-redshift ETGs have  been observed by~\citet{FER05},
who found that about one-third of field ETGs at $z \sim 0.7$ have blue
cores,   in  contrast   to  only   $10\%$  at   lower   redshift  (see
also~\citealt{MAE01,  MEN04,  FER09}).   The  possibility  that  color
gradients  could  be  due merely  to  dust  in  a well  mixed  stellar
population has been  considered (e.g. Goudfrooij \& de  Jong 1995) but
many ETGs exhibit radial gradients of metal absorption features, which
cannot  be attributed  to  dust (e.g.  Gonzalez~1993;  Mehlert et  al.
2000). Furthermore, many  ETGs are undetected by IRAS  at 60 $\mu$ and
100 $\mu$. Cosmological simulations do not yet have sufficient spatial
resolution   to   provide    reliable   color   gradients.    However,
chemo-dynamical simulations  of ETGs are starting  to yield meaningful
results   for  metallicity   gradients,   reproducing  some   observed
population gradients (see Kobayashi 2004).

The   last  decade   has   witnessed  the   appearance  of   important
observational  studies  using photometric  and  spectroscopic data  to
constrain     age,     metallicity     and    elemental     abundances
(~\citealt{JORG:99, PBM:01, TeF:02,  Tra:00}).  They conclude that age
and  metallicity   are  anti-correlated   for  galaxies  of   a  given
luminosity. Furthermore,  Poggianti et  al.~(2001b) find that  in high
density regimes  a large  fraction of systems  shows no signs  of star
formation in  their centers.  Another important  finding relating star
formation and  environment comes from the  work of~\citet{Kunt:02} who
analyzed  a sample  of early-type  systems in  the low  density regime
(field) showing  that these galaxies  are younger and  more metal-rich
than  their counterparts  in  clusters.  Most  of  these studies  used
samples of at  most a few hundred galaxies. Since  these works are the
nearby  counterparts of  the studies  of the  evolution  of early-type
galaxies   (ETGs)   as   a    function   of   redshift   in   clusters
(e.g.~\citealt{vDS:03})           and          in          low-density
environments~\citep{Treu:99}, it  is necessary to  establish large and
well-defined samples  covering a broad range {  of} environments as
we do here.

This is the fourth paper of a series studying the global properties of
ETGs in the  local Universe ($z<0.1$). We focus  on stellar population
gradients and  how they  depend on galaxy  properties. We  use optical
data ($griz$ photometry and spectra) from the Sloan Digital Sky Survey
Data  Release 6  \citep[SDSS,][]{ade08}  and near-infrared  photometry
from   the  UKIRT  Infrared   Deep  Sky   Survey  Data   Release  Four
\citep[UKIDSS,][]{Law07}.  The  selection of  ETGs is described  in La
Barbera et al.~(2010a; hereafter Paper  I), and we refer the reader to
that paper for all the  details on sample selection and the procedures
used to  derive the galaxy  parameters. { The results  presented in
  this paper  apply not  only to ellipticals,  but also  to lenticular
  galaxies as well as those dominated  by a red central bulge (but see
  Sec.~3.3 and Sec.~8.3).}

This  paper is  organized as  follows:  Sec. 2  briefly describes  the
sample  used while in  Sec.  3  we present  how stellar  and dynamical
masses  were  computed  and  stellar population  properties  measured.
Measurement of  color gradients are  described in Sec. 4,  while their
distributions are  discussed in Sec. 5.   Sec 6 details  how we derive
stellar population gradients from color gradients. Sec. 7 presents our
approach to constrain the  variation of stellar population parameters,
and  discusses  the systematics  involved.   Sec.   8 illustrates  the
dependence of  the effective color  gradient on galaxy  parameters and
Sec.  9  examines  the   age  versus  metallicity  gradients  in  {
  ETGs}. Finally, in Sec.  10 we summarize our most important results.
Throughout the paper, we adopt a cosmology with $\rm H_0 \! = \!  75\,
km  \,  s^{-1} \,  Mpc^{-1}$,  $\Omega_{\rm m}  \!   =  \!  0.3$,  and
$\Omega_{\Lambda} \!  = \! 0.7$.

\section{The sample}
\label{sec:samples}

{  The sample  of ETGs  is  selected from  SDSS-DR6, following  the
  procedure  described   in  La Barbera et al. (2008b), La Barbera \& de Carvalho (2009),  and  Paper
  I. We select  galaxies in the redshift range of  0.05 to 0.095, with
  $^{0.1}M_{r}{<}-20$,  where $^{0.1}M_{r}$  is  the k-corrected  SDSS
  Petrosian magnitude in r-band.  The k-correction is determined using
  the  software   $kcorrect$  (version  $4\_1\_4$;  ~\citealt{BL03a}),
  through  a   restframe  r-band  filter  blue-shifted   by  a  factor
  $(1+z_0)$.  We adopt $z_0=0.1$  as in (e.g.) ~\citealt{Hogg04}.  The
  lower  redshift  limit of  the  sample  is  chosen to  minimize  the
  aperture  bias~\citep{GOMEZ03},  while   the  upper  redshift  limit
  guarantees  not  only  a   high  level  of  completeness  (according
  to~\citealt{SAR06}), but  also allows us to  define a volume-limited
  sample of  {\it bright} early-type  systems.  At the  upper redshift
  limit  of $z=0.095$,  the magnitude  cut of  $-20$  also corresponds
  approximately to the magnitude  limit to which the SDSS spectroscopy
  is  complete  (i.e.  a  Petrosian  magnitude  of  $m_r \sim  17.8$).
  Following   ~\citet{BER03},   {we  define   ETGs   using  the   SDSS
    spectroscopic  parameter $eClass$,  which  indicates the  spectral
    type of a  galaxy on the basis of  a principal component analysis,
    and the SDSS photometric parameter $fracDev_r$, which measures the
    fraction  of  the galaxy  light  that is  better  fitted  by a  de
    Vaucouleurs  (rather than  an  exponential) law.   ETGs are  those
    systems with $eClass \!  < \!  0$ and $fracDev_r \!  > \!  0.8$. }
  We   select  only   galaxies  with   central   velocity  dispersion,
  $\sigma_0$, available from SDSS-DR6, in  the range of $70$ and $420$
  km\,s$^{-1}$, and with no spectroscopic warning on (i.e.  $zWarning$
  attribute set to zero).  These constraints are chosen to select only
  reliable velocity  dispersion measurements from  SDSS.  Applying the
  above criteria leads to a sample of $39,993$ ETGs.

Out of the $39,993$ ETGs defining the SPIDER sample, $5,080$ also have
photometry in  the $YJHK$ wavebands from the  UKIDSS-Large Area Survey
(see~\papdata).}   All  galaxies  have  two measures  of  the  central
velocity  dispersion,  one from  SDSS-DR6  and  an alternate  estimate
obtained    by    fitting   SDSS    spectra    with   the    STARLIGHT
software~\citep{CID05}.  In  all wavebands, the  structural parameters
including effective  radius, \re,  the mean surface  brightness within
that radius, \mie, and the  Sersic index, \sn, have been homogeneously
measured using  2DPHOT~\citep{LBdC08}.  Total magnitudes  are computed
from the effective parameters in each filter.

For  this work,  we select  ETGs from  the SPIDER  survey  as follows.
First,  we  exclude  galaxies  which  are poorly  {  fitted}  by  a
two-dimensional Sersic  model. This is done by  selecting only objects
with  $\chi^2<2$  in all  wavebands,  where  $\chi^2$  is the  rms  of
residuals  between  the   galaxy  image  in  a  given   band  and  the
corresponding   best-fitting   two-dimensional   Sersic  model.    The
threshold  $\chi^2=2$  excludes  those  objects which  are  more  than
5~$\sigma$  above  the  peak  of  the  $\chi^2$  distribution  in  all
passbands.   As shown  in~\papdata, the  distributions of  $\chi^2$ in
$grizYJHK$ are peaked around one, with a width of $\sim 0.2$.  Second,
since we estimate the internal color gradients from \re \, and \sn, we
exclude  galaxies with  large uncertainties  on these  parameters.  As
shown  in~\papdata, the  errors on  \sn \,  are nearly  independent of
$S/N$ ratio,  and are always less  than $\sim 0.2$~dex.   On the other
hand, the  errors on \re \,  strongly depend on the  $S/N$.  To remove
galaxies with  large uncertainties without overly  reducing the sample
size, we select  only those ETGs with a \lre  \, uncertainty less than
$0.5$~dex in  all wavebands. These  restrictions yield a  subsample of
$37,068$  galaxies  with  parameters  in $griz$.   Of  these,  $4,546$
objects also have photometry in  $YJHK$.  Hereafter, we refer to these
as the optical and optical+NIR samples of ETGs, respectively.

\section{Galaxy parameters}
\label{sec:pars}
We  analyze the  dependence of  internal  color gradients  in ETGs  on
galaxy  parameters,  i.e.   structural  parameters,  optical  and  NIR
magnitudes,  stellar  and  dynamical  masses, and  stellar  population
properties   (age,   metallicity,   and  $\alpha$-enhancement).    The
derivation  of  magnitudes  and  structural  parameters  is  described
in~\papdata.  Below, we describe the estimate of stellar and dynamical
masses  (Sec.~\ref{sec:stellar_masses}  and~\ref{sec:dyn_masses}),  as
well    as    that    of    the    stellar    population    parameters
(Sec.~\ref{sec:stellar_pop}).

\subsection{Stellar masses from SED fitting}
\label{sec:stellar_masses}
Galaxy stellar masses can be  estimated by fitting their observed SEDs
with stellar  population synthesis models.  \citet{bri00} demonstrated
that  such  techniques,  combining  optical and  near-IR  fluxes  with
spectroscopic redshifts, could be used  to derive masses with a factor
of two uncertainty to $z  \sim1 $. Local studies by \citet{bell03} and
\citet{Cole01}  examined  the  stellar  mass  function  using  similar
methods. \citet{sal05,sal07} have shown that galaxy properties derived
from  SED fitting  with  broadband colors  are  consistent with  those
derived from  spectroscopic data. Thus,  we use SED fitting  to derive
galaxy stellar masses.

We employ  the LePhare  code (S.  Arnouts \& O.   Ilbert) used  by the
COSMOS  survey  \citep{ilb09}  to  fit  a  suite  of  \citet[hereafter
  BC03]{BrC03} model  SEDs to our  observed 8-band photometry  and the
known spectroscopic redshift. LePhare performs a $\chi^2$ minimization
between the  observed fluxes (and  their associated errors)  and those
synthesized from the models.  Because early-types are unlikely to have
significant  ongoing   star  formation,  we   use  a  simple   set  of
\citet{BrC03}  SEDs, with 6  different star-formation  e-folding times
$\tau$,  5  internal   reddenings  $E(B-V)$,  and  four  metallicities
$Z/Z_{\odot}$, for a total of 120 possible models. The values of these
parameters   are  given   in  Tab.~\ref{tab:sedmods}.   We   assume  a
\citet{cal00}   extinction   law   and  \citet{cha03}   initial   mass
function. The reddening  E(B-V) is limited to 0.5  magnitudes to avoid
incorrect fitting  of observationally red galaxies  as highly reddened
blue galaxies; such objects should  be absent from our sample. Because
each  BC03  template is  normalized  to one  solar  mass,  we use  the
observed $K$-band flux to  determine the normalization of the best-fit
template which  then gives the stellar  mass. We use  $K$-band for the
mass scaling since  it is only weakly affected  by dust extinction and
is  quite  insensitive  to  the  presence  of  young,  luminous  stars
\citep{kau98,bun05}. 

\begin{deluxetable}{c|c|c}
\small
 \tablecaption{BC03 Galaxy SED Parameters}
\tabletypesize{\scriptsize}
\tablewidth{0pt}
\tablecolumns{3}
\tablehead{$\tau$~(Gyr) & $E(B-V)$ & $Z/Z_{\odot}$ }
\startdata
1 & 0 & 0.2 \\
2 & 0.1 & 0.4 \\
3 & 0.2 & 1.0 \\
5 & 0.3 & 3.0 \\
10 & 0.4 & \\
15 & 0.5 & \\
\enddata
\label{tab:sedmods}
\end{deluxetable}

Galaxy  fluxes are  estimated in  an  adaptive aperture  of radius  $3
\times r_{K,i}$ in  all wavebands, where $r_{K,i}$ is  the i-band Kron
radius (see~\papdata).   We use these  aperture fluxes as they  have a
better accuracy than those obtained from (2DPHOT) total magnitudes. In
order  to  derive  total  stellar  masses, $M_{\star}$,  we  apply  an
aperture correction. We  calculate the fraction of total  to $3 \times
r_{K,i}$  aperture flux  in K-band,  and then  multiply  the estimated
stellar  mass by this  fraction. This  procedure assumes  that stellar
mass in a galaxy is distributed in the same way as the K-band light.

Rather  than  use  an   individual  best-fitting  SED  (based  on  the
$\chi^{2}$), we  use a mass  computed from the model  corresponding to
the median $e^{-\chi^{2}/2}$. This  reduces stochastic mass errors due
to individual models which happen  to have very low $\chi^{2}$ values,
since  there remains  some  degeneracy in  the  SED fits  even with  8
filters. A small  number of galaxies (1465 of  39993 with optical data
only, and 286  of 4546 with NIR data)  are not well fit by  any of the
SEDs,  resulting  in  $e^{-\chi^{2}/2}=0$  and thus  no  stellar  mass
estimate. These objects, comprising 3.6\% and 6.3\% of the optical and
optical+NIR  samples, respectively,  are excluded  from  analyzes that
require the stellar masses.

We tested our stellar masses by  comparing to those derived by a group
at  Max  Plank  using  SDSS photometry  alone.   \footnote{These  mass
  estimates            can            be           found            at
  http://www.mpa-garching.mpg.de/SDSS/DR7/Data/stellarmass.html.}  The
overall  agreement   is  excellent.   We  find  a   modest  offset  in
$M_{*,SPIDER} -  M_{*,MPI}$ of 0.14 dex;  this is because  we use Kron
magnitudes rather than fiber magnitudes; Kron magnitudes are $\sim0.5$
mag  brighter,  which  translates   into  a  difference  $\sim0.2$  in
$log(M)$. The  $rms$ scatter between  the two mass estimators  is only
0.1 dex, a  factor of $\sim2$ smaller than  most systematic errors and
biases  introduced  by  different  choices  of  IMF,  extinction  law,
etc. More details  of the SED fitting and  potential sources of errors
and  biases in  the stellar  mass estimates  will be  presented  in an
upcoming   paper  in  this   series  devoted   to  this   issue  alone
{~(Swindle et al.~2010, in preparation)}.

{  We note  that both  the SED  fitting and  the analysis  of color
  gradients (Sec.~\ref{sec:grads}) are  performed with a different set
  of  models (i.e.   BC03) than  that used  for the  spectral analysis
  (i.e.   MILES,  see  Sec.~\ref{sec:stellar_pop}).  The  BC03  models
  provide  a  high  resolution  of  1~$\AA$ in  the  wavelength  range
  3322~$\AA$  to  9300~$\AA$  (griz),  with  a  median  resolution  of
  10~$\AA$ in zYJHK. We do not use the MILES library here as we do for
  other tests  in this  paper due to  its limited  wavelength coverage
  (3525~$\AA$ to 7500~$\AA$).  With  7500~$\AA$ lying near the central
  wavelength of the i-band, ignoring the izYJHK data inevitably yields
  failures  in  SED  identification,   provided  the  range  of  model
  parameters  given in  Tab.~1.   This uncertainty,  coupled with  the
  benefit of  the K-band flux for mass-scaling,  motivates this choice
  of  models for  the SED  fitting.  The  technical  justification for
  using the MILES library in  analyzing spectra and color gradients is
  also detailed in Sec.~\ref{sec:stellar_pop}.}

\subsection{Dynamical masses}
\label{sec:dyn_masses}
\begin{figure}
\begin{center}
\epsscale{0.8}
\plotone{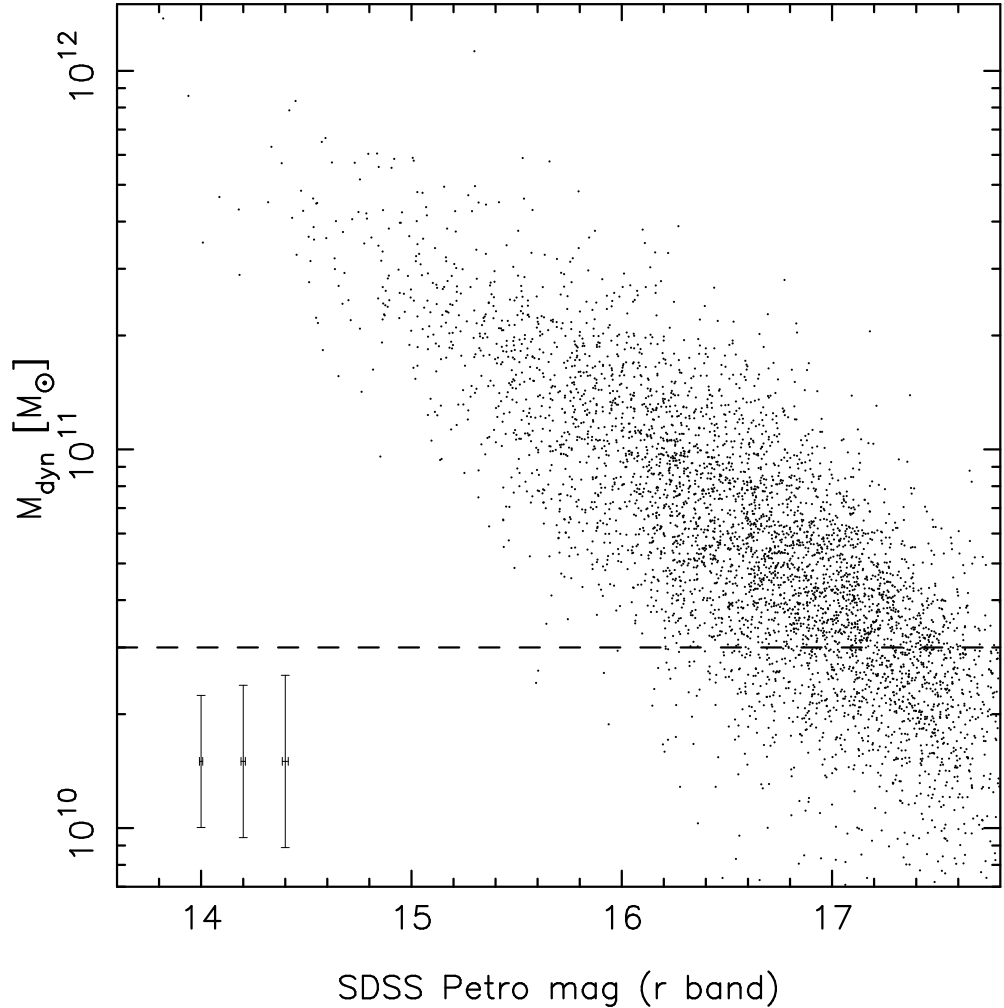}
\caption{Dynamical  mass, for  ETGs in  the optical+NIR  sample,  as a
  function  of the  r-band Petrosian  apparent magnitude,  $m_p$, from
  SDSS.   Considering the scatter  in the  relation, and  assuming the
  sample to be complete down to  $m_p \sim 17.8$, we conclude that the
  ETG's  sample is  complete, with  respect to  $M_{dyn}$,  only above
  $\sim 3 \times 10^{10} M_{\odot}$.  This mass limit is marked by the
  horizontal  dashed line in  the plot.   The vertical  and horizontal
  error bars in the lower-left  of the Figure correspond (from left to
  right) to  {\bf 25-, median, and 75-percentile}  errors on $M_{dyn}$
  and $m_p$.
\label{fig:mpetr_mdyn}
}
\end{center}
\end{figure}

Dynamical masses, $M_{dyn}$, are  obtained from galaxy effective radii
and velocity dispersions.  Applying the virial theorem, we write
\begin{equation}
M_{dyn}= A \times \sigma_0^2 r_{e,K}
\end{equation}
where $r_{e,K}$ is the effective radius  in K-band, and $A$ is a scale
factor that accounts for the internal structure of ETGs. Most previous
works estimated $A$ by assuming a single universal model for the light
distribution in ETGs, such  as the King model~\citep{JFK96}.  However,
this  does not account  for the  variety of  ETG light  profile shapes
\citep[e.g.][]{CCD93}, as parametrized by the Sersic index. To account
for this,  following a similar procedure to  that of~\citet{DOF08}, we
estimate $A$ using de-projected  Sersic galaxy models.  The models and
the de-projection procedure are detailed in~\citet{CLB09}.  The models
describe spherical, non-rotating,  isotropic galactic systems. We also
assume that  the spatial distribution  of dark matter follows  that of
the stellar  component.  Under these assumptions,  each model consists
of a  single Sersic component  characterized by three  parameters: the
half-mass   radius,   the   total   mass,   and   the   Sersic   index
(see~\citet{CLB09} for details).  The use of more complex models (i.e.
those with two-components) is well beyond the scope of this paper, and
we  postpone it  to  a  forthcoming contribution.   For  each ETG,  we
construct the  corresponding Sersic model, using  the effective radius
and Sersic index of the galaxy in K-band. We estimate $A$ by computing
the ratio of the  quantities $M_{dyn}$ and $\sigma_0^2 \times r_{e,K}$
for the given model.  We compute $\sigma_0$ by projecting the model in
two-dimensions,  within a  circular aperture  of radius  $1.5''$, i.e.
the same  aperture as the  SDSS fibers.  Since  we do not  have K-band
structural  parameters for the  optical sample  of ETGs,  $M_{dyn}$ is
only   estimated    for   galaxies   in    the   optical+NIR   sample.
Fig.~\ref{fig:mpetr_mdyn}  shows  the   dynamical  masses  versus  the
Petrosian magnitudes,  from which we  see that our  limiting magnitude
corresponds  to  a  lower  mass   limit  of  $\sim  3  \times  10^{10}
M_{\odot}$.

\subsection{Stellar population properties}
\label{sec:stellar_pop}
As described in Paper I,  we have re-measured velocity dispersions for
ETGs  in  the   SPIDER  sample  using  STARLIGHT~\citep{CID05}.   This
software   determines  the  linear   combination  of   Simple  Stellar
Population (SSP)  models which, broadened with a  given $\sigma$, best
matches the observed galaxy spectrum.   Hereafter, we refer to the set
of SSP models  provided as input to STARLIGHT as  the SSP {\it basis},
and  to  the  output  linear  combination of  SSPs  as  the  synthetic
spectrum.   For the purpose  of remeasuring  the $\sigma_0$'s,  we run
STARLIGHT with  a {\it basis} of  SSPs from the  MILES galaxy spectral
library (Vazdekis  et al.   2010), covering a  wide range of  ages and
metallicities, but with fixed solar abundance ratio ([$\alpha$/Fe]=0).

Here  we  also use  STARLIGHT  to  estimate  the $Age$,  $[Z/H]$,  and
$[\alpha/Fe]$ of each ETG. In  contrast to the more common practice of
using   index-index  diagrams   to  estimate   SSP-equivalent  stellar
population   parameters~\footnote{We  refer  to   {\it  SSP-equivalent
    parameters}  as  those of  the  SSP  model  that best-matches  the
  position of a given  galaxy in the index-index diagrams.}, STARLIGHT
uses all of  the information present in the data, by  fitting a set of
SSP models  to the entire  galaxy spectrum.  A detailed  comparison of
stellar population  parameter estimates from  different techniques for
the SPIDER sample is currently under way, and will be the subject of a
forthcoming paper in this series.  Here, we provide only the essential
information  on how  we use  STARLIGHT to  derive $Age$,  $[Z/H]$, and
$[\alpha/Fe]$.   For   each  galaxy,   we  run  STARLIGHT   using  the
alpha-enhanced     MILES    (hereafter     $\alpha$-MILES)    spectral
library~(Cervantes et. al.~2007). As detailed in App.~\ref{app:MILES},
the $\alpha$-MILES  library extends the MILES  galaxy spectral library
to SSP  models with non-solar  abundance ratios, in  the $[\alpha/Fe]$
range of $-0.2$ to $+0.6$, allowing  the spectra of ETGs to be modeled
with  better accuracy  than that  achieved with  solar-abundance MILES
models.   The  $\alpha$-MILES  SEDs  cover  the  same  spectral  range
($3525-7500$ \AA), with the same spectral resolution (2.3 \AA), as the
(solar  abundance) MILES  library.   Hence, they  are  well suited  to
analyzing  SDSS  spectra, whose  spectral  resolution  is $\sim$  2.36
\AA\  FWHM,   in  the  wavelength  interval   4800-5350~\AA.   To  run
STARLIGHT, we select a {\it  basis} of $176$ $\alpha$-MILES SSPs, with
ages ranging from $1$ to  $\sim 14$~Gyr in eleven steps, metallicities
of    $[Z/H]=-0.68,   -0.38,   0.,    0.2$,   and    enhancements   of
$[\alpha/Fe]=+0.0, +0.20, +0.40, +0.60$.  This grid in $Age$, $[Z/H]$,
and  $[\alpha/Fe]$  allows  us  to  cover  a  wide  range  of  stellar
population  properties without  overly increasing  the  execution time
required  to  run  STARLIGHT~\footnote{For  $1,600$ (out  of  $39,993$
  SPIDER) ETGs, we  found that all SSP models in  the {\it basis}, but
  those with $[\alpha/Fe]=+0.0$, received zero-weight in the STARLIGHT
  synthetic spectra. For these objects, we enlarged the {\it basis} by
  including  also $\alpha$-MILES  SSPs  with $[\alpha/Fe]=-0.20$,  and
  re-ran STARLIGHT  accordingly.}.  For  each galaxy, we  first smooth
the {\it basis} models to match the wavelength-dependent resolution of
the SDSS  spectrum (see  Paper I for  details). STARLIGHT  outputs the
light percentage,  $X$, of  each SSP {\it  basis} model in  the output
synthetic spectrum.   Hence, for a  given property $Y$  (e.g.  $Age$),
one can estimate its {\it luminosity-weighted} value, $Y_L$, as
\begin{equation}
Y_L = \frac{\sum Y \times X}{\sum X},
\label{eq:lum_weight_SP}
\end{equation}
where the summation is performed  over all the SSP {\it basis} models.
The $Age$ is  obtained directly from Eq.~\ref{eq:lum_weight_SP}. Since
metallicity and enhancement are logarithmic quantities, we first apply
Eq.~\ref{eq:lum_weight_SP}    by    setting    $Y_L=10^{[Z/H]}$    and
$Y_L=10^{[\alpha/Fe]}$,  and then  compute the  logarithm of  $Y_L$ in
both cases. The uncertainties on $Age$, $[Z/H]$, and $[\alpha/Fe]$ are
computed by  comparing the estimates  of these quantities  for $2,313$
galaxies  with  repeated  observations  in SDSS,  following  the  same
procedure described in Paper I to obtain the errors on $\sigma_0$. The
average uncertainties are  $\sim 20 \%$ on $Age$,  and $\sim 0.05$ for
$[Z/H]$ and $[\alpha/Fe]$.

\begin{figure*}
\begin{center}
\plotone{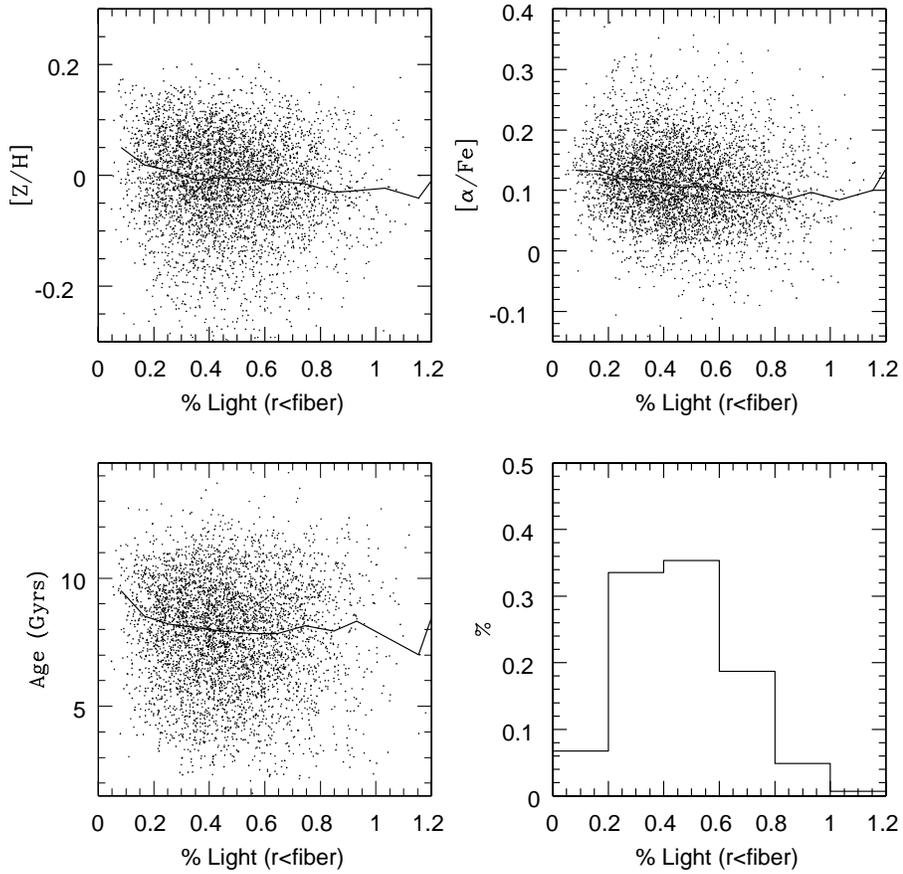}
\caption{Correlation between $Age$ (lower left), $[Z/H]$ (upper left),
  and $[\alpha/Fe]$ (upper right) and the fraction of light inside the
  fiber  {\it  wrt} the  total  light,  L(r$< r_{\rm  fiber}$)/L$_{\rm
    tot}$. The  histogram of  the fraction is  exhibited in  the lower
  right panel. The  solid line displayed in each  panel corresponds to
  the median behavior.
\label{figfrac}
}
\end{center}
\end{figure*}

{ An important  potential bias when using the  spectra from SDSS is
  the effect  of a  fixed spectroscopic fiber  aperture on  the $Age$,
  $[Z/H]$, and  $[\alpha/Fe]$ estimates. The fraction  of galaxy light
  inside the fiber will vary with galaxy size and redshift. Because we
  select  the  SPIDER sample  based  on  the  $eClass$ parameter,  the
  galaxies  are selected  to have  passive spectra  in  their centers.
  This means  that, as discussed  in Paper I, our  putative early-type
  sample could  be contaminated by  early spiral systems,  i.e. spiral
  galaxies with  a prominent  bulge component. Indeed,  in Paper  I we
  show that  this contamination is about  15\%, falling to  ~5\% for a
  higher  quality  sample  defined   on  the  basis  of  visual  image
  classification.  As we discuss at the end of Sec.~8.3 (Fig.~16), the
  trends  with  stellar population  properties  remain unchanged  when
  restricting   the   SPIDER  sample   to   the  lower   contamination
  subsample. The  converse might also  hold, i.e. we might  be missing
  ETGs   with   strong    emission   features   localized   to   their
  cores.  However,  very  few   ETGs  are  expected  to  have  spectra
  exhibiting such strong central emission.

In Fig.~\ref{figfrac} we display how $Age$, $[Z/H]$, and $[\alpha/Fe]$
vary with the  fraction of light inside the fiber  {\it wrt} the total
light,   L(r$<   r_{\rm  fiber}$)/L$_{\rm   tot}$,   along  with   the
distribution of this fraction, showing that $\sim$65\% of the galaxies
have a  ratio between $0.2$  and $0.6$. The median  trends, negligible
compared  to the scatter  in $Age$,  $[Z/H]$, and  $[\alpha/Fe]$, show
that the derived stellar population parameters are not affected by the
aperture   effect,    reinforcing   what   we    have   presented   in
App.~\ref{sec:apcor}, i.e.   that the average $r_{e}$  does not change
much {\it  wrt} the stellar  population parameters, implying  that the
correlation of color gradients  with stellar population properties has
no dependence on $r_{e}$.}

The values  of $Age$,  $[Z/H]$, and $[\alpha/Fe]$  used here  are {\em
not} SSP-equivalent parameters, as  commonly used in the literature,
but  are  luminosity-weighted  quantities  inferred  from  the  galaxy
spectra.  As mentioned above,  a detailed comparison of SSP-equivalent
and luminosity-weighted parameters will  be performed in a forthcoming
contribution.

\section{Color gradient estimates}
\label{sec:cg_estimates}
Internal  color gradients  are estimated  using the  same  approach as
~\citet[][hereafter LdC09]{LdC:09}, using  the two-dimensional fits to
the  $grizYJHK$  galaxy  images  with seeing-convolved  Sersic  models
(\papdata).  Color  gradients are defined as the  logarithmic slope of
the galaxy radial  color profile, $\nabla_{W-X}={d \!(W-X)}/{d \!(\log
  \rho)}$, where $X$ and $W$ are  any two wavebands, and $\rho$ is the
distance to  the galaxy center.   From $grizYJHK$ data we  can measure
seven  independent color  gradients.  As  in LdC09,  we  calculate the
color gradients between  the $g$ band and all  other wavebands, in the
form of $\nabla_{g-X}$. For each galaxy and for each band, we estimate
the mean surface brightness of  the de-convolved Sersic model on a set
of concentric  ellipses, whose  ellipticities and position  angles are
fixed to those from the $r$-band Sersic fit.  The ellipses are equally
spaced in equivalent radius $\rho$ by $0.01 r_{e,r}$ , where $r_{e,r}$
is the the r-band effective radius.   The color index $g-X$ at a given
radius $\rho$ is obtained by subtracting the mean surface brightnesses
at the  corresponding ellipse.   Each color profile  is fitted  in the
radial  range of $\rho_{min}  = 0.1  r_e$ to  $\rho_{max} =  r_e$ (see
e.g.  ~\citealt{PVJ90})  using  an  orthogonal least  squares  fitting
procedure.  The  slope of the best-fitting profile  gives the gradient
$\nabla_{g-X}$.    For  the   optical   sample  of   ETGs,  only   the
$\nabla_{g-r}$, $\nabla_{g-i}$, and $\nabla_{g-z}$ color gradients are
estimated. As  discussed in LdC09,  small changes in the  radial range
used  in  this  computation  do  not significantly  change  the  color
gradient estimates.

To estimate the uncertainties  on $\nabla_{g-X}$ we repeated the above
procedure by  shifting \re  \, and  \sn \, in  each band  according to
their  measurement  errors,  taking  into  account  the  corresponding
covariance term (\papdata).  We repeat this using $N=1000$ shifts, and
the errors on $\nabla_{g-X}$ are  then obtained from the widths of the
distributions of $\nabla_{g-X}$.

\section{Distribution of color gradients}
\label{sec:dist_cgs}
\begin{figure*}
\begin{center}
\plotone{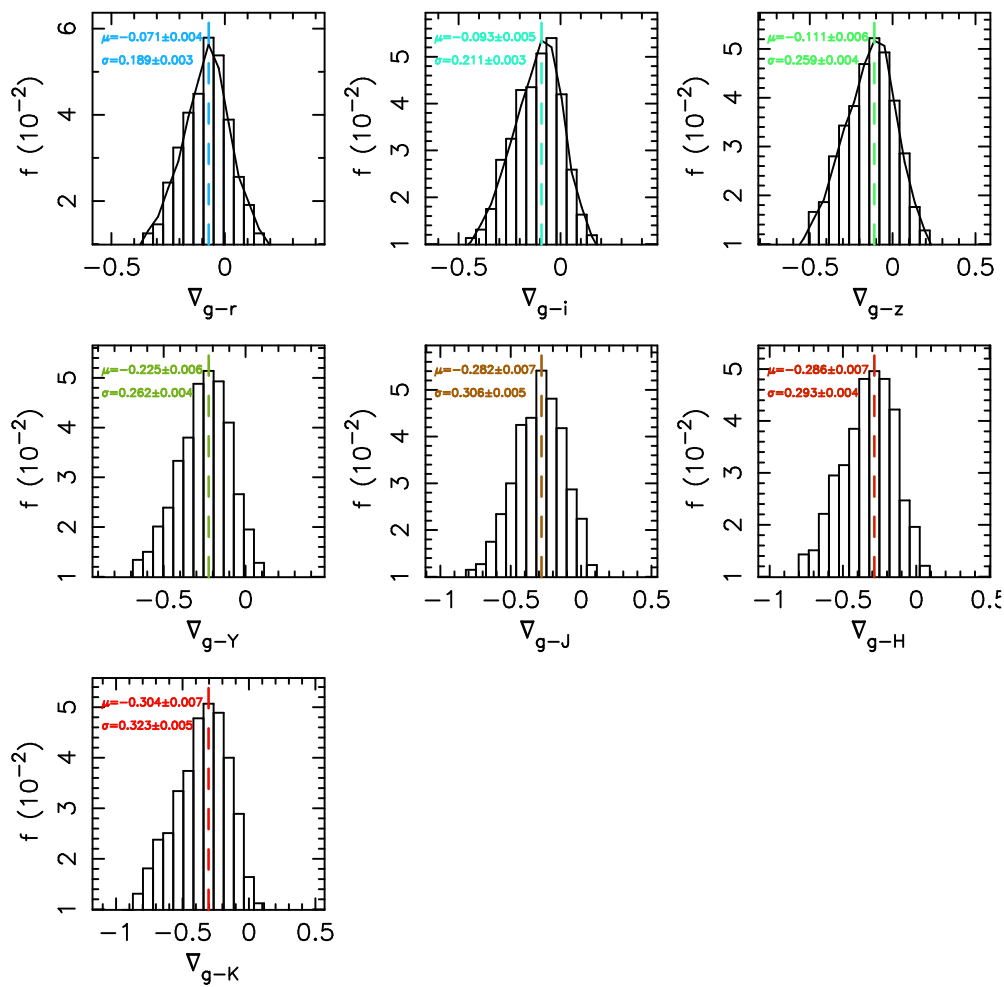}
\caption{Distribution  of optical+NIR  color gradients  of  the SPIDER
  ETGs.  From left to right and top to bottom, the histograms show the
  distributions of  $\nabla_{g-X}$ color gradients,  with $X=rizYJHK$,
  for the  optical+NIR sample  of ETGs. On  the y-axis, we  report the
  fraction $f$ of galaxies  in the different $\nabla_{g-X}$ bins.  The
  peak and width  of each distribution are reported  in the upper-left
  corner  of each panel,  together with  the corresponding  $1 \sigma$
  uncertainties.  The  vertical dashed lines mark the  position of the
  peaks.   The solid black  curves in  the three  top panels  show the
  distributions of  $\nabla_{g-r}$, $\nabla_{g-i}$, and $\nabla_{g-z}$
  color gradients, for the optical sample of ETGs.
\label{cgrad_hist}
}
\end{center}
\end{figure*}

Fig.~\ref{cgrad_hist} plots the distributions of $g-X$ color gradients
for  both the  optical  and  optical+NIR samples  of  ETGs.  For  each
distribution, we  compute the corresponding peak and  width ($\mu$ and
$\sigma$) using  bi-weight statistics~\citep{Beers:90}. The  values of
$\mu$  and $\sigma$, with  the corresponding  bootstrap uncertainties,
are    reported   in    Fig.~\ref{cgrad_hist},   as    well    as   in
Tabs.~\ref{tab:stat_cgrad_optNIR} and~\ref{tab:stat_cgrad_opt} for the
optical+NIR and  the optical samples  of ETGs, respectively.   All the
distributions  are sharply  peaked. The  peak value  of $\nabla_{g-X}$
smoothly decreases  from about  $-0.07$ in $g-r$  to $-0.3$  in $g-K$,
i.e. the typical color  gradient of ETGs becomes increasingly negative
as we enlarge  the wavelength baseline over which  it is computed.  It
is remarkable that the  peak values of $\nabla_{g-r}$, $\nabla_{g-i}$,
and  $\nabla_{g-z}$,   as  well  as  the   corresponding  widths,  are
consistent  between the  optical  (i.e. Tab.~\ref{tab:stat_cgrad_opt})
and  optical+NIR (i.e.   Tab.~\ref{tab:stat_cgrad_optNIR})  samples of
ETGs. Fig.~\ref{cgrad_hist} also shows  that the distribution of $g-r$
through $g-z$ color gradients is consistent between the optical (solid
curve in  the Figure) and the  optical+NIR samples.  The  width of the
$\nabla_{g-X}$  distributions smoothly  increases  from $g-r$  through
$g-K$. This trend  might be due to larger  errors on the near-infrared
structural parameters  than on  the optical ones  (\papdata), implying
that  the typical  uncertainty on  color gradients  is larger  for the
optical--NIR    rather     than    optical--optical    $\nabla_{g-X}$.
Tabs.~\ref{tab:stat_cgrad_optNIR}     and    ~\ref{tab:stat_cgrad_opt}
provide   the   median   uncertainties   on   the   color   gradients,
$\sigma^{err}$,  obtained  by  propagating  the errors  on  structural
parameters  (Sec.~\ref{sec:cg_estimates}).  We estimate  the intrinsic
width of the  $\nabla_{g-X}$ distributions, $\sigma^i$, by subtracting
in  quadrature $\sigma^{err}$  from  $\sigma$, i.e.   $\sigma^i=\left(
\sigma^2-(\sigma^{err})^2\right)^{1/2}$,  with   the  constraint  that
$\sigma^i \ge 0$. The corresponding intrinsic scatters are reported in
Tabs.~\ref{tab:stat_cgrad_optNIR} and ~\ref{tab:stat_cgrad_opt}.  {
  It should be  noted that this procedure is only  a simplistic way of
  estimating  the internal  scatter,  as it  holds  only for  Gaussian
  distributions,  which  is  not  the  case  for  the  color  gradient
  distributions  (see Fig.~3).}  The  $\sigma^i$ {  increases} from
$g-r$ through  $g-K$, implying  that the distribution  of optical--NIR
color gradients has a  significant intrinsic dispersion. The origin of
this  dispersion is  analyzed in  Sec.~\ref{sec:cg_galpars},  where we
study  the  dependence of  stellar  population  gradients  in ETGs  on
photometric  and  spectroscopic properties.   We  also  note that  the
fraction  of  galaxies  having  inverted  (positive)  color  gradients
becomes less and less pronounced when we move from optical--optical to
optical--NIR wavebands.  For $\nabla_{g-K}$, we find that only a small
fraction  of  ETGs  ($\sim  8\%$)  exhibit  positive  color  gradients
($\nabla_{g-K}>0$).   {  From  the  peak  value  of  $\nabla_{g-K}$
  ($-0.304$)   and   the  intrinsic   width   of  the   $\nabla_{g-K}$
  distribution ($0.187$, see Tab.~2), we estimate that only $\sim 5\%$
  of galaxies in the optical+NIR  sample are expected to have positive
  $g-K$  gradients.  Hence,  the  presence of  a  (small) fraction  of
  objects with  positive gradients is  a real feature, and  not merely
  explained by  galaxies with negative gradients  being scattered into
  the positive region by the measurement errors.}

The  peak   values  of  $\nabla_{g-X}$  are   consistent,  within  the
uncertainties,  with   those  estimated  by La Barbera  \& de Carvalho (2009) (hereafter
LdC09) for the  entire SPIDER sample of $5,080$  ETGs with optical+NIR
data available.  { A  discrepancy} is found for $\nabla_{g-z}$, for
which  LdC09 measured  a peak  value of  $0.089 \pm  0.006$,  while we
measure  $-0.111 \pm  0.006$.   This difference  does  not affect  the
results of LdC09, since our value of $-0.111$ is fully consistent with
those  expected from  the  best-fitting stellar  population models  of
color  gradients in LdC09  (see their  Fig.~1).  The  main differences
between the  sample of  optical+NIR color gradients  in this  work and
that  of LdC09  are  that (i)  we  select only  the  ETGs with  better
$\chi^2$  and  more  accurate  structural  parameters,  and  (ii)  the
computation of  color gradients in  LdC09 was slightly  different than
that performed  here.  We  derive the color  profiles using,  for each
band, the  two-dimensional Sersic model whose axis  ratio and position
angle parameters  are those measured in  that band. In  LdC09, we {
  assumed the} two dimensional Sersic  models in all bands to have the
same axis ratios  and position angles as those  in the $r$-band.  This
also explains  why the color  gradient distributions are  broader than
those measured in LdC09 (see their Tab.~1).

\begin{deluxetable}{l|c|c|c|c}
\small
 \tablecaption{Statistics of color gradients for the optical+NIR sample.}
\tabletypesize{\scriptsize}
\tablewidth{0pt}
\tablecolumns{5}
\tablehead{ color &  $\mu$ & $\sigma$ & $ \sigma^{err}$ & $\sigma^{i}$ \\
    gradient &  & & & }
\startdata
$\nabla_{g-r}$ & $ -0.071 \pm 0.004$ & $0.189 \pm 0.003$ & $0.209$ & $0.000  $ \\
$\nabla_{g-i}$ & $ -0.093 \pm 0.005$ & $0.211 \pm 0.003$ & $0.192$ & $0.087  $ \\
$\nabla_{g-z}$ & $ -0.111 \pm 0.006$ & $0.259 \pm 0.004$ & $0.220$ & $0.136  $ \\
$\nabla_{g-Y}$ & $ -0.225 \pm 0.006$ & $0.262 \pm 0.004$ & $0.228$ & $0.128  $ \\
$\nabla_{g-J}$ & $ -0.282 \pm 0.007$ & $0.306 \pm 0.005$ & $0.268$ & $0.147  $ \\
$\nabla_{g-H}$ & $ -0.286 \pm 0.007$ & $0.293 \pm 0.004$ & $0.277$ & $0.094  $ \\
$\nabla_{g-K}$ & $ -0.304 \pm 0.007$ & $0.323 \pm 0.005$ & $0.264$ & $0.187  $ \\
\enddata
\label{tab:stat_cgrad_optNIR}
\end{deluxetable}

\begin{deluxetable}{l|c|c|c|c}
\small
 \tablecaption{Statistics of color gradients for the optical sample.}
\tabletypesize{\scriptsize}
\tablewidth{0pt}
\tablecolumns{5}
\tablehead{ color &  $\mu$ & $\sigma$ & $ \sigma^{err}$ & $\sigma^{i}$ \\
    gradient &  & & & }
\startdata
$\nabla_{g-r}$ & $ -0.074 \pm 0.001$ & $0.195 \pm 0.001$ & $0.210$ & $0.000  $ \\
$\nabla_{g-i}$ & $ -0.093 \pm 0.001$ & $0.213 \pm 0.001$ & $0.195$ & $0.087  $ \\
$\nabla_{g-z}$ & $ -0.112 \pm 0.002$ & $0.261 \pm 0.002$ & $0.222$ & $0.137  $ \\
\enddata
\label{tab:stat_cgrad_opt}
\end{deluxetable}

\section{From color to stellar population gradients}
The  color   gradient  reflects   the  radial  variation   of  stellar
populations in a  galaxy. We can combine different  estimates of color
gradients  from various  pairs  of available  wavebands  to infer  the
underlying  stellar population gradient  in an  ETG.  The  approach we
follow is presented, in general terms, in Sec.~\ref{sec:approach}.  In
Sec.~\ref{sec:age_met_constrain} we describe  how this approach can be
applied when the variation  of stellar population properties is mainly
due  to  age and  metallicity.  We discuss  the  role  of the  filters
available to SPIDER in  reducing the age--metallicity degeneracy.  The
results  of  this section  are  then  applied  to the  color  gradient
estimates  in Sec.~\ref{sec:grads} to  constrain radial  variations of
age and metallicity within ETGs.

\subsection{Measuring variations of stellar population parameters}
\label{sec:approach}
We  consider  two   stellar  populations  characterized  by  different
physical parameters,  such as age,  metallicity, $\alpha-$enhancement,
etc. In general, the problem is how one can use the difference between
the  color  indices  of  the  two stellar  populations  to  infer  the
corresponding difference(s)  in physical parameters.   The color index
difference  may be:  (i) radial  - the  internal color  gradient; (ii)
between  galaxies  with  different  masses  -  the  Fundamental  Plane
coefficients  in different  wavebands  (see La  Barbera et  al.~2010b;
hereafter  Paper  II);  (iii)  in   luminosity  -  the  slope  of  the
color-magnitude relation; or (iv) in the residuals in luminosity about
a given  scaling relation. We  first describe the general  approach we
adopt, and then, in  Sec.~\ref{sec:grads}, we focus on its application
to point (i).

We indicate as  $\Delta_i$ the variation of a  given physical property
between the two stellar populations, where the index $i$ runs over the
considered  set  of  properties.   The two  stellar  populations  have
differences in  colors $\delta_j$, where  $j$ spans the set  of colors
defined by  the available wavebands.  One can  introduce the following
approximations:
\begin{eqnarray}
 \delta_j & = & \sum_i A_{ij} \cdot \Delta_i  \label{deltaeq} 
\label{eq:deltaj}
\end{eqnarray}
where $A_{ij}$  is the  partial derivative of  the j$-th$  color index
with respect  to the i$-th$  stellar population parameter.   The above
equation holds if (a) the  color indices are well behaved functions of
the stellar  population parameters and either (b1)  the $\Delta_i$ are
small  enough to  assume  a  linear approximation  or  (b2) the  color
indices are linear functions of the stellar population parameters. The
quantities $A_{ij}$ have to be  evaluated at the average values of the
stellar  population   parameters  of  the   two  stellar  populations.
Eqs.~\ref{eq:deltaj} can be solved  in a $\chi^2$ sense, by minimizing
the expression:
\begin{eqnarray}
 \chi^2 & = & \sum_i \left( \delta_i - \sum_k A_{ki} \cdot \Delta_k \right)^2. \label{chi2} 
\end{eqnarray}
Requiring that the derivatives  of $\chi^2$ with respect to $\Delta_j$
vanish, one obtains the following system of equations:
\begin{eqnarray}
 \sum_k \left( \sum_i A_{ki} A_{ji} \right) \Delta_k & = & \sum_i A_{ji} \delta_i. \label{chi2sol} 
\end{eqnarray}
Setting  $M_{jk}=\left(  \sum_i  A_{ki}  A_{ji} \right)  \cdot  \left(
\sum_i A_{ji}  \delta_i \right)^{-1}$, we  obtain a new system  of $N$
linear  equations in  the  $N$ unknown  stellar population  quantities
$\Delta_j$:
\begin{eqnarray}
 \sum_j M_{ij} \cdot \Delta_j  = 1 \label{deltaeqsol} 
\label{deltaeq2}
\end{eqnarray}
where the  quantities $M_{ij}$ are  completely defined by  the partial
derivatives $A_{ij}$ and the measured color differences $\delta_i$. If
the matrix $M_{ij}$ is non-degenerate, we can invert the linear system
of Eqs.~\ref{deltaeqsol} and derive  a unique solution for the unknown
quantities $\Delta_i$ (e.g. differences in age, metallicity, ...) from
the measured color differences $\delta_j$.

\subsection{Constraining age and metallicity variations}
\label{sec:age_met_constrain}
In this section, we investigate how the color indices in SPIDER can be
used to constrain variations of stellar population properties in ETGs,
specifically  the effects of  age and  metallicity.  Since  we analyze
color       gradients      in       the       form      $\nabla_{g-X}$
(Sec.~\ref{sec:cg_estimates}), we  consider here color  indices of the
form $g-X$, where $X$ is  one of $rizYJHK$.  The $g$-band includes the
4000 \AA~ break  at the median redshift of  the SPIDER samples, making
the $g-X$ colors  most sensitive to the effects  of stellar population
variations.

For the optical sample, each ETG is characterized by three independent
colors, $g-r$,  $g-i$, and $g-z$.  Assuming that  the relevant stellar
population   parameters  are   the  age   $t$  and   metallicity  $Z$,
equations~\ref{deltaeq} can be re-written as
\begin{eqnarray}
 \delta_{g-r} & = & \frac{\partial (g-r)}{\partial {\log t}} \cdot \Delta_{t} + \frac{\partial (g-r)}{\partial {\log Z}} \cdot \Delta_{Z}  \label{age_met_system_a} \\ 
 \delta_{g-i} & = & \frac{\partial (g-i)}{\partial {\log t}} \cdot \Delta_{t} + \frac{\partial (g-i)}{\partial {\log Z}} \cdot \Delta_{Z}  \label{age_met_system_b} \\ 
 \delta_{g-z} & = & \frac{\partial (g-z)}{\partial {\log t}} \cdot \Delta_{t} + \frac{\partial (g-z)}{\partial {\log Z}} \cdot \Delta_{Z}. \label{age_met_system_c}
\label{eq:gz_eqs}
\end{eqnarray}
where $\Delta_t$ and $\Delta_Z$ are the logarithmic differences of age
and metallicity  between two stellar populations. As  noted above (see
beginning of  Sec.~\ref{sec:approach}), the quantities $\delta_{g-r}$,
$\delta_{g-i}$, and  $\delta_{g-z}$ could be the  differences of color
indices  between  galaxies  having  different parameters  (e.g  mass),
although  here  we focus  on  the  case  where $\delta_{g-X}$  is  the
difference in  color index  per radial decade  in a galaxy,  i.e.  the
internal color gradient  ($\delta_{g-X} \equiv \nabla_{g-X}$). In this
case, the quantities $\Delta_t$  and $\Delta_z$ provide the difference
in  age and  metallicity  per radial  decade  (i.e.  $\Delta_t  \equiv
\nabla_t$      and      $\Delta_z      \equiv      \nabla_z$,      see
Sec.~\ref{sec:calc_age_met_grads}).     Using    the   formalism    of
Sec.~\ref{sec:approach}, $\Delta_{t}$ and  $\Delta_{Z}$ are derived by
solving the linear system:
\begin{eqnarray}
 M_{t,t} \Delta_{t} + M_{t,Z} \Delta_{Z} & = & 1 \label{age_met_solve_a} \\
 M_{Z,t} \Delta_{t} + M_{Z,Z} \Delta_{Z} & = & 1 \label{age_met_solve_b}
\label{eq:M_gz_eqs}
\end{eqnarray}
where  the  matrix $  M$  is  computed  directly from  the  partial
derivative of color indices with respect to $\log t$ and $\log Z$, and
the  measured color  differences  $\delta_{g-r}$, $\delta_{g-i}$,  and
$\delta_{g-z}$.   For the optical+NIR  sample of  ETGs, the  system of
Eqs.~\ref{age_met_system_a}--~\ref{age_met_system_c}   includes   four
extra equations,  corresponding to the $g-Y$, $g-J$,  $g-H$, and $g-K$
color indices.  The quantities $\Delta_{t}$ and $\Delta_{Z}$ are still
obtained         by        solving        the         system        of
Eqs.~~\ref{age_met_solve_a}--\ref{age_met_solve_b},  where  the matrix
$ M$  is defined  by using the  color differences in  all available
wavebands.

\begin{figure}
\begin{center}
\plotone{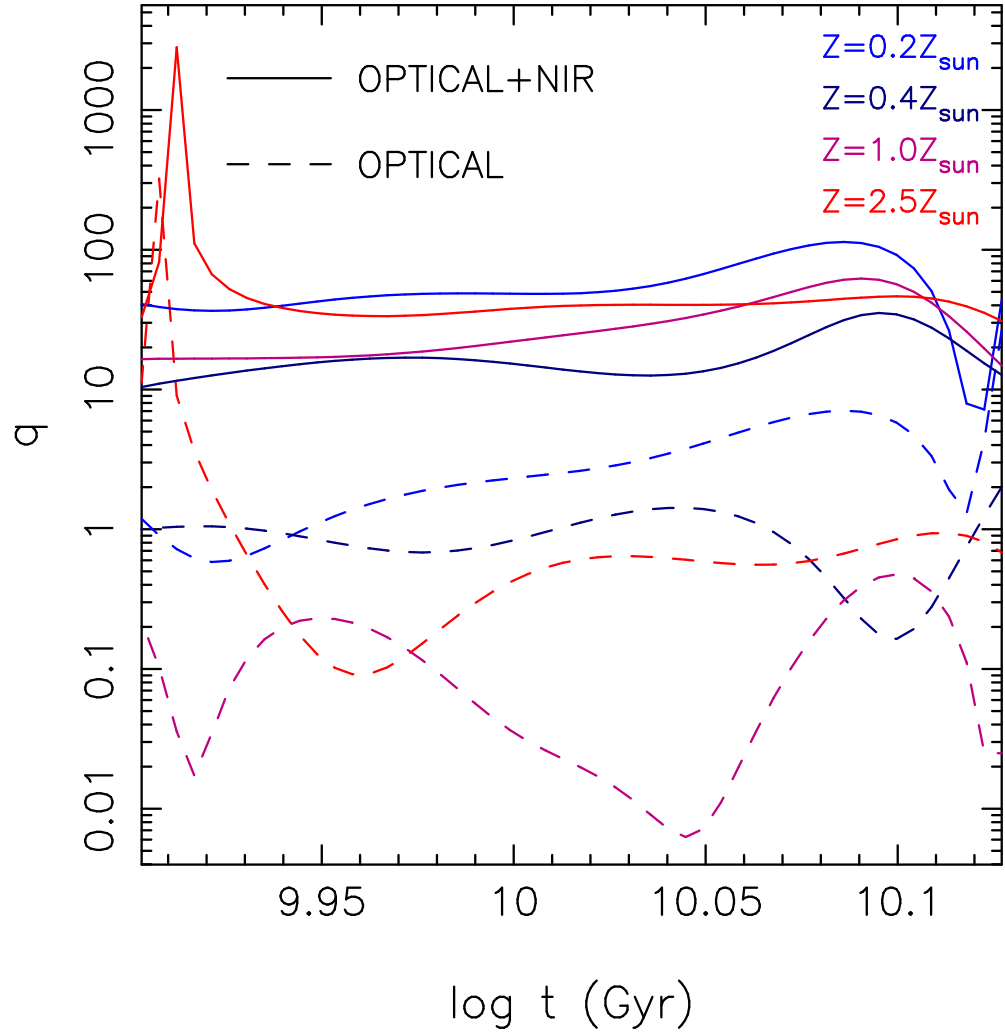}
\caption{Degeneracy, $q$,  in deriving age  and metallicity variations
  from  color indices  as a  function of  the logarithmic  age  of SSP
  models from the BC03  synthesis code. Different colors correspond to
  different metallicities $Z$, as  shown in the upper--right corner of
  the plot.   Dashed and  solid curves correspond  to the  cases where
  optical and optical+NIR colors are used, respectively.
\label{qval}
}
\end{center}
\end{figure}

Since  the NIR  light is  far less  sensitive to  metallicity (through
line-blanketing) than the optical, and  is also less sensitive to age,
we expect that including the NIR wavebands helps to (partly) break the
age-metallicity degeneracy,  i.e. the degeneracy of the  matrix $M$ is
reduced when including all the  $g-X$ color indices.  To examine this,
we  set  $\delta_i   =  0.01$  as  a  typical   uncertainty  in  color
differences. { We adopt this value  of $\delta_i$ as it is an upper
  bound  for  the typical  uncertainty  on  the  mean value  of  color
  gradients.  In  fact, the typical error  on the color  gradient of a
  given  galaxy is  $\sim 0.2$--$0.3$  (Tab.~2).  For  the optical+NIR
  sample of ETGs, this implies a typical uncertainty on the mean color
  gradient  of $\sim  0.003$--$0.005$~\footnote{  These values  are
    obtained by dividing the  typical errors on color gradients ($\sim
    0.2$--$0.3$), by  the root square  of the size of  the optical+NIR
    sample  of ETGs.   Notice that  the resulting  expected  errors on
    color gradients are slightly smaller than those on the peak values
    of $\nabla_{g-X}$ ($\sim 0.004$--$0.007$, see Tab.~2), as the peak
    values  are estimated  by  the bi-weight  (rather  than the  mean)
    statistics.  }.}   In order to  estimate the derivatives  of color
indices with  respect to  age and metallicity,  we adopt  a polynomial
approximation  of the {  magnitude} of  a given  stellar population
model with respect to $t$  and $Z$.  Details of this approximation are
provided in App.~\ref{sec:polfit}.  Here,  we consider SSP models from
the~\citet{BrC03} synthesis  code (hereafter  BC03) with a  Scalo IMF.
We compute the  matrix $ M$ for different values of  $t$ and $Z$ of
these  SSP models,  in the  range where  the  polynomial approximation
holds, i.e. $ 5\!<\! t (Gyr)\!<\!13.5$ and $ 0.2 \! < \!Z/Z_{\odot} \!
< \!  2.5$.  Fig.~\ref{qval} plots  the determinant of the matrix $M$,
$q=det(   M)$,  as   a   function  of   $\log   t$  for   different
metallicities. The case where only optical colors are used is shown by
the dashed  curves.  $q$  is close to  one for  non-solar metallicity,
while it is always smaller than  one (by almost an order of magnitude)
in the case of $Z=Z_{\odot}$.  This implies that for solar metallicity
the  matrix $  M$  is  nearly degenerate  and  hence, as  expected,
optical   colors  alone   are  not   able  to   effectively  constrain
age--metallicity variations.   The solid  curves show the  cases where
all the optical+NIR colors are  used.  $q$ is then always greater than
$\sim10$.  In particular, for solar  metallicity, $q$ is two orders of
magnitude greater than with  optical colors alone, implying that using
all of the $grizYJHK$ wavebands  places much better constraints on age
and metallicity variations.

\begin{figure}
\begin{center}
\plotone{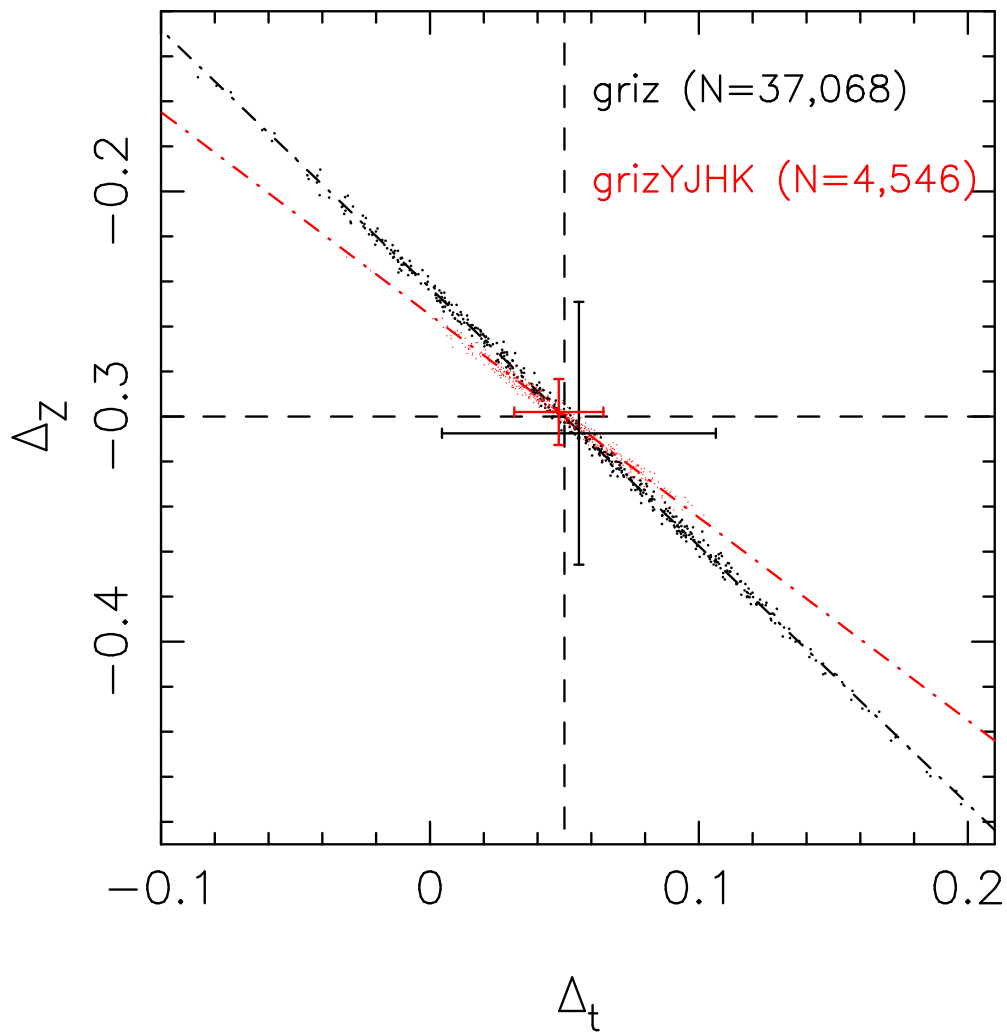}
\caption{Measured   $\Delta_t$   and   $\Delta_Z$   from   Monte-Carlo
  simulations, where one shifts color indices according to the typical
  uncertainties  on color  gradient estimates,  and recovers  the true
  $\Delta_t$   and   $\Delta_Z$   via   the  approach   described   in
  Sec.~\ref{sec:age_met_constrain}.   Different  colors correspond  to
  cases where optical (black)  and optical+NIR (red) color indices are
  considered. The  error bars mark the $1~\sigma$  standard errors for
  the  means   of  $\Delta_t$   and  $\Delta_Z$  recovered   from  the
  Monte-Carlo realizations. { The horizontal and
vertical  dashed  lines  mark   the  true  values  of  $\Delta_t$  and
$\Delta_Z$.   The dash-dotted lines  are obtained  from least-squares
orthogonal fitting of $\Delta_Z$ vs.  $\Delta_t$.}
\label{age_met_var}
}
\end{center}
\end{figure}

To illustrate  this point more quantitatively,  we perform Monte-Carlo
simulations, setting  $\Delta_t=0.05$ and $\Delta_Z=-0.3$,  { which
  correspond to  two stellar populations  with a difference in  age of
  $\sim 11 \%$ and a factor  of two difference in metallicity. } These
values of $\Delta_t$ and $\Delta_Z$ are close to those of the mean age
and  metallicity internal gradient  of ETGs  measured by  LdC09. Using
{   Eqs.~\ref{eq:deltaj}},  we   compute  the   color  differences,
$\delta_i$,  between the  two stellar  populations for  all  the seven
$g-X$ colors.  The quantities  $A_{ij}$ are computed for $t=10Gyr$ and
$Z=Z_{\odot}$.   For each  iteration,  the color  indices are  shifted
according to a normal  distribution with a width of $\sigma_{\delta}$,
and  then we  estimate $\Delta_t$  and $\Delta_Z$  with  the procedure
outlined  above.   The  procedure  is  repeated  twice,  adopting  (i)
$\sigma_{\delta}=0.002$  and using  only the  $griz$ colors,  and (ii)
with  $\sigma_{\delta}={0.005}$  and all  the  $grizYJHK$ bands.   The
above $\sigma_{\delta}$'s  represent the typical  uncertainties on the
mean internal  color gradients in  the optical and  optical+NIR SPIDER
samples  of   ETGs  ({  see   the  errors  on  $\mu$   reported  in
  Tab.~\ref{tab:stat_cgrad_opt}  and Tab.~\ref{tab:stat_cgrad_optNIR},
  respectively)}.   Fig.~\ref{age_met_var} shows the  distributions of
inferred $\Delta_t$ and $\Delta_Z$.  In  both cases, there is a strong
correlation between $\Delta_t$ and $\Delta_Z$, implying that even with
the  full  $grizYJHK$ filter  set,  one  cannot  completely break  the
age-metallicity degeneracy.  However,  the amplitude of the correlated
variation is much smaller in the  case where the optical and NIR bands
are  adopted. In  fact, although  for case  (ii)  $\sigma_{\delta}$ is
larger than  in case (i),  the corresponding error bars  on $\Delta_t$
and $\Delta_Z$  are much  smaller, only $\sim  0.01$ for both  $t$ and
$Z$.  In other terms, optical+NIR colors constrain the size of the age
and     metallicity     differences     more    effectively.      From
Fig.\ref{age_met_var}, we  conclude that even a small  age gradient of
$\sim 11 \%$ can be detected at $\sim 2.5 \sigma$ with the optical+NIR
dataset.

\section{Stellar population gradients}
\label{sec:grads}
In this section, we present our approach to constraining the variation
of stellar  population parameters with  distance from the  ETG center.
We consider only the effects of age and metallicity for simplicity, as
the proper characterization of  a given stellar population may require
other    quantities   (e.g.    the    shape   of    the   IMF).     In
App.~\ref{sec:age_met_bias}, we  describe the role  of dust extinction
gradients in  ETGs, discussing their  possible impact on  our results.
We first  describe how  we estimate the  internal age  and metallicity
gradients           (Sec.~\ref{sec:calc_age_met_grads}).            In
Sec.~\ref{sec:eff_grad}, we show how  our approach allows us to define
an {\it effective} color gradient, $\nabla_\star$, that results from a
combination of all the  available color gradients.  The systematics in
\nablas      \,     estimates      are      then     discussed      in
Sec.~\ref{sec:err_eff_grad}.

\subsection{Inferring the age and metallicity gradients}
\label{sec:calc_age_met_grads}
We adopt  the procedure described  in Sec.~\ref{sec:age_met_constrain}
to  infer the  radial gradients  in age  ($\nabla_t$)  and metallicity
($\nabla_Z$).     Setting    $\delta_{g-X}    =    \nabla_{g-X}$    in
Eqs.~\ref{age_met_system_a}--~\ref{age_met_system_c},  the  quantities
$\Delta_t$ and  $\Delta_Z$ become  the logarithmic differences  in age
and  metallicity per radial  decade inside  a galaxy,  i.e.  $\Delta_t
\equiv      \nabla_t$     and     $\Delta_Z      \equiv     \nabla_Z$.
Eqs.~\ref{age_met_system_a}--~\ref{age_met_system_c}      are     then
rewritten as follows:
\begin{eqnarray}
 \nabla_{g-X} & = & \frac{\partial (g-X)}{\partial {\log t}} \cdot \nabla_{t} +
\frac{\partial (g-X)}{\partial {\log Z}} \cdot \nabla_{Z},
\label{eq:grad_gX}
\end{eqnarray}
with  $X=rizYJHK$.  As  shown in  Sec.~\ref{sec:age_met_constrain}, in
order to derive $\nabla_t$ and $\nabla_Z$, one has to solve the linear
system:
\begin{eqnarray}
M_{t,t} \nabla_{t} + M_{t,Z} \nabla_{Z} & = & 1 \label{eq:age_met_solve_a2} \\
M_{Z,t} \nabla_{t} + M_{Z,Z} \nabla_{Z} & = & 1 \label{eq:age_met_solve_b2}
\end{eqnarray}
where  the matrix $  M$ is  defined as  in Sec.~\ref{sec:approach},
setting  $\delta_j =  \nabla_{g-X}$.  The  computation of  $  M$ is
performed using  different stellar population models.  For each model,
we  use Eqs.~\ref{eq:age_met_solve_a2}  and ~\ref{eq:age_met_solve_b2}
to derive the corresponding $\nabla_t$ and $\nabla_Z$. This procedure,
as  presented  in Sec.  6,  minimizes  the age-metallicity  degeneracy
depending on how many color terms are available and how they track the
most conspicuous spectral features (e.g.  $g-r$ probes the 4000 \AA \,
break,  yielding important information  on the  age of  the underlying
stellar population).

We consider SSP models from three different sources: BC03, \citet{M05}
(M05), and  Charlot and Bruzual  (2009, in preparation;  CB10).  These
models are based on  different synthesis techniques and have different
IMFs.  The M05 model uses the fuel consumption approach instead of the
isochrone synthesis of  BC03 and CB10. The CB10  code implements a new
AGB  phase  treatment~\citep{MG07}.    The  IMFs  are:  Scalo  (BC03),
Chabrier (M05), and Salpeter (CB10). Moreover, we also use a composite
stellar population  model from BC03 having  exponential star formation
rate  (SFR)   with  an  e-folding  time   of  $\tau=1$~Gyr  (hereafter
$BC03_{\tau=1Gyr}$).   The models  are convolved  with  the $grizYJHK$
throughput  curves,  and  the  polynomial approximation  described  in
App.~\ref{sec:polfit}  is used  to calculate  the  partial derivatives
entering  in $M$.   The derivatives  are computed  assuming an  age of
$t=9.27$~Gyr, corresponding to a  formation redshift of $z=2.5$ at the
median redshift  of the SPIDER sample ($<z>_{\rm  median} \sim 0.07$),
and  solar metallicity  ($Z=Z_{\odot}$). For  each  stellar population
model and each galaxy, we perform $N=1000$ iterations where in each of
them  we  perturb  the  observed  color  gradients  according  to  the
corresponding uncertainties.  The errors  on \nablat \, and \nablaz \,
are  then  the  widths  of  their  distributions  resulting  from  all
iterations. Note  that in  the above procedure  all of the  models are
computed at the median redshift of our sample ($z=0.0725$), neglecting
the     (small)    redshift     range    of     the     sample.     In
App.~\ref{sec:age_met_bias}, we show  that this approximation does not
affect our results.

\begin{figure}
\begin{center}
\plotone{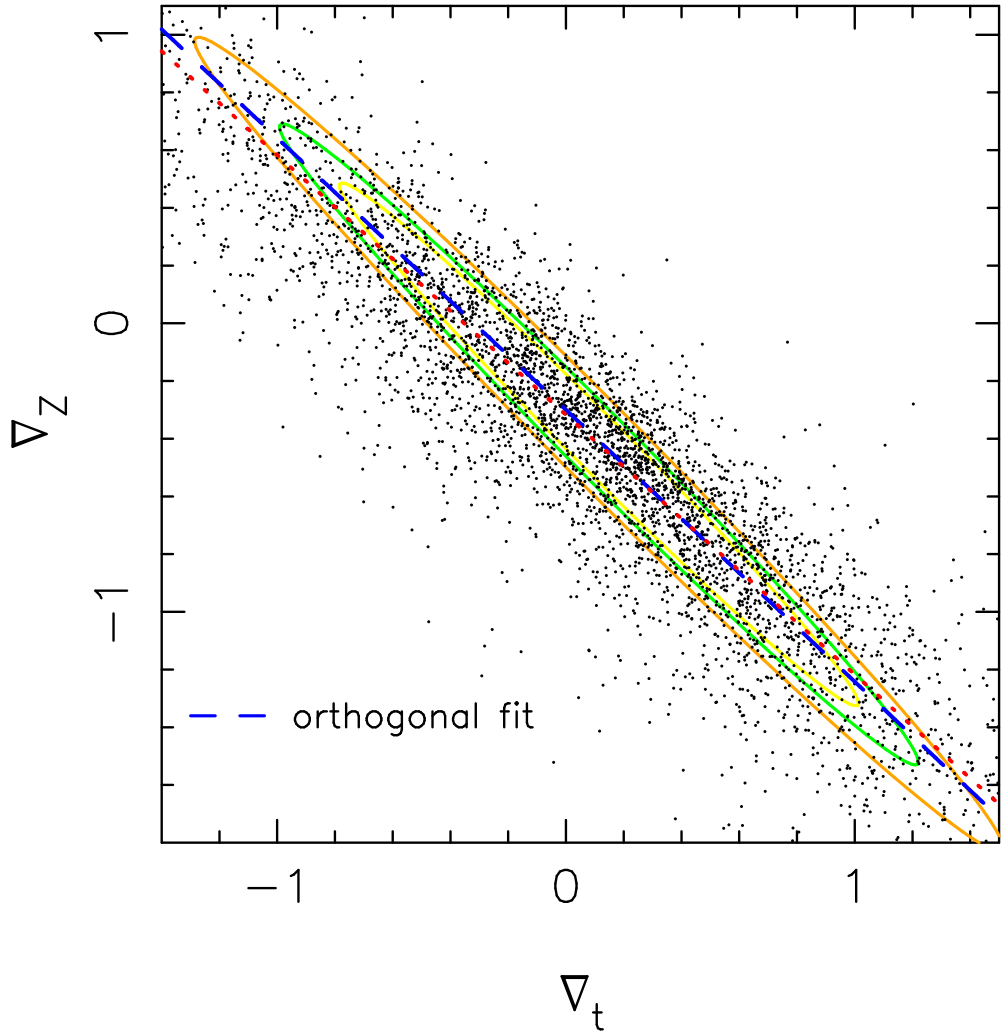}
\caption{Correlation of metallicity and age gradients as estimated for
  the  optical+NIR sample  of ETGs,  with  the BC03  SSP models.   The
  correlation  results from the  age--metallicity degeneracy  of color
  indices. The dashed line is obtained from a least-squares orthogonal
  fit of the data, with its  slope, $\alpha$, being used to define the
  {\it  effective} color  gradient  \nablas. The  dotted  red line  is
  obtained by fixing the slope of the best-fit line to that of the red
  line in  Fig.~\ref{age_met_var}. { The ellipses  plot the typical
    measurement error contours that  correspond to the median, 25- and
    75-percentile uncertainties on \nablaz.  }
\label{fig:nabla_age_met_BC03}
}
\end{center}
\end{figure}

\subsection{The effective color gradient $\nabla_\star$}
\label{sec:eff_grad}
As  discussed  in  Sec.~\ref{sec:age_met_constrain},  even  using  the
entire   $grizYJHK$   set   of   wavebands   we   cannot   break   the
age--metallicity    degeneracy.    This    is    further   shown    in
Fig.~\ref{fig:nabla_age_met_BC03},  where  we   plot  the  age  versus
metallicity  gradients  obtained with  the  above  procedure, for  the
optical+NIR  sample, with  the BC03  SSP models.   The \nablaz  \, and
\nablat \, estimates  are strongly correlated. As shown  from the blue
dashed  line in  Fig.~\ref{fig:nabla_age_met_BC03},  obtained from  an
orthogonal  least-squares  fit  of  \nablaz  \,  versus  \nablat,  the
correlation is well described by a linear relation:
\begin{equation}
\nabla_Z \propto \alpha \cdot \nabla_t.
\label{eq:rel_nabla_tz}
\end{equation}
The  correlation  is  very  similar  to that  between  $\Delta_Z$  and
$\Delta_t$ in Fig.~\ref{age_met_var}. This  is shown by the red dotted
line  in  Fig.~\ref{fig:nabla_age_met_BC03}, which  is  obtained by  a
linear fit of \nablaz \, versus \nablat \, with slope fixed to that of
the red dot-dashed line  in Fig.~\ref{age_met_var}.  The slopes of the
blue dashed and  red dotted lines differ by  only $4\%$, implying that
the strong correlation  of $\nabla_t$ and $\nabla_Z$ is  due mainly to
correlated errors on  differences in age and metallicity,  such as the
one between $\Delta_Z$ and $\Delta_t$ shown in Fig.~\ref{age_met_var}.
In other  terms, individual  \nablaz \, and  \nablat \,  estimates are
affected by large, correlated statistical uncertainties because of the
age--metallicity   degeneracy  {   (see  the   error   ellipses  in
  Fig.~\ref{fig:nabla_age_met_BC03})}.     However,   as    shown   in
Sec.~\ref{sec:age_met_constrain},  using optical+NIR data  allows {\it
  unbiased  estimates of  the mean  \nablaz \,  and \nablat  \,  to be
  obtained, provided  that the number  of galaxies is large  enough to
  make statistical uncertainties on the means small enough relative to
  the absolute values of the  differences in age and metallicity (i.e.
  \nablat \, and  \nablaz)}. This allows us to  bin galaxies according
to  different properties and  perform a  meaningful comparison  of the
mean   \nablat  \,   and  \nablaz   \,  among   different   bins  (see
Sec.~\ref{sec:age_met_grads}).

\begin{figure*}
\begin{center}
\plotone{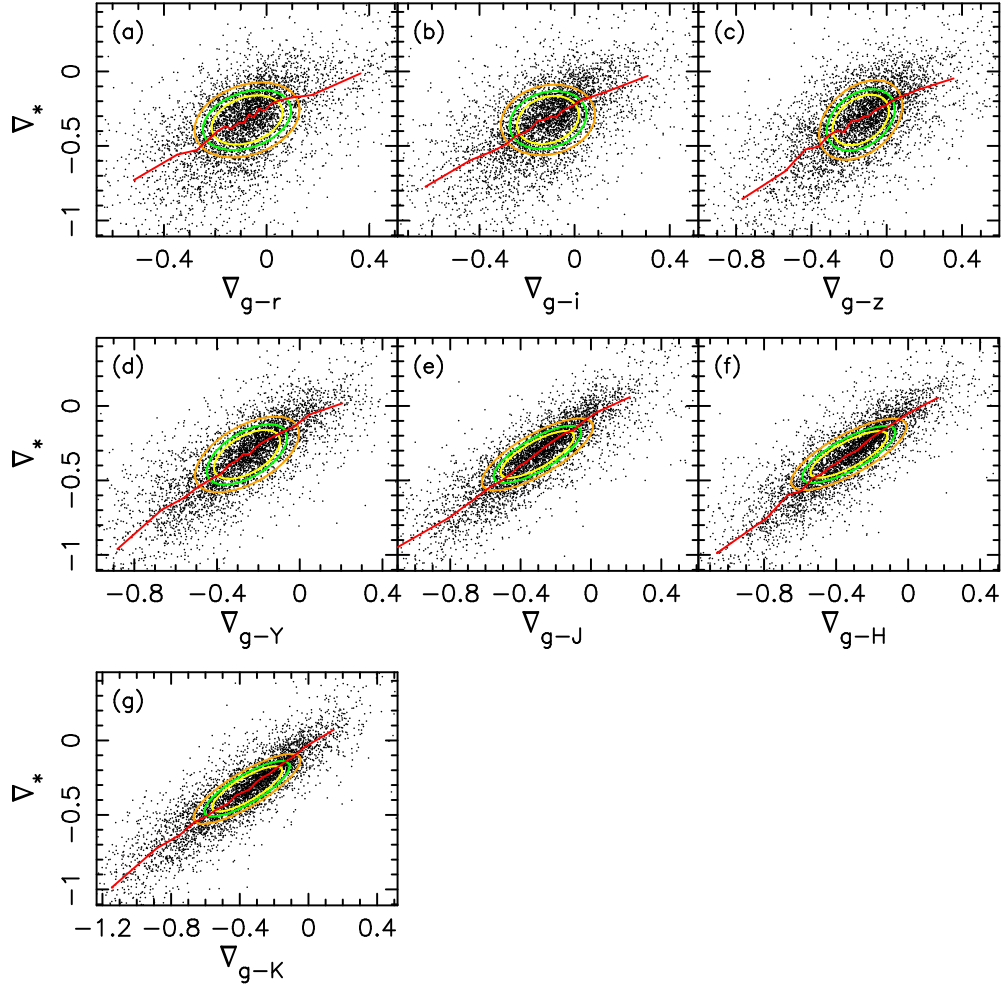}
\caption{Correlation  of the effective  color gradient,  \nablas, with
  the $g-X$  color gradients, $\nabla_{g-X}$. From left  to right, and
  top to bottom, the $X$ ranges  from $g$ through $K$. For each panel,
  the red curve is obtained by median binning the \nablas \, $wrt$ the
  corresponding color gradient, { while the ellipses plot the typical 
 measurement error contours that correspond to the median,  25-  and 75-percentile  
uncertainties on \nablas.  }
\label{fig:nablas_cgrads}
}
\end{center}
\end{figure*}

In  order   to  avoid  the  correlated   uncertainties  on  individual
$\nabla_t$ and $\nabla_Z$ estimates,  we can define an {\it effective}
color gradient:
\begin{equation}
 \nabla_{\star}= \nabla_Z - \alpha \cdot \nabla_t.
~\label{eq:nablas}
\end{equation} 
From a  geometrical viewpoint, $\nabla_{\star}$  measures the distance
of  a given  point  in the  $(  \nabla_Z, \nabla_t)$  plane along  the
$\nabla_Z$  direction  to   the  line  defining  the  age--metallicity
degeneracy. By definition, the  correlated variation of \nablaz \, and
\nablat,  which is  described by  Eq.~\ref{eq:rel_nabla_tz},  does not
change  the  \nablas.  Hence,  the  effective  color  gradient is  not
affected by  the age--metallicity degeneracy.  For  galaxies having no
age gradients, \nablas \,  reduces to the galaxy metallicity gradient.
Moreover, since $\alpha$  is negative (see below), Eq.~\ref{eq:nablas}
can be rewritten as
\begin{equation}
 \nabla_{\star}= \nabla_Z + |\alpha| \cdot \nabla_t.
\label{eq:nablas_mod}
\end{equation}  
This shows that when  \nablaz$=0$, \nablas \, is directly proportional
to  the age gradient.   More generally,  Eq.~\ref{eq:nablas_mod} shows
that \nablas \, behaves like  a color gradient.  A bluer outer stellar
population relative to the inner one  in a galaxy (because of either a
metallicity or age gradient) yields  a lower \nablas.  This is further
shown in  Fig.~\ref{fig:nablas_cgrads}, where we plot \nablas  \, as a
function of all  the color gradients (see Sec.~\ref{sec:cg_estimates})
for  the   optical+NIR  sample  of   ETGs.  On  average,   \nablas  \,
monotonically  increases as  a  function of  each $\nabla_{g-X}$.   In
essence, $\nabla_{\star}$ can  be seen as a color  gradient that takes
into account all of the colors.

Using \nablas \, provides  three important advantages over traditional
color gradients. First,  because \nablas \, combines all  of the color
{ gradient},  its statistical uncertainty  is significantly smaller
than that  on any  color { gradient}  using one pair  of passbands.
Second,  different samples (bins)  of galaxies  can be  compared using
just  one quantity (\nablas),  rather than  separately for  each color
{      gradient}.      Finally,      as     we      prove     below
(Sec.~\ref{sec:err_eff_grad}),  the  effective  color  gradient  is  a
completely model-independent  quantity (in the  same way {  as} the
observed color  gradients), which is not  the case for  \nablaz \, and
\nablat.

For  each  stellar  population   model,  we  derive  $\alpha$  via  an
orthogonal  least-square fit  of \nablaz  \,  vs. \nablat  \, for  all
galaxies. Then  we calculate  $\nabla_{\star}$ for each  galaxy, using
Eq.~\ref{eq:nablas}.  The value  of $\alpha$ changes between different
models, as  well as for  a given model  when using a different  set of
wavebands  (i.e.   optical  vs.  optical+NIR).   For  the  optical+NIR
wavebands, we obtain $\alpha=-0.943$ (BC03), $-0.663$ (CB10), $-1.341$
($BC03_{\tau=1Gyr}$), and~$-0.713$ (M05).   For the optical wavebands,
we   obtain   $\alpha=-1.280,    -1.186,   -1.756,   $   and~$-1.007$,
respectively.

\subsection{Systematics in $\nabla_\star$}
\label{sec:err_eff_grad}
\begin{figure}
\begin{center}
\plotone{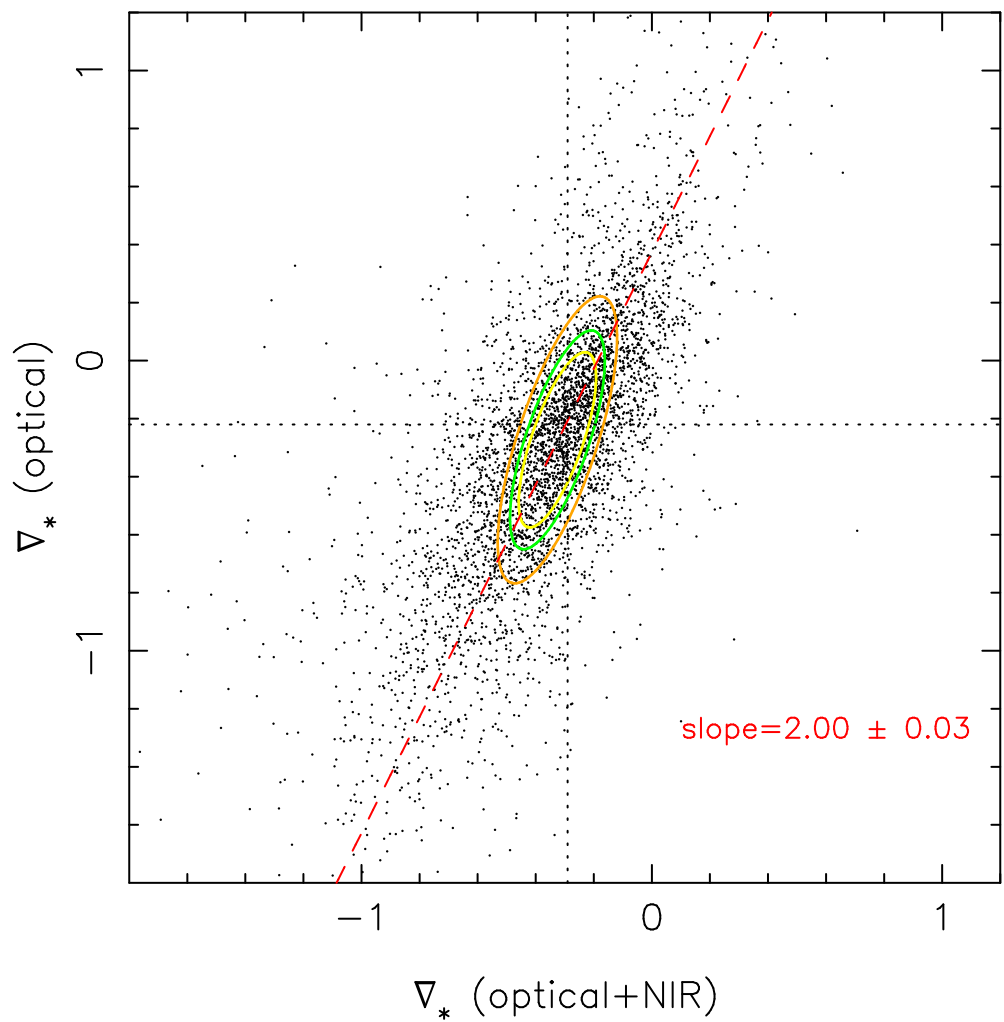}
\caption{Comparison  of  the effective  color  gradient derived  using
  either optical  or optical+NIR wavebands. The dashed  red line shows
  the  orthogonal best-fitting  relation  between the  two \nablas  \,
  estimates. The slope of the relation is reported in the lower--right
  corner of  the plot. The  horizontal and vertical dotted  lines mark
  the  peaks of the  distribution of  \nablas \,  for the  optical and
  optical+NIR data,  respectively. { The ellipses  plot the typical
    measurement error contours that  correspond to the median, 25- and
    75-percentile  uncertainties  on   the  optical+NIR  \nablas.  The
    ellipses  are constructed  by assuming,  for simplicity,  that the
    covariance term of the errors  is proportional to that of the data
    themselves.}
\label{fig:conf_gradstar_optNIR}
}
\end{center}
\end{figure}

\begin{figure}
\begin{center}
\plotone{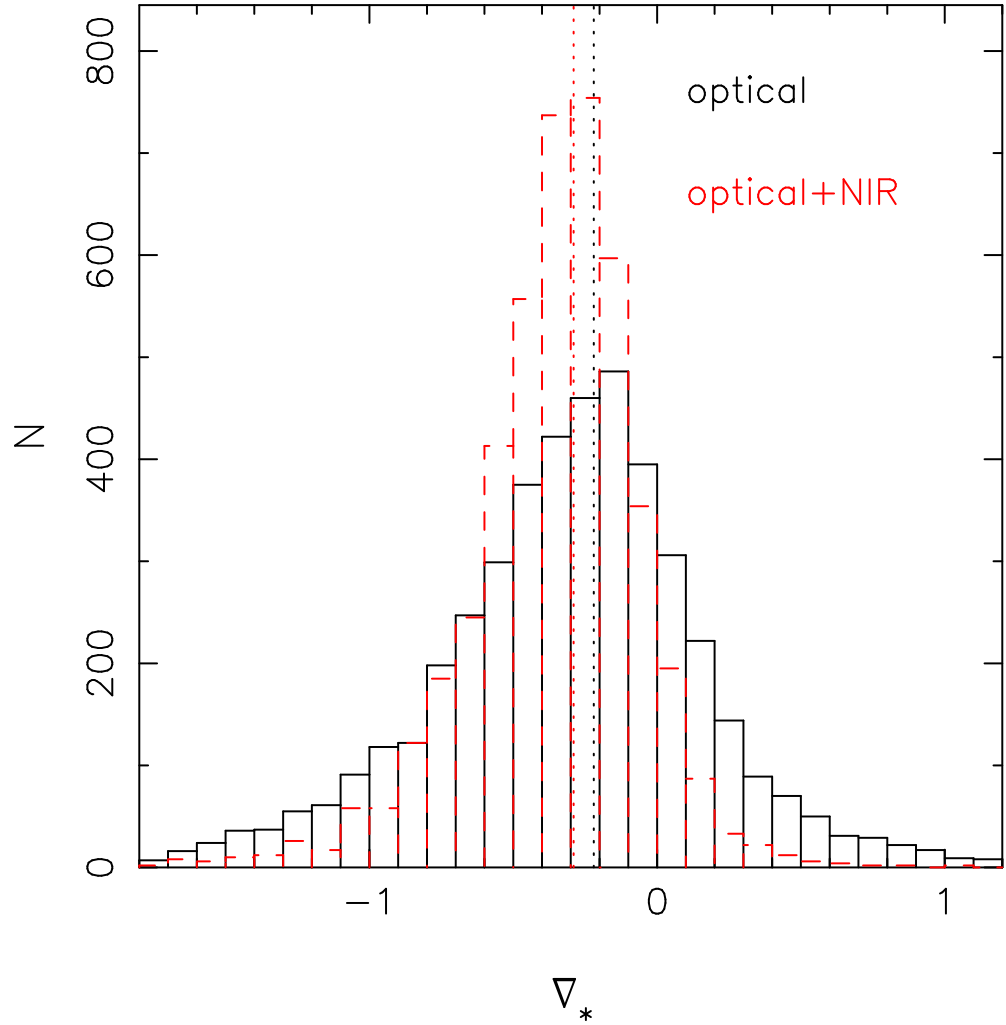}
\caption{Comparison of the  distribution of effective color gradients,
  obtained using either optical (black) or optical+NIR (red) data. The
  peaks of  the two  distributions are marked  by the  vertical dotted
  lines. Note how the distribution becomes narrower when including the
  NIR wavebands.
\label{fig:conf_dist_grads}
}
\end{center}
\end{figure}

We examine how  sensitive \nablas \ is to varying  the set of utilized
wavebands.   This  is  shown  in  Fig.~\ref{fig:conf_gradstar_optNIR},
where  we compare  the \nablas  \ measured  with optical+NIR  and only
optical bands.   These two values  of \nablas \ are  linearly related,
indicating  that  optical data  alone  can  still  provide a  reliable
estimate of  the {\it effective}  color gradient.  This  is consistent
with the fact  that \nablas \, is not  affected by the age-metallicity
degeneracy  (influencing   the  optical  more   than  the  optical+NIR
gradients) and is also proportional to each of the color gradients. On
the  other hand,  the  slope  of the  correlation  between \nablas  \,
(optical) and  \nablas \, (optical+NIR) is  significantly greater than
one  at  $2.00  \pm  0.03$.  This  reflects  the  reduced  statistical
uncertainty on \nablas \ when incorporating the NIR information.
This  is  further  shown  in Fig.~\ref{fig:conf_dist_grads}  where  we
compare  the  histograms of  \nablas  \,  obtained  using optical  and
optical+NIR  data.   The  width  of  the  \nablas  \  distribution  is
significantly larger when using optical $\it {wrt} $ optical+NIR data.
Also,  a  small  offset  in  the  peak of  the  two  distributions  is
present.  These results  are quantified  in Tab.~\ref{tab:stat_nablas}
where  we report  the  statistics,  i.e. the  peak  ($\mu$) and  width
($\sigma$)       as      estimated      using       the      bi-weight
estimator~\citep{Beers:90},   for   different   samples  and   stellar
population  models. In  the  same  Table, we  also  report the  median
uncertainty, $\sigma_{\star}^{err}$,  as well as  the intrinsic width,
$\sigma_{\star}^i$, defined in the same way as the quantity $\sigma^i$
in Sec.~\ref{sec:dist_cgs}.  From these data we conclude  that for the
optical+NIR  sample, the  peaks of  the \nablas  \,  distributions are
consistent for different stellar  population models, i.e.  the \nablas
\, estimated from optical+NIR is model-independent.  In contrast, when
using only optical data, a significant systematic different is present
between the peak values of different models, though this difference is
quite  small ($<  0.02$) $wrt$  the  absolute peak  value of  \nablas.
Furthermore, in agreement  with Fig.~\ref{fig:conf_dist_grads}, we see
that $\sigma$ is significantly larger when using optical data ($\sigma
\sim 0.4$) as opposed to  using optical+NIR data ($\sigma \sim 0.25$),
i.e.  \nablas \, is affected by significantly larger uncertainties (by
a factor  of $\sim  1.6$) when using  only optical data.  Finally, the
peak values  of the  optical and optical+NIR  \nablas \,  are slightly
different, in the  sense that the {\it effective}  color gradients are
on average more  negative by $\sim 0.06$ for  the optical+NIR than for
the optical  samples.  The  differences in $\mu$  might be due  to the
different   shapes  of   the  optical   and  optical+NIR   \nablas  \,
distributions,  rather than  some intrinsic  difference.  In  fact, as
seen  in Fig.~\ref{fig:conf_dist_grads},  the distribution  of \nablas
(optical) is clearly asymmetric.
\begin{deluxetable}{l|c|c|c|c|c|c|c|c}
\small
 \tablecaption{Statistics of the {\it effective} color gradients.}
\tabletypesize{\scriptsize}
\tablewidth{0pt}
\tablecolumns{9}
\tablehead{    model & \multicolumn{4}{c}{optical+NIR sample} & \multicolumn{4}{c}{optical sample} \\
          &  $\mu$ & $\sigma$ & $\sigma_{\star}^{err}$ & $\sigma^{i}_{\star}$ & $ \mu $ & $\sigma$ & $\sigma_{\star}^{err}$ & $\sigma^{i}_{\star}$}
\startdata
$BC03            $ & $ -0.283 \pm   0.007$ & $  0.246 \pm   0.005$ & $  0.163$ & $  0.184$ & $  -0.219 \pm   0.004$ & $   0.407 \pm   0.002 $ & $  0.377$ & $  0.152 $ \\
$M05             $ & $ -0.286 \pm   0.007$ & $  0.246 \pm   0.005$ & $  0.164$ & $  0.182$ & $  -0.209 \pm   0.003$ & $   0.388 \pm   0.002 $ & $  0.355$ & $  0.157 $ \\
$CB10            $ & $ -0.296 \pm   0.007$ & $  0.254 \pm   0.005$ & $  0.181$ & $  0.178$ & $  -0.235 \pm   0.004$ & $   0.436 \pm   0.003 $ & $  0.406$ & $  0.160 $ \\
$BC03_{\tau=1Gyr}$ & $ -0.291 \pm   0.008$ & $  0.257 \pm   0.005$ & $  0.162$ & $  0.200$ & $  -0.225 \pm   0.004$ & $   0.417 \pm   0.002 $ & $  0.390$ & $  0.147 $ \\
\enddata
\label{tab:stat_nablas}
\end{deluxetable}

\begin{figure*}
\begin{center}
\plotone{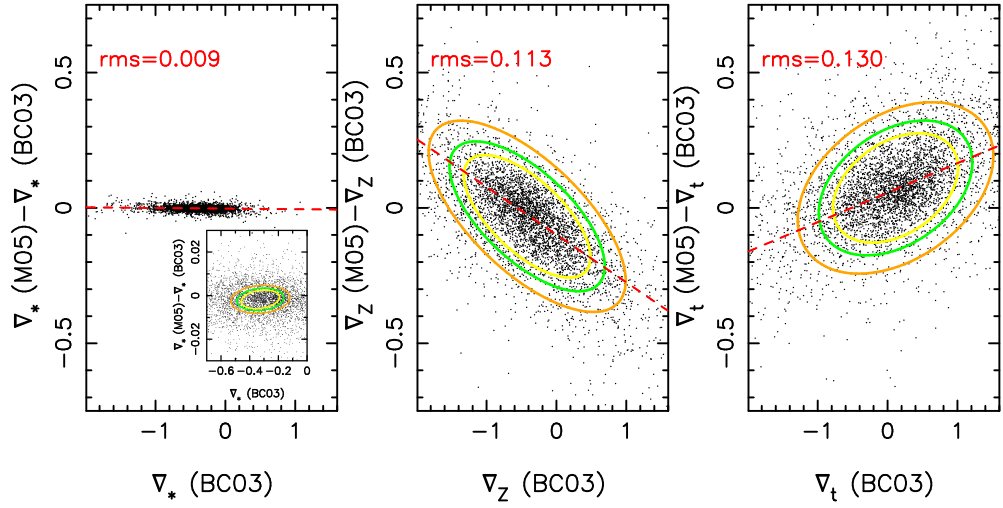}
\caption{Comparison of effective color (left), metallicity(middle) and
  age gradients (right) derived by using M05 and BC03 SSP models. Each
  panel plots  the gradient difference in  the sense of  M05-BC03 as a
  function of BC03 gradients.  The red dashed lines are the orthogonal
  best-fitting relations. { The ellipses plot the typical 
 measurement error contours that correspond to the median,  25-  and 
75-percentile  uncertainties on the x-axis variables. The covariance terms of the
errors are derived as for Fig.~\ref{fig:conf_gradstar_optNIR}. The left inset panel zooms into the center of the distribution to make the error ellipses visible.}
\label{fig:conf_gradstar_models}
}
\end{center}
\end{figure*}

The robustness of  \nablas, \nablaz ~and \nablat ~with  respect to the
adopted    stellar     population    model    is     illustrated    in
Fig.~\ref{fig:conf_gradstar_models},  where we compare  the $\nabla$'s
obtained using the optical+NIR sample of ETGs for BC03 and M05. We see
that the scatter in \nablas  \, is significantly reduced in comparison
to that  for \nablaz  \, or  \nablat.  As noted  above, \nablas  \, is
model-independent, while this is obviously not the case for \nablaz \,
and \nablat  \,. This  gives extra  support to using  \nablas \,  as a
useful and  robust representation of  the radial gradients  of stellar
population properties in galaxies.

\section{Dependence of the effective color gradient on galaxy parameters}
\label{sec:cg_galpars}
From  Table 4  we see  that the  stellar population  gradients exhibit
significant  intrinsic  dispersion,  motivating  an  analysis  of  the
correlations of  \nablas \, versus other galaxy  parameters.  We study
correlations  of  \nablas   \,  with  several  photometric  quantities
(Sec.~\ref{sec:corr_strpar}),  different   estimates  of  galaxy  mass
(Sec.~\ref{sec:corr_mass}), and  spectroscopic parameters like central
velocity  dispersion  and  stellar  population  indicators  like  age,
metallicity       and        $[\alpha/Fe]$       abundance       ratio
(Sec.~\ref{sec:corr_sppar}).    Because  of   the   large  statistical
uncertainty on individual measures of stellar population gradients, we
bin the  data and analyze  trends with the  mean \nablas \, .  In each
case, we  verified that  changing the number  of galaxies in  each bin
does not affect  our results.  The results presented  in the following
subsections refer  to \nablas \,  computed using the BC03  SSP models.
Similar findings hold when considering other stellar population models
described in Sec.~\ref{sec:grads}.

\begin{figure*}
\begin{center}
\plotone{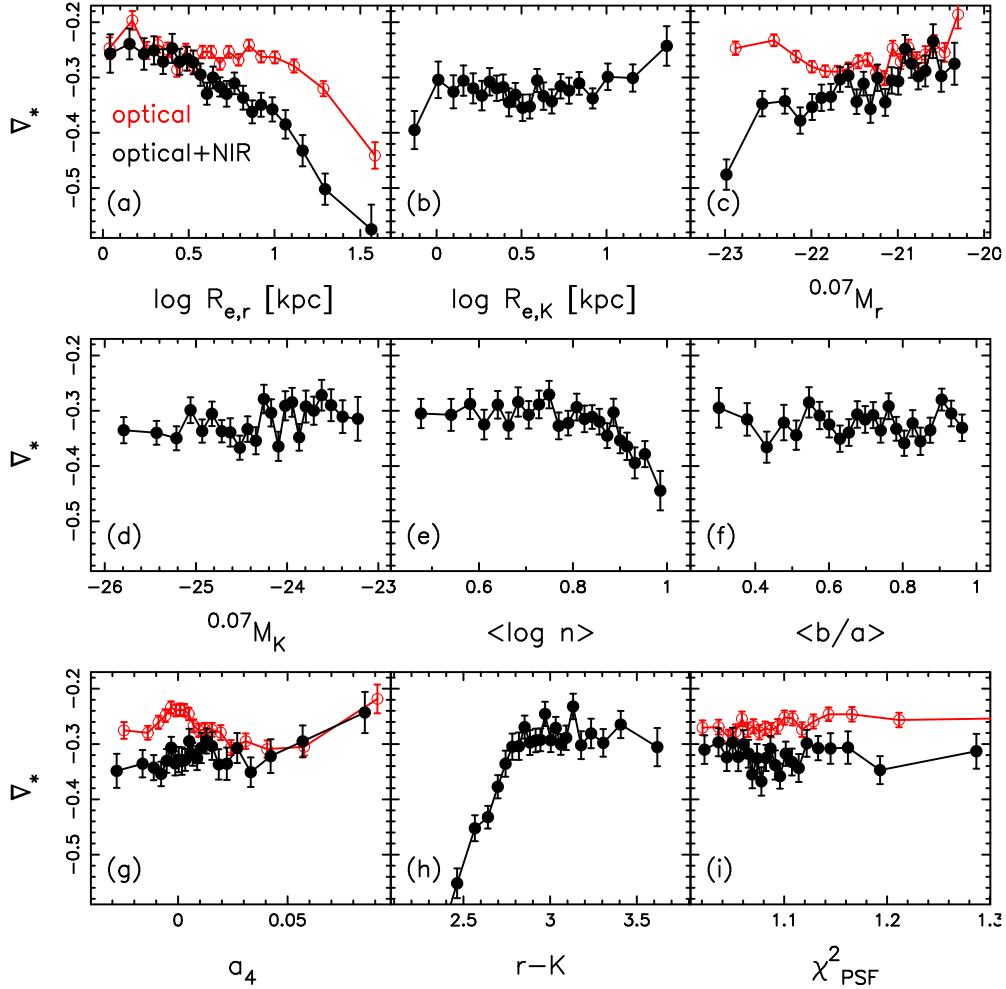}
\caption{Dependence  of  effective  stellar  population  gradients  on
  galaxy  structural  parameters.   Each  plot has  been  obtained  by
  binning the  \nablas \, with respect  to a given  quantity $p$.  The
  following quantities are plotted: the r-band effective radius (panel
  a),  K-band effective  radius (panel  b), r-band  absolute magnitude
  (panel  c),  K-band absolute  magnitude  (panel  d),  { logarithmic median} of  the
  $grizYJHK$ Sersic  indices (panel e), median of  the $grizYJHK$ axis
  ratios (panel  f), the median of  the $a_4$ parameters  in the $gri$
  passbands (panel g), $r-K$ color  index (panel h), median of the PSF
  modeling $\chi^2_{PSF}$ in the  $grizYJHK$ passbands (panel i).  For
  each panel, the circles are  the median \nablas \, in different bins
  of  $p$.    Error  bars  are  1~$\sigma$  standard   errors  on  the
  medians. The  circles are connected  by solid lines. The  ETG sample
  with optical+NIR data is plotted in black. For panels (a), (c), (g),
  and (i),  the binning is  also done for  the optical sample  of ETGs
  (red color).
\label{fig:grads_par}
}
\end{center}
\end{figure*} 

\subsection{Effective  gradients vs. photometric properties}
\label{sec:corr_strpar}
{  We bin  \nablas \,  with  respect to  the following  photometric
  parameters: r- and K-band effective half-light radii, ($R_{e,r}$ and
  $R_{e,K}$);  r-  and   K-band  total  magnitudes  ($^{0.07}M_r$  and
  $^{0.07}M_K$), obtained through  a PSF-convolved Sersic modeling of
  the  galaxy image;  logarithmic  median Sersic  index ($<\log  n>$);
  median   axis  ratio   (  $<b/a>$);   the  $a_4$   parameter,  which
  characterizes the galaxy isophotal shape (boxiness - a$_{4} <$ 0 and
  diskyness -  a$_{4}>$ 0, see~\papdata); the  optical-NIR color index
  ($r-K$) and  the median PSF  fitting $\rm\chi^2$ ($\rm\chi^2_{PSF}$,
  Fig.~\ref{fig:grads_par}).   For  each  galaxy,  median  values  are
  computed  using data at  the available  wavebands, i.e.   $griz$ and
  $grizYJHK$.  We  adopted the $r$- and $K$-band  radii and magnitudes
  as representative of  the sizes and luminosities of  galaxies in the
  optical  and NIR  spectral regions,  respectively.  The  $r-K$ color
  index is  estimated using 2DPHOT total galaxy  magnitudes.  Each bin
  includes the  same number  of galaxies, i.e.   $N=200$ ETGs  for the
  optical+NIR sample and $N=2000$ ETGs for the optical sample.}

Fig.~\ref{fig:grads_par} shows the relations between \nablas\, and the
photometric  quantities listed  above. Panels  (a) and  (c)  show that
\nablas \, changes significantly  with the r-band effective radius and
$r$-band total  absolute magnitude in  the sense that larger  and more
optically luminous ETGs have  stronger (more negative) effective color
gradients.   This  result is  somewhat  expected  as  a more  negative
\nablas  \, implies  that bluer  (either less  metal rich  or younger)
stars in the galaxy  are preferentially distributed towards the galaxy
periphery. We  note that  the trend with  $^{0.07}M_r$ is  weaker than
that with $R_{e,r}$.  When we  restrict ourselves to only optical data
the trends with radius and  magnitude are less pronounced (see below).
Moreover, \nablas\, has no dependence  on radius or luminosity when we
use $K$-band  ({ Panels b and  d}).  This finding  is of particular
interest,  as the  NIR light  follows  more closely  the stellar  mass
distribution than the  optical light. It implies that  for bright ETGs
the  effective color gradient  does not  vary significantly  along the
stellar mass  sequence.  A  mild correlation exists  between \nablas\,
and Sersic  index, as shown in  Panel (e).  For  $\log n \widetilde{>}
0.75$  ($n \widetilde{>}  5.6$)  ETGs  with higher  $n$  tend to  have
stronger  effective  color  gradients,  likely because  galaxies  with
higher   $n$   have   higher   luminosity   due   to   the   (optical)
luminosity--Sersic  index relation  of  ETGs~\citep{CCD93}. Panel  (f)
shows no trend whatsoever of \nablas \, with axis ratio while with the
boxy/disky parameter $a_4$, shown in  Panel (g), a weak correlation is
present especially  when optical+NIR  data is used.   In panel  (h) we
show that  a strong  correlation exists between  \nablas \,  and total
galaxy colors. For $r-K< 2.9$, galaxies with bluer colors also tend to
have stronger  gradients. This is  not a spurious result  arising from
the  correlation  of  galaxy  colors  and magnitudes.   In  fact,  the
color-magnitude  relation implies that  brighter galaxies  should have
redder  colors.  This,  together with  the  trend of  \nablas \,  with
$^{0.07}M_r$, would  produce a  trend opposite to  that seen  in Panel
(h).  Moreover,  as shown  in~\papdata, no significant  correlation is
found between total magnitudes and total (rather than aperture) galaxy
colors.   Finally,  Panel (i)  probes  possible  systematics in  color
gradient  estimates  related  to  the  PSF modeling.   It  shows  that
\nablas\, is not correlated with  the $\chi^2$ from fitting the galaxy
image with a PSF convolved S\'ersic light distribution. Hence, our PSF
models are accurate enough to provide, on average, unbiased \nablas \,
estimates.

In summary, we find that the effective color gradient in (bright) ETGs
do not  correlate with  (NIR) luminosities and  galaxy radii,  while a
correlation exists  with galaxy colors. The former  finding is further
investigated  in the  next  section,  where we  analyze  the trend  of
\nablas  \, with  galaxy mass.   The  correlation of  \nablas \,  with
colors is  further studied in Sec.~\ref{sec:corr_sppar},  where we bin
\nablas \, with respect to the stellar population parameters of ETGs.

\begin{figure}
\begin{center}
\plotone{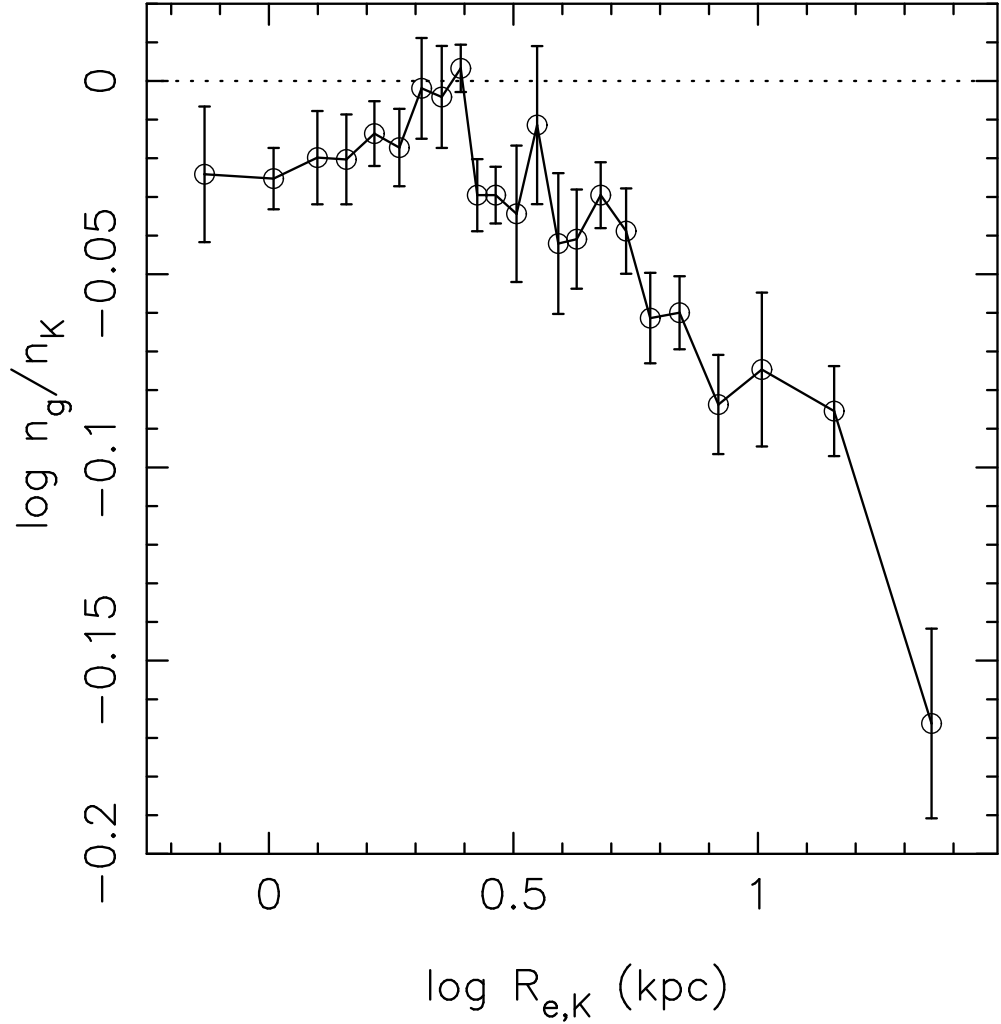}
\caption{Logarithmic  ratio of  $g$ to  $K$-band Sersic  indices  as a
  function of $  \log R_e$ in K-band.  The  solid line connects the
  data-points  obtained by  median binning  the distribution  of $\log
  n_{g}/n_{K}$  with respect  to  $ \log  R_{e,K}$,  with each  bin
  including  the same  number (N=200)  of points.   The  trend implies
  larger galaxies to have a profile shape more concentrated in the NIR
  than  in the optical  (i.e. lower  $n_{g}/n_{K}$).  The  dotted line
  marks  zero Sersic  index  ratio. Error  bars  denote the  $1\sigma$
  errors on the median in each bin.~\label{fig:ng_nk}}
\end{center}
\end{figure}

The lack of correlation between  \nablas \, and $ R_{e,K}$ (Panel b
of Fig.~\ref{fig:grads_par})  deserves further comments.   In Paper II
we  show  that the  ratio  of  optical  ($g$-band) to  NIR  ($K$-band)
effective radii, $ R_{e,g}/R_{e,K}$ decreases as a function of $
R_{e,K}$, with  the largest galaxies having  $ R_{e,g}/R_{e,K} \sim
1$ (see Fig.~7 of Paper II).  This variation implies that the slope of
the  Kormendy relation  for ETGs  increases from  $g$ through  $K$.  A
similar  result,  at  optical  wavebands ($g-r$),  has  been  recently
reported  by~\citet{RBH:10}.  Using the  ratio of  optical radii  as a
proxy for  the color gradient,  the decrease of  $ R_{e,g}/R_{e,K}$
with $ R_{e,K}$ implies  that larger galaxies should have shallower
gradients (see also Tortora et al.~2010), in contrast with the lack of
correlation  in Panel b  of Fig.~\ref{fig:grads_par}.   However, color
gradients are  determined not only  from the ratio of  effective radii
but also  from that of the  Sersic indices. In fact,  the variation of
the  surface brightness  profile between  two wavebands  is determined
from  both   $  R_e$  and  $n$.    Fig.~\ref{fig:ng_nk}  plots  the
logarithmic  binned ratio  of $g$  to $K$-band  Sersic  indices, $\log
n_g/n_K$, as a function of $ \log R_{e,K}$.  We find that $n_g/n_K$
decreases as  a function  of $ R_{e,K}$,  i.e. larger  galaxies are
more concentrated  in the NIR than lower-$  R_{e,K}$ systems.  This
compensates for the trend of $ R_{e,g}/R_{e,K}$ with $ R_{e,K}$,
making \nablas \,  nearly constant with NIR radius.   Thus there is no
inconsistency  between   the  lack  of  correlation  in   Panel  b  of
Fig.~\ref{fig:grads_par} and the correlation  shown in Fig.~7 of Paper
II.

\begin{figure}
\begin{center}
\plotone{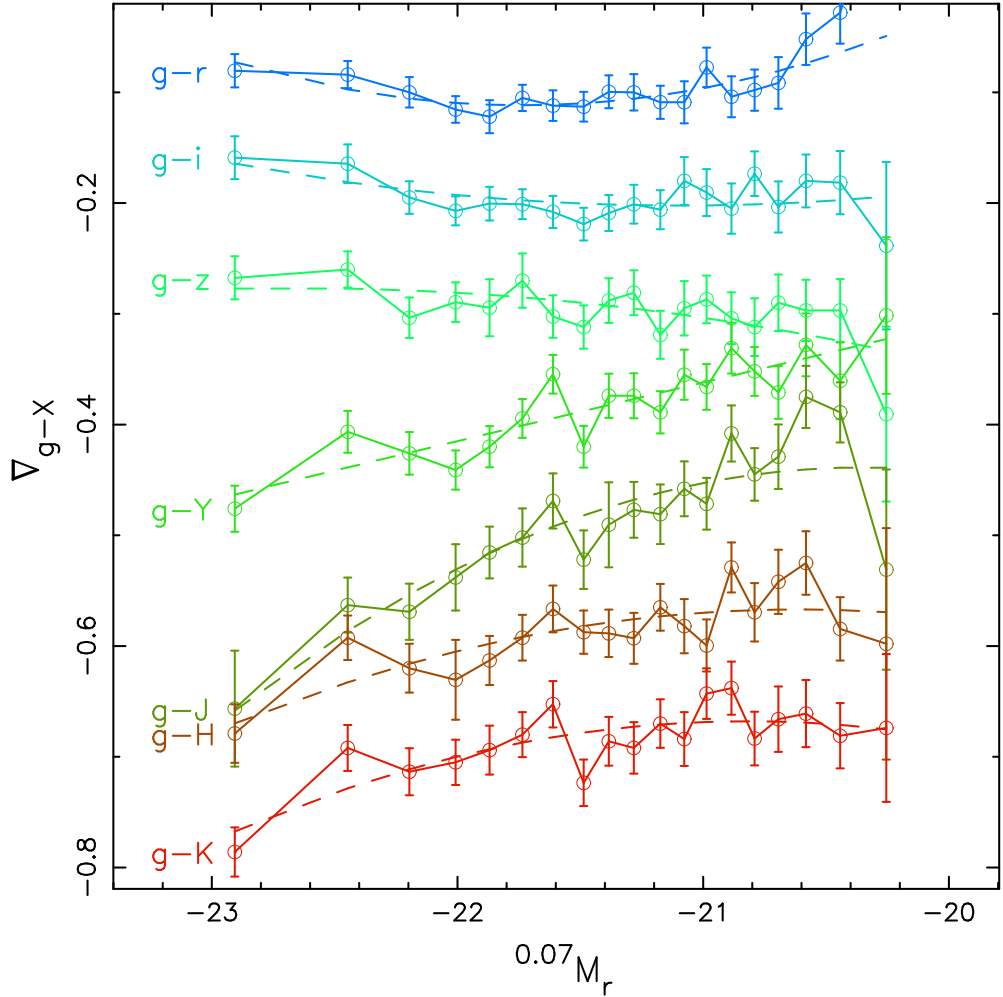}
\caption{Median color gradient, $\nabla_{g-X}$, with $X=rizYJHK$, as a
  function  of  r-band absolute  magnitude,  $^{0.07}M_r$. Error  bars
  denote $1~\sigma$  standard errors on the medians.  The dashed lines
  are obtained by an second-order polynomial fitting of $\nabla_{g-X}$
  versus    $^{0.07}M_r$,    with    $\nabla_{g-X}$    as    dependent
  variable. Different  colors correspond to  different wavebands, from
  $g$ through $K$  (top to bottom), as indicated by  the labels on the
  left part of  the plot. For better displaying  the different curves,
  the  $\nabla_{g-X}$  are  re-normalized  to have  median  values  of
  $-0.1,-0.2,-0.3,-0.4,-0.5,-0.6,-0.7$,  from  $g$  through  $K$.  The
  actual    medians    are   given    by    the    peak   values    in
  Tab.~\ref{tab:stat_cgrad_optNIR}.
\label{fig:nabla_gX_mag}
}
\end{center}
\end{figure}

As mentioned  above, we  find that color  gradients exhibit  a complex
behavior with galaxy luminosity. When estimated from optical+NIR color
gradients, \nablas  \, becomes more negative as  the optical magnitude
becomes brighter.   On the contrary,  for the optical sample  of ETGs,
i.e. when \nablas \, is estimated from $\nabla_{g-r}$, $\nabla_{g-i}$,
and $\nabla_{g-z}$,  this steepening  disappears, and \nablas  \, does
not vary with $M_r$ (Panel c of Fig.~\ref{fig:grads_par}).  The reason
for the  different behavior of  \nablas, when estimated  using optical
{\em     vs.}      optical+NIR      data     is     illustrated     in
Fig.~\ref{fig:nabla_gX_mag}, where we  plot the median color gradient,
$\nabla_{g-X}$, as  a function  of the optical  magnitude for  all the
available wavebands  ($X=rizYJHK$). The trends  of $\nabla_{g-X}$ with
$^{0.07}M_r$ are modeled  with second-order polynomials (dashed curves
in the Figure). The most remarkable feature is that the color gradient
variation with  magnitude depends on the  waveband. The $\nabla_{g-r}$
gradient  {  exhibits}  a  { double-valued}  behavior,  becoming
flatter  in   both  more  and   less  luminous  galaxies.    {  For
  $\nabla_{g-z}$}, the { double-valued} behavior disappears and the
$\nabla$ decreases  monotonically as  a function of  $^{0.07}M_r$.  At
redder wavebands, the trend reverses, and $ \nabla_{g-X}$ becomes a
monotonically  increasing  function  of  $^{0.07}M_r$.   As  shown  in
Fig.~\ref{fig:nablas_cgrads},   the   effective   color  gradient   is
proportional to each  of the individual color gradients  from which it
is  estimated.   When using  optical  data  alone,  \nablas \,  mainly
reflects   the  behavior   of   $\nabla_{g-r}$,  $\nabla_{g-i}$,   and
$\nabla_{g-z}$. As  a result, \nablas \, exhibits  no strong variation
with  luminosity,  similar  to  the {  double-valued}  behavior  of
$\nabla_{g-r}$.  When  \nablas \, is estimated  from optical+NIR data,
the  optical-NIR gradients  dominate  the trend,  with  \nablas \,  an
increasing function  of $^{0.07}M_r$. The main conclusion  is that the
behavior of  color gradients with  luminosity depends on  the waveband
where   the   luminosity   is   estimated   (Panels   c   and   d   of
Fig.~\ref{fig:grads_par}), as  well as the wavebands  used to estimate
the gradient  itself.  This complex  dependence may explain,  at least
partly,  the discrepant  results  found in  the  literature about  the
relation between color gradients and luminosity (mass).  For instance,
\citet{PVJ90} first reported a  surprising lack of correlation between
color   gradients  and  galaxy   luminosity,  noting   that  brightest
ellipticals do not  exhibit less steep gradients, as  expected if they
were  the   debris  of  repeated   mergers  of  lower   mass  systems.
\citet{dP05}  also reported  no correlation  between the  size  of the
color gradients and galaxy luminosity.  \citet{TaO03} found that color
gradients in  cluster ETGs become steeper (more  negative) in brighter
galaxies. The  same result was found by~\citet{BaP:94}  for the bulges
of  spiral galaxies.   On the  other hand,  \citet{LdC:05}  found that
color    gradients    do   not    depend    on   galaxy    luminosity,
and~\citet{Choi:07} found that optical color gradients are essentially
constant over a wide range of  galaxy luminosity, with a weak trend of
becoming   flatter   at   both   fainter  and   brighter   magnitudes.
Recently,~\citet[][hereafter RBH10]{RBH:10},  using the ratio  of $g$-
to $r$-band  effective radii from  the SDSS as  a proxy for  the color
gradient,  also  found that  ETGs  with  intermediate luminosity  have
stronger  gradients  than  those  at  low and  high  luminosities,  in
agreement  with  the  {  double-valued}  behavior  we  observe  for
$\nabla_{g-r}$.

\subsection{Effective color gradient vs. galaxy mass}
\label{sec:corr_mass}
\begin{figure}
\begin{center}
\plotone{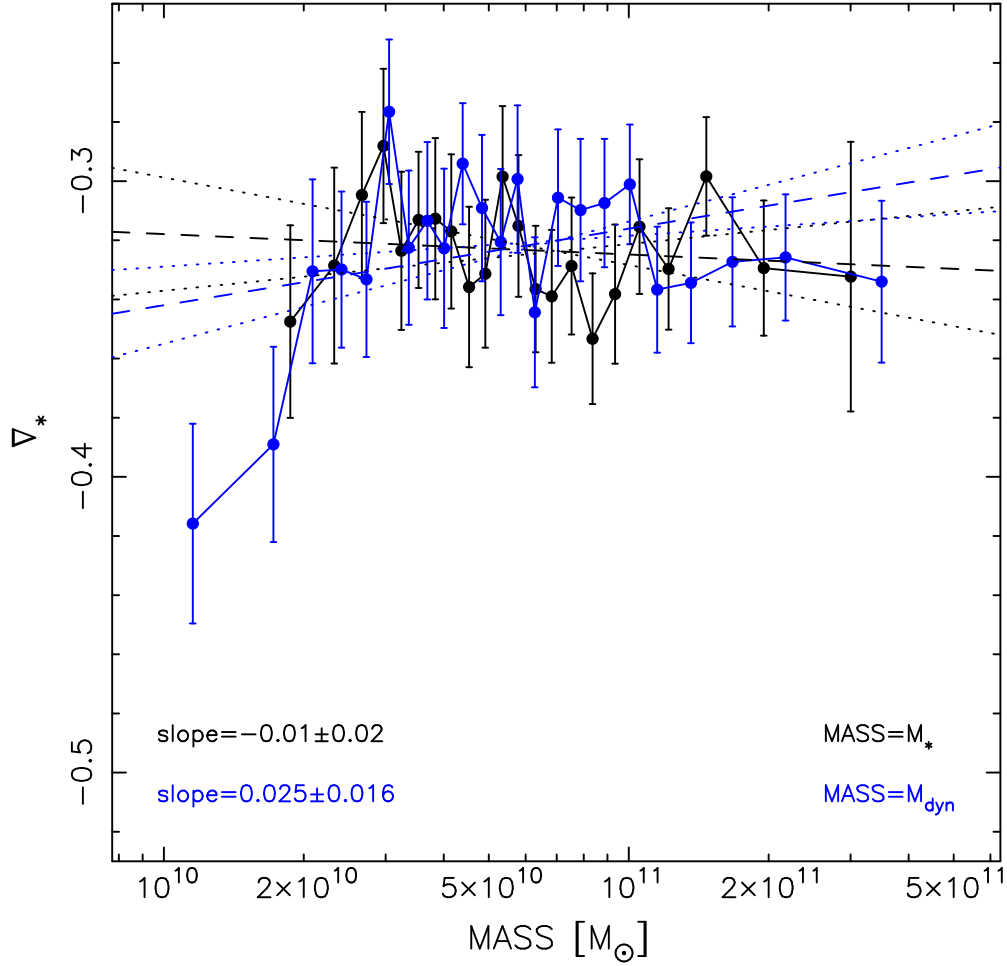}
\caption{Correlation of the  median {\it effective} stellar population
  gradient, \nablas,  with stellar (black) and  dynamical (blue) mass,
  for the  optical+NIR sample of  ETGs.  Error bars  denote 1~$\sigma$
  uncertainties on the median value in each bin. Each bin includes the
  same  number   of  galaxies   ($N=200$).   Dashed  lines   mark  the
  best-fitting relations, with the corresponding slopes being reported
  in  the lower-left  corner  of  the plot.  The  dotted lines  define
  1~$\sigma$ confidence contours on the best-fit lines.
\label{fig:grads_masses}
}
\end{center}
\end{figure}

Fig.~\ref{fig:grads_masses} plots the median  \nablas \, as a function
of  galaxy stellar  ($M_\star$) and  dynamical ($M_{dyn}$)  mass.  The
binning  is performed  as in  Fig.~\ref{fig:grads_par}, with  each bin
having  the same  number of  galaxies. To  quantify the  dependence of
\nablas \, on mass, we perform linear least-squares fits of \nablas \,
with respect to  $M_\star$ and $M_{dyn}$, with the  slopes reported in
the  figure. We find  no trend  of the  effective color  gradient with
mass.   In particular,  the  slope  of \nablas  \,  vs.  $M_{dyn}$  is
consistent with zero at the $1.5~\sigma$ significance level, while the
slope of \nablas  \, vs. $M_{\star}$ is consistent  with zero at $\sim
0.5~\sigma$.  For $M_{dyn} <  2 \times 10^{11} M_{\odot}$, the \nablas
\, becomes  more negative than at  higher mass, but  this effect might
not  be {  real, as}  in  this mass  regime, the  SPIDER sample  is
incomplete   due   to    the   $r$-band   magnitude   selection   (see
Fig.~\ref{fig:mpetr_mdyn}).  {  We  verified  that the  trend  with
  stellar  mass   remains  unchanged  when   using  different  stellar
  population   models  (e.g.   CB10)  to   perform  the   SED  fitting
  (Sec.~\ref{sec:stellar_masses})}.

\subsection{Effective color gradient vs. velocity dispersion and stellar population properties}
\label{sec:corr_sppar}

\begin{figure*}
\begin{center}
\plotone{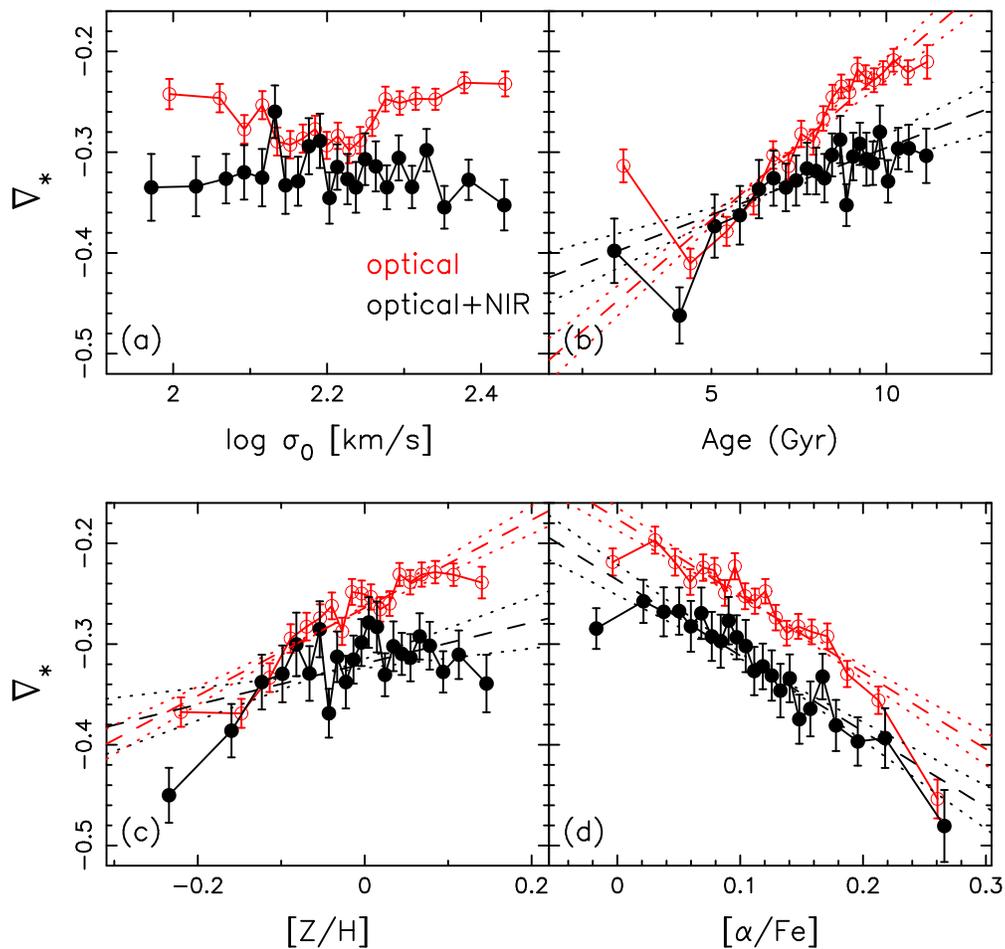}
\caption{The   same   as    Fig.~\ref{fig:grads_par}   but   for   the
  spectroscopic parameters.  From left to right and top to bottom, the
  median  \nablas  \, are  plotted  as  a  function of  the  following
  quantities:   the  central  velocity   dispersion  (panel   a),  the
  luminosity-weighted  age   (panel   b),  the   luminosity-weighted
  metallicity  (panel c),  and the  luminosity  weighted $[\alpha/Fe]$
  abundance ratio  (panel d). The  optical+NIR and optical  samples of
  ETGs are plotted  with black and red colors,  respectively, as shown
  in the lower-right corner of panel (a).
\label{fig:grads_spar}
}
\end{center}
\end{figure*}

Fig.~\ref{fig:grads_spar} plots  the median effective  color gradient,
\nablas\,,  as  a  function   of  central  velocity  dispersion,  age,
metallicity,  and  $\alpha$-enhancement,  for  both  the  optical  and
optical+NIR samples. For the optical, the plot of \nablas \, vs. $\log
\sigma_0$  seems  to  suggest  a {  double-valued}  behavior,  with
\nablas  \,  reaching  a  minimum  around $\log  \sigma_0  \sim  2.2$.
However, this is not confirmed by the optical+NIR data, where no trend
appears.      This     result     supports     the     findings     of
Secs.~\ref{sec:corr_strpar}  and~\ref{sec:corr_mass}, that  \nablas \,
is  not driven  by galaxy  mass.  On  the contrary,  we find  a strong
dependence of  \nablas \, on stellar population  properties, i.e.  the
age, $[Z/H]$, and $[\alpha/Fe]$. In order to quantify these trends, we
perform a robust linear fit of \nablas \, as a function of each of the
three  parameters,  minimizing  the   sum  of  absolute  residuals  in
\nablas. The  uncertainties on  fitting coefficients are  estimated by
the  width of their  distribution among  $1000$ iterations,  where the
median \nablas  \, in  each bin are  shifted according to  their error
bars, and the fitting process is repeated. The slope and offset of the
best-fit lines  are reported in  Tab.~\ref{tab:fit_grads_sppar}.  From
Panel  (b), we  see  that the  effective  color gradient  is flat  for
galaxies with  older stellar populations.  The  trend is statistically
significant  for both the  optical and  optical+NIR samples,  with the
slopes of \nablas  \, vs.  $Age$ relations greater  than zero at $\sim
3$ and  $\sim 9$~$\sigma$, respectively.   We note that  galaxies with
ages younger than  $\sim 4$~Gyr seems not to follow  the same trend as
galaxies with older  ages.  As discussed below, this  inversion in the
trend  of \nablas  \,  with $Age$  is  caused by  galaxies with  faint
spiral-like  morphological  features   and  less  accurate  structural
parameters.   The behavior  of  \nablas \,  with  metallicity is  less
clear. For the optical sample, the slope of the \nablas \, vs. $[Z/H]$
best-fit relation  is larger than  zero at over  7~$\sigma$.  However,
the trend is only weakly detected in the optical+NIR sample, where the
slope  is only marginally  larger than  zero, at  $\sim 2.2$~$\sigma$.
{   We   find   that}   \nablas   \,  strongly   depends   on   the
$\alpha$-enhancement parameter, in the sense that galaxies with larger
$[\alpha/Fe]$  also  have shallower  effective  color gradients.   The
slope of  the best-fitting line is significantly  different from zero,
at more than 5~$\sigma$, for both the optical and optical+NIR samples.

\begin{deluxetable}{l|c|c|c|c}
\small
 \tablecaption{Coefficients of the linear  fit of \nablas \, versus stellar
   population properties.} 
\tabletypesize{\scriptsize}
\tablewidth{0pt}
\tablecolumns{5}
\tablehead{    Property & \multicolumn{2}{c}{optical+NIR sample} & \multicolumn{2}{c}{optical sample} \\
          &  offset & slope & offset & slope }
\startdata
$\log Age \, (Gyr)$ & $ -0.529 \pm   0.058$ & $  0.227 \pm   0.063$ & $ -0.714 \pm   0.050$ & $  0.498 \pm   0.055$ \\ 
$[Z/H]    $ & $ -0.321 \pm   0.007$ & $  0.225 \pm   0.100$ & $ -0.263 \pm   0.004$ & $  0.430 \pm   0.057$ \\ 
$[\alpha/Fe]   $ & $ -0.239 \pm   0.016$ & $ -0.733 \pm   0.125$ & $ -0.174 \pm   0.010$ & $ -0.785 \pm   0.086$ \\ 
\enddata
\label{tab:fit_grads_sppar}
\end{deluxetable}

\begin{figure*}
\begin{center}
\plotone{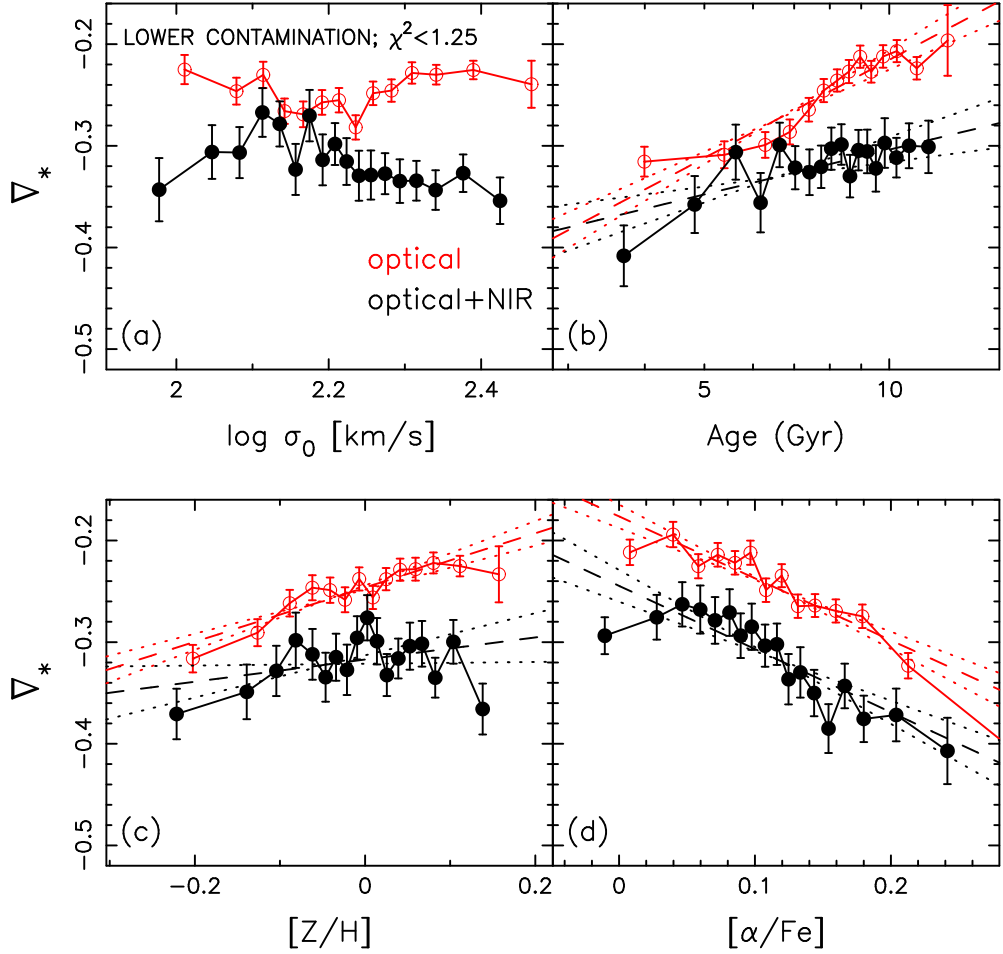}
\caption{The same as Fig.~\ref{fig:grads_spar} but for the optical+NIR
  and optical  samples of ETGs with lower  contamination from galaxies
  with faint  spiral-like morphological  features, and lower  value of
  the two-dimensional fitting $\chi^2$ in r-band ($\chi^2<1.5$).
\label{fig:grads_spar_best}
}
\end{center}
\end{figure*}

As discussed in~\papdata,  the SPIDER sample of ETGs  is affected by a
small  contamination from  late-type galaxies  with a  prominent bulge
component.  To analyze  the impact of this contamination,  we define a
purer  sample of  ETGs,  where  contamination is  reduced  to at  most
5~$\%$. To analyze if the trends of \nablas \, with stellar population
properties are affected by  the presence of late-type contaminants, we
define two subsamples of ETGs, by matching the optical and optical+NIR
samples  with the  lower  contamination sample  of~\papdata.  We  also
select  only galaxies  with better  quality structural  parameters, by
removing objects whose two-dimensional  fitting $\chi^2$ in any of the
available wavebands is larger than $1.25$ (see~\papdata).  The optical
and  optical+NIR   subsamples  include  $3,928$   and  $31,523$  ETGs,
respectively. Fig.~\ref{fig:grads_spar_best} plots  the same trends as
in    Fig.~\ref{fig:grads_spar}   but    for   the    better   quality
subsamples. The  coefficients of the best-fitting lines  of \nablas \,
vs.    age,   metallicity,    and   enhancement,   are   reported   in
Tab.~\ref{tab:fit_grads_sppar_nocont}.  The  better quality subsamples
exhibit  trends fully  consistent  with those  obtained  for the  full
samples.   In the case  of \nablas  \, vs.  age, we  do not  find that
galaxies with younger ages ($<4$~Gyr) have flatter gradients, as found
for the full samples.

A further issue is that of the fixed aperture size ($1.5''$ SDSS fiber
radius) where the stellar  population properties are measured. Because
of  radial  gradients in  stellar  population  properties, this  fixed
aperture  size might  produce a  correlation of  the  estimated $Age$,
$[Z/H]$, and  $[\alpha/Fe]$, parameters with galaxy  radius. Since the
\nablas's  correlate  with  the   (optical)  $R_e$  (see  Panel  a  of
Fig.~\ref{fig:grads_par}),  the aperture  effect might  also  bias the
correlation   of    the   $\nabla$'s   with    $Age$,   $[Z/H]$,   and
$[\alpha/Fe]$.   However,  as   shown  in   App.~\ref{sec:apcor},  the
variation   of   $R_e$   along   $\sigma_0$,   $Age$,   $[Z/H]$,   and
$[\alpha/Fe]$, is negligible  with respect to the full  range of $R_e$
values (Panel a of Fig.~\ref{fig:grads_par}), implying that the trends
in Fig.~\ref{fig:grads_spar}  are not affected at all  by the aperture
effect.

\begin{deluxetable}{l|c|c|c|c}
\small
 \tablecaption{The same  as Tab.~\ref{tab:fit_grads_sppar} for  the samples
   with  lower  contamination and  better  quality  of the  structural
   parameters.} 
\tabletypesize{\scriptsize}
\tablewidth{0pt}
\tablecolumns{5}
\tablehead{    Property & \multicolumn{2}{c}{optical+NIR sample} & \multicolumn{2}{c}{optical sample} \\
          &  offset & slope & offset & slope }
\startdata
$\log Age \, (Gyr)$ & $ -0.465 \pm 0.062$ & $  0.159 \pm   0.067$ & $ -0.539 \pm   0.046$ & $  0.325 \pm   0.050$ \\ 
$[Z/H]         $ & $ -0.319 \pm   0.008$ & $  0.121 \pm   0.101$ & $ -0.246 \pm   0.005$ & $  0.269 \pm   0.059$ \\ 
$[\alpha/Fe]   $ & $ -0.244 \pm   0.017$ & $ -0.626 \pm   0.130$ & $ -0.174 \pm   0.012$ & $ -0.621 \pm   0.094$ \\ 
\enddata
\label{tab:fit_grads_sppar_nocont}
\end{deluxetable}


\section{Age versus metallicity gradients in ETGs}
\label{sec:age_met_grads}
In  order to  separate  the  contribution of  age  and metallicity  to
$\nabla_\star$, we restrict the  analysis to the optical+NIR sample of
ETGs,   as  optical   data   alone  are   completely  ineffective   at
disentangling         the          two         components         (see
Sec.~\ref{sec:age_met_constrain}). First, we analyze the statistics of
the    distributions     of    age    and     metallicity    gradients
(Sec.~\ref{sec:age_met_stat}).    Sec.~\ref{sec:age_met_trends}  deals
with the trends of \nablat \, and \nablaz \, with respect to different
proxies of galaxy  mass, as well as stellar  population parameters. In
Sec.~\ref{sec:age_met_mass}, the  same trends are  analyzed separately
for low-  and high-mass galaxies.   In App.~\ref{sec:age_met_bias}, we
also discuss  some possible sources of  bias in the trends  of age and
metallicity  gradients,  such  as   the  role  of  internal  reddening
gradients in ETGs.

\subsection{Statistics of age and metallicity gradients}
\label{sec:age_met_stat}
In   Tab.~\ref{tab:stat_nablas_tz},  we   report  the   peak   of  the
distributions    of   age   ($\mu_{_{\nabla_t}}$)    and   metallicity
($\mu_{_{\nabla_Z}}$) gradients for  all the stellar population models
described in  Sec.~\ref{sec:grads}.  In agreement  with previous works
(e.g.~\citealt{PVJ90,  TaO03,  TaO05,  LBM03}),  we  find  that  color
gradients  imply the presence  of a  negative metallicity  gradient in
ETGs ($\nabla_Z <  0$), with the outer stellar  populations being less
metal-rich than the inner ones. The peak value of $\nabla_Z$ spans the
range of $-0.45$ to $-0.3$, depending on the stellar population model.
In agreement  with our  previous work (LdC09),  we find that  the peak
$\nabla_t$  is significantly  greater than  zero, i.e.   ETGs  have on
average  younger stars in  the center  than the  outskirts. $\nabla_t$
ranges from $0.05$ to $0.2$,  being significantly larger than zero for
all models.  Hence,  as discussed in LdC09, the  existence of a small,
but    significantly    positive   age    gradient    is   a    robust
(model-independent)  result.   From   the  values  of  $\nabla_Z$  and
$\nabla_t$ we see that, for all models, metallicity is the main driver
of color gradients.

\begin{deluxetable}{l|c|c}
\small
\tablecaption{Peak values of the  distributions of radial gradients in age
   ($\nabla_t$) and metallicity ($\nabla_Z$) of ETGs.}  
\tabletypesize{\scriptsize}
\tablewidth{0pt}
\tablecolumns{3}
\tablehead{    model & $\mu_{_{\nabla_Z}}$ &  $\mu_{_{\nabla_t}}$}
\startdata
$BC03            $ & $ -0.401 \pm   0.021$ & $  0.130 \pm   0.023$ \\
$M05             $ & $ -0.417 \pm   0.017$ & $  0.198 \pm   0.025$ \\
$CB10            $ & $ -0.309 \pm   0.010$ & $  0.052 \pm   0.012$ \\
$BC03_{\tau=1Gyr}$ & $ -0.456 \pm   0.023$ & $  0.134 \pm   0.019$ \\
\enddata
\label{tab:stat_nablas_tz}
\end{deluxetable}

\begin{figure}
\begin{center}
\plotone{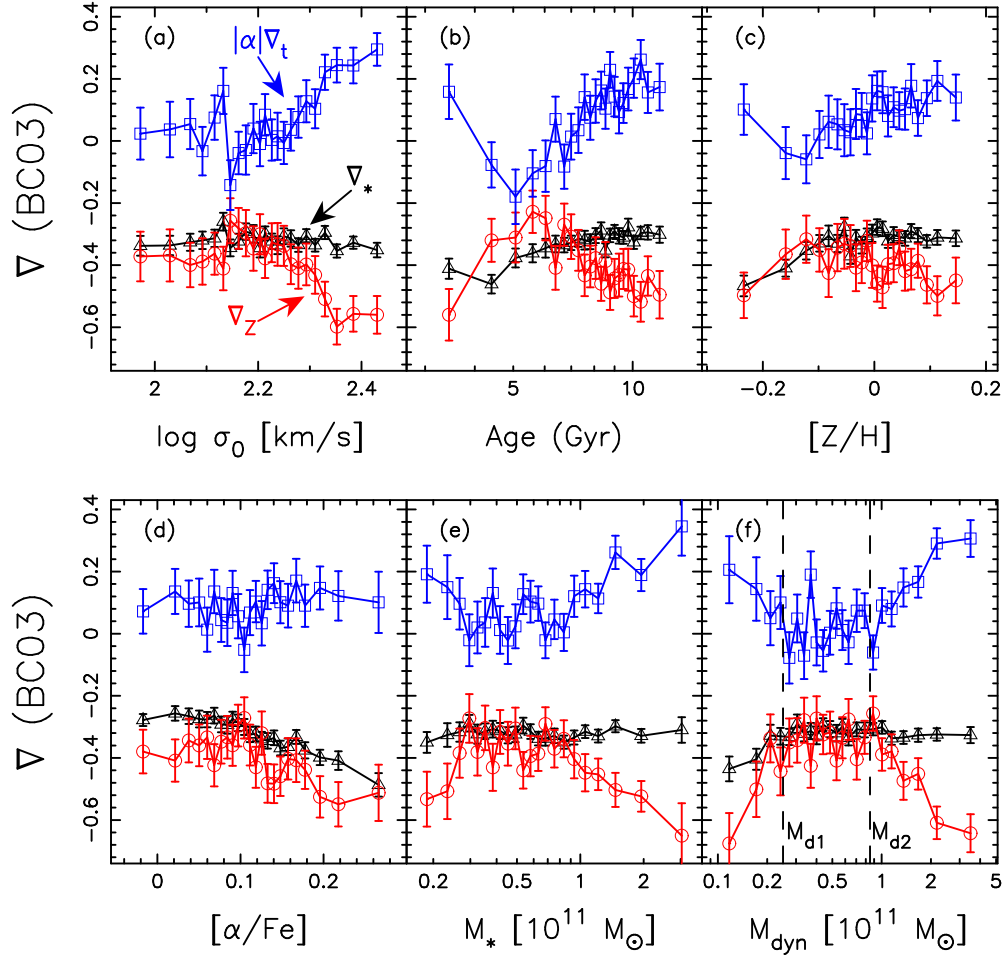}
\caption{Median  values  of  the  age  (blue)  and  metallicity  (red)
  gradients in ETGs as a  function of (a) central velocity dispersion,
  (b-d)  age, metallicity,  and $[\alpha/Fe]$,  parameters,  and (e-f)
  stellar and  dynamical masses.  The  plots refer to  the optical+NIR
  sample of ETGs.  Error bars denote 1~$\sigma$ standard errors on the
  medians.  For  each bin  of a given  quantity, the  median effective
  color gradient, $\nabla_\star$, is  also plotted in black color. Age
  gradients  are  multiplied  by   the  factor  $\alpha$,  defined  in
  Sec.~\ref{sec:grads},  so that,  for each  bin, the  effective color
  gradient can  be approximately  obtained by summing  up the  age and
  metallicity gradients in that bin (see Eq.~\ref{eq:nablas_mod}). The
  vertical dashed lines in panel (f) mark the $M_{dyn}$ used to define
  the {\it low-} ($M_{d1} \le M_{dyn} \le M_{d2}$) and {\it high-}mass
  ($M_{dyn} > M_{d2}$) subsamples of ETGs (see the text).
\label{fig:allgrads_pars}
}
\end{center}
\end{figure}

\begin{figure}
\begin{center}
\plotone{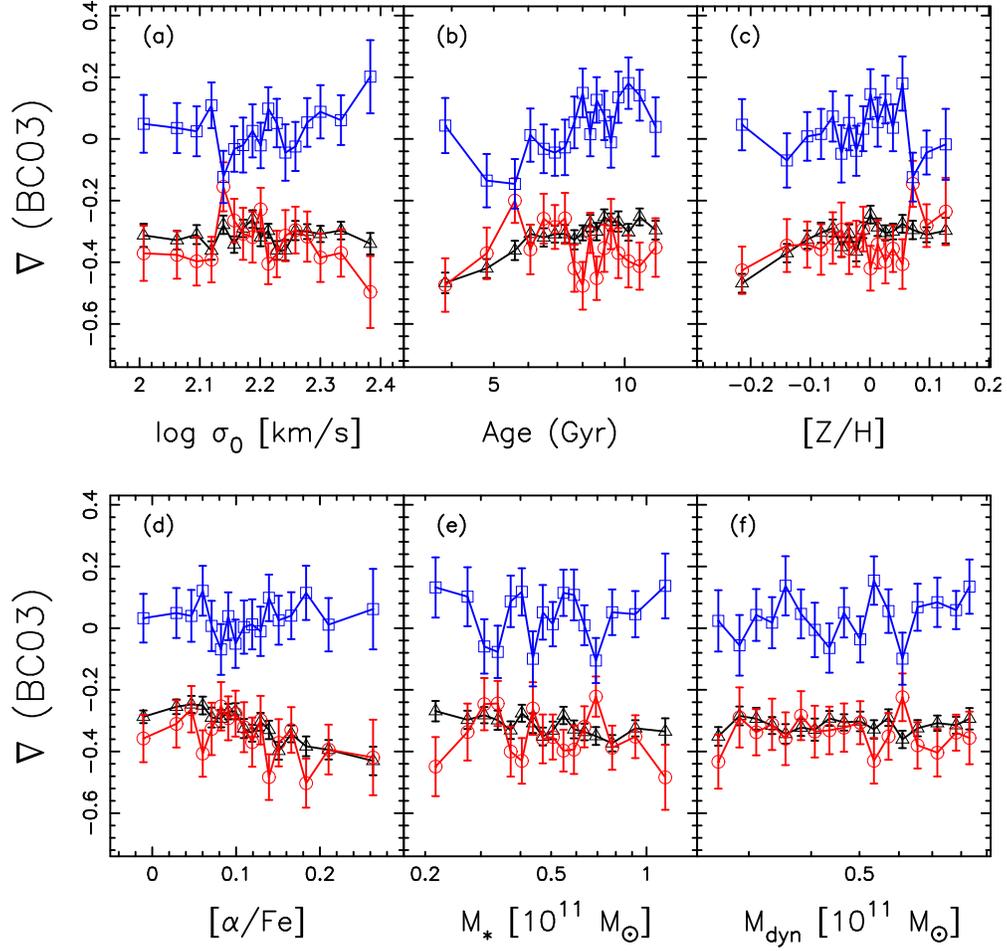}
\caption{The  same as  Fig.~\ref{fig:allgrads_pars} but  plotting only
  galaxies in the {\it low-mass} bin ($ 2.5 \times 10^{10} M_\odot \le
  M_{dyn} \le 8.5 \times 10^{10} M_\odot$).
\label{fig:allgrads_pars_low}
}
\end{center}
\end{figure}

\begin{figure}
\begin{center}
\plotone{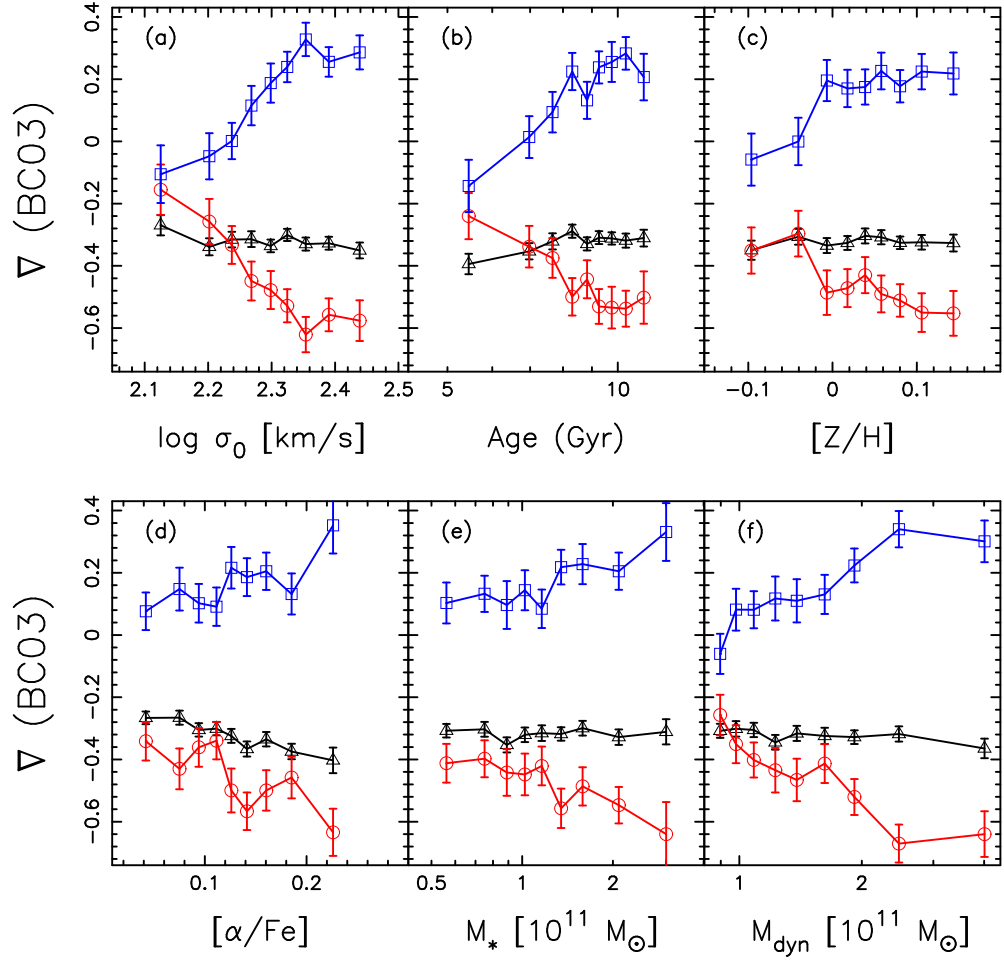}
\caption{The  same as  Fig.~\ref{fig:allgrads_pars} but  plotting only
  galaxies  in the  {\it  high-mass}  bin ($  M_{dyn}  > 8.5  \times
  10^{10} M_\odot$).
\label{fig:allgrads_pars_high}
}
\end{center}
\end{figure}

\subsection{Trends of age and metallicity gradients}
\label{sec:age_met_trends}

Fig.~\ref{fig:allgrads_pars} plots the median $\nabla_Z$ (red circles)
and $\nabla_t$ (blue squares)  in different bins of stellar population
parameters (age,  metallicity, and enhancement),  velocity dispersion,
stellar and dynamical mass. For each  bin of a given quantity, we also
show  the  median  $\nabla_\star$   in  that  bin  (black  triangles).
$\nabla_\star$, $\nabla_Z$ and $\nabla_t$ are estimated using BC03 SSP
models.  Age gradients  are multiplied  by the  absolute value  of the
$\alpha$ parameter (see Sec.~\ref{sec:grads}),  so that, for each bin,
the    sum     of    age    and     metallicity    gradients    gives,
approximately\footnote{For each galaxy, $\nabla_\star$ is equal to the
  sum    of   age    and   metallicity    gradients,    according   to
  Eq.~\ref{eq:nablas}.    On   the   other   hand,   the   values   of
  $\nabla_\star$,  $\nabla_t$, and $\nabla_Z$  in a  given bin  do not
  satisfy  exactly  Eq.~\ref{eq:nablas},   as  they  are  obtained  by
  computing the  median value of the  distributions of $\nabla_\star$,
  $\nabla_t$, and $\nabla_Z$ in that bin.}, $\nabla_\star$ in that bin
(see  Eq.~\ref{eq:nablas}).  This  allows us  to analyze  the relative
contribution of age  and metallicity to the trends  of color gradients
$wrt$ the different quantities:

\subsubsection{Trends with mass proxies}
For  {\it massive}  galaxies,  i.e.  \ls$\widetilde{>}2.2$  ($\sigma_0
\sim 160  \, km \,  s^{-1}$), $M_\star \widetilde{>} 6  \times 10^{10}
M_\odot$, and  $M_{dyn} \widetilde{>}  9 \times 10^{10}  M_\odot$, the
lack  of  correlation between  $\nabla_\star$  and either  $\sigma_0$,
$M_\star$,     or     $M_{dyn}$     (see     Secs.~\ref{sec:corr_mass}
and~\ref{sec:corr_sppar})  is  the result  of  opposing variations  of
metallicity and  age gradients $wrt$ such quantities.   The \nablaz \,
{\it decreases} while the \nablat \, {\it increases} as the mass proxy
increases.  The  two trends cancel  each other in the  resulting color
gradients.   The  origin  of  this  opposite behavior  is  visible  in
Fig.~\ref{fig:nabla_gX_mag}.   More luminous  (massive)  galaxies have
flatter  optical-optical  color  gradients  (e.g.~$\nabla_{g-r}$)  and
steeper optical-NIR gradients than  lower mass systems. Because of the
age--metallicity  degeneracy,  a  flatter  $\nabla_{g-r}$ may  be  due
either  to  a  positive  age  gradient  in  concert  with  a  negative
metallicity gradient, or an intrinsically weaker metallicity gradient.
The  latter case  is excluded  by the  optical-NIR gradients,  as this
would make the $\nabla_{g-K}$ flatten  with mass, while we observe the
opposite  trend.  Fig.~\ref{fig:allgrads_pars}  also shows  that  low-
(relative  to  high-)  mass  galaxies exhibit  different  trends.  For
\ls$\widetilde{<}2.2$  and  $M_\star  \widetilde{<} 6  \times  10^{10}
M_\odot$, no significant correlation of \nablaz \, and \nablat \, with
mass  proxy is  seen, while  for  low-$M_{dyn}$ objects  ($< 9  \times
10^{10} M_\odot$) a more  complex behavior is observed.  To illustrate
this,  we  define  two  characteristic dynamical  masses,  $M_{d1}=2.5
\times  10^{10}  M_\odot$  and  $M_{d2}=8.5 \times  10^{10}  M_\odot$,
marked   by   the   vertical    dashed   lines   in   Panel   (f)   of
Fig.~\ref{fig:allgrads_pars}.  For $M_{dyn} \widetilde{<} M_{d2}$ down
to  $M_{dyn} \sim  M_{d1}$,  there is  no  significant correlation  of
\nablaz  \,  and  \nablat   \,  with  $M_{dyn}$,  while  for  $M_{dyn}
\widetilde{<}   M_{d1}$  the  metallicity   gradient  tends   to  {\it
  decrease},  while  the age  gradient  tends  to  {\it increase},  as
$M_{dyn}$ decreases.   However, in this mass regime  the SPIDER sample
becomes incomplete  because of  the $r$-band magnitude  selection (see
Sec.~\ref{sec:samples}).   We  find  that  all the  above  trends  are
essentially unchanged regardless of the stellar population models used
to    estimate    the    age    and   metallicity    gradients    (see
Sec.~\ref{sec:grads}).

\subsubsection{Trends  with age}
For  galaxies  older  than  $\sim  5$~Gyr, we  find  that  \nablaz  \,
decreases,  while \nablat  \, increases  as  a function  of the  $Age$
parameter.  The trend of \nablat  \, is stronger than that of \nablaz,
making  the   color  gradient,  \nablas,  increase   with  $Age$  (see
Sec.~\ref{sec:corr_sppar}). For the M05 and $BC03_{\tau=1Gyr}$ models,
the behavior  of \nablat \, and \nablaz  \, with $Age$ is  the same as
for  BC03 SSP  models.  On  the  other hand,  for $CB10$,  we find  no
significant variation of \nablaz \, with $Age$, while \nablat \, still
increases  with the  $Age$.  This  confirms  that the  trend of  color
gradients with age, for  $Age \widetilde{>} 5$~Gyr, are mainly because
of \nablat.  For galaxies with $Age < 5$~Gyr, the trends of \nablaz \,
and  \nablat \, reverse.   Lower ages  imply more  negative (positive)
metallicity  (age)  gradients.  Even  in  this  case,  the \nablat  \,
dominates the trend, and the resulting color gradient increases as the
$Age$  decreases. This  result  holds for  all  the different  stellar
population models.

\subsubsection{Trends with  metallicity}
For galaxies more  metal-rich than $[Z/H] \sim -0.12$,  the \nablaz \,
tends  to  decrease,  while  the  \nablat  \,  tends  to  increase  as
metallicity increases.  The two trends cancel each  other producing no
variation of the median color  gradient, \nablas, as a function of the
$[Z/H]$ parameter, consistent with the \nablas--$[Z/H]$ slope reported
in Tab.~\ref{tab:fit_grads_sppar_nocont} for the optical+NIR sample of
ETGs. The  trends of \nablaz \, and  \, \nablat \, seem  to reverse at
low metallicity,  i.e.  $[Z/H] \widetilde{<} -0.12$ ,  similar to that
observed for  $Age$. In this  regime, the variation of  color gradient
with $[Z/H]$ is  dominated by \nablaz, which becomes  more negative as
$[Z/H]$ decreases. We found the  behavior of \nablat \, and \nablaz \,
with $Z/H$ to be independent of the adopted stellar population model.

\subsubsection{Trends  with the enhancement}
In contrast to the trends with age and metallicity, we do not find any
significant    variation    of    $\nabla_t$    with    $[\alpha/Fe]$.
Fig.~\ref{fig:allgrads_pars}  reveals that  the strong  correlation of
\nablas  \, with enhancement  is because  of the  metallicity gradient
decreasing with $[\alpha/Fe]$.

\subsection{Low- and high-mass ETGs}
\label{sec:age_met_mass}
Since the trend of age  and metallicity gradients $wrt$ different mass
proxies depends on the range of galaxy mass, we bin the sample of ETGs
according  to $M_{dyn}$,  and analyze  the  trends of  \nablaz \,  and
\nablat \, in  each bin.  We define two bins,  including {\it low-} ($
M_{d1}  <  M_{dyn}  <  M_{d2}$),  and {\it  high-}mass  ($M_{dyn}  \ge
M_{d2}$)  galaxies, where  $M_{d1}$  and $M_{d2}$  are defined  above.
Although   the  sample   of  ETGs   becomes  incomplete   at  $M_{dyn}
\widetilde{<}  3 \times 10^{10}  M_\odot$, we  adopt a  slightly lower
mass limit  of $M_{dyn} \sim 2.5  \times 10^{10} M_\odot$,  as this is
approximately  the $M_{dyn}$  at which  the  trend of  \nablat \,  and
\nablaz   \,   with   mass   changes   behavior  (see   Panel   f   of
Fig.~\ref{fig:allgrads_pars}).

Figs.~\ref{fig:allgrads_pars_low} and~\ref{fig:allgrads_pars_high} are
the same  as Fig.~\ref{fig:allgrads_pars},  but plot only  galaxies in
the low-  and high-mass bins,  respectively. In order to  quantify the
trends of \nablat, \nablaz, and \nablas \, $wrt$ a given parameter, we
fit each trend with a  linear relation, minimizing the sum of absolute
residuals of  median gradients (see  Sec.~\ref{sec:corr_strpar}).  The
slopes     and     their     uncertainties     are     reported     in
Tabs.~\ref{tab:fit_nablas_low}  and~\ref{tab:fit_nablas_high}  for the
low-      and       high-mass      bins,      respectively.       From
Figs.~\ref{fig:allgrads_pars_low} and~\ref{fig:allgrads_pars_high}, as
well  as Tabs.~\ref{tab:fit_nablas_low} and~\ref{tab:fit_nablas_high},
we see that:
\begin{description}
\item[\it   --  Low-mass.]   The   {\it  effective}   color  gradient,
  $\nabla_\star$,  does  not  show  any significant  correlation  with
  either  $\sigma_0$  or  $M_{dyn}$,  while  it  tends  to  marginally
  decrease with  stellar mass, as  the \nablas--$M_\star$ \,  slope in
  Tab.~\ref{tab:fit_nablas_low} is  less than zero  ($-0.12 \pm 0.05$)
  at  the  $2.2~\sigma$ confidence  level.   No  significant trend  of
  \nablaz  \,  and  \nablat  \,  is  detected  $wrt$  any  mass  proxy
  ($M_{dyn}$, $M_{\star}$,  and $\sigma_0$).   On the other  hand, for
  stellar  population parameters,  we  find that  \nablas \,  strongly
  correlates with $Age$ and  $[\alpha/Fe]$.  The {\it effective} color
  gradient increases  for larger $Age$,  and becomes more  negative as
  the $[\alpha/Fe]$  increases.  The slopes of  the \nablas--$Age$ and
  \nablas--$[\alpha/Fe]$  linear fits  differ from  zero at  more than
  $4~\sigma$,  proving that  the  trends are  highly significant.  The
  \nablas \, also tends to increase with metallicity, but the slope of
  the \nablas--$[Z/H]$  fit is larger than zero  at only $2.2~\sigma$.
  As noted above, the trend of  \nablas \, with $Age$ is mainly due to
  \nablat, while that with $[\alpha/Fe]$ is driven by \nablaz.

 \item[\it -- High-mass.] For  high-mass galaxies, the {\it effective}
   color gradient does not  vary significantly with $Age$, metallicity
   nor with  any mass proxy. In  fact, the slopes  of the correlations
   involving  \nablas  \,  in Tab.~\ref{tab:fit_nablas_high}  are  all
   consistent   with  zero   at   less  than   $2~\sigma$.   For   the
   $[\alpha/Fe]$,  there   is  a  significant  trend   of  \nablas  \,
   decreasing  as  the  enhancement   increases.   The  slope  of  the
   \nablas--$[\alpha/Fe]$  linear fit ($-0.84  \pm 0.2$)  differs from
   zero at over  $4~\sigma$, and is consistent with  that obtained for
   low-mass ETGs ($-0.68 \pm 0.17$, see Tab.~\ref{tab:fit_nablas_low})
   and   for    the   entire    sample   ($-0.73   \pm    0.13$,   see
   Tab.~\ref{tab:fit_grads_sppar}).   For   all  quantities,  we  find
   anti-correlated variations of \nablaz  \, and \nablat, which cancel
   each  other  in the  trends  of  \nablas.   For $[\alpha/Fe]$,  the
   metallicity gradient dominates, resulting in the strong correlation
   of \nablas \, and $[\alpha/Fe]$.
\end{description}

From         Figs.~\ref{fig:allgrads_pars},~\ref{fig:allgrads_pars_low}
and~\ref{fig:allgrads_pars_high}, we  see that positive  age gradients
in ETGs are  more associated with high rather  than low mass galaxies.
At  high mass,  the  age  gradient strongly  increases  as a  function
$\sigma_0$, $M_{dyn}$, and to a lesser extent with stellar mass, while
these  trends are not  observed for  low-mass galaxies.   However, for
both low- and high-mass, \nablat \, becomes significantly positive for
galaxies older than $\sim 7$~Gyr.   This implies that age gradients in
ETGs are mainly driven by  mass, and only secondarily by $Age$.  Panel
(b) of  Fig.~\ref{fig:allgrads_pars} also shows  that for $Age<5$~Gyr,
the trend of  \nablat \, with $Age$ reverses:  the younger the age,the
more positive the age gradient becomes. Consistent with the downsizing
picture  (\citealt{Cowie:96}), galaxies  at  $Age<5$~Gyr are  low-mass
systems, as  shown by the  inversion in the \nablat-$Age$  trend being
observed    only    for    the    low-mass    bin    (Panel    b    of
Fig.~\ref{fig:allgrads_pars_low}).

\begin{figure}
\begin{center}
\epsscale{0.5}
\plotone{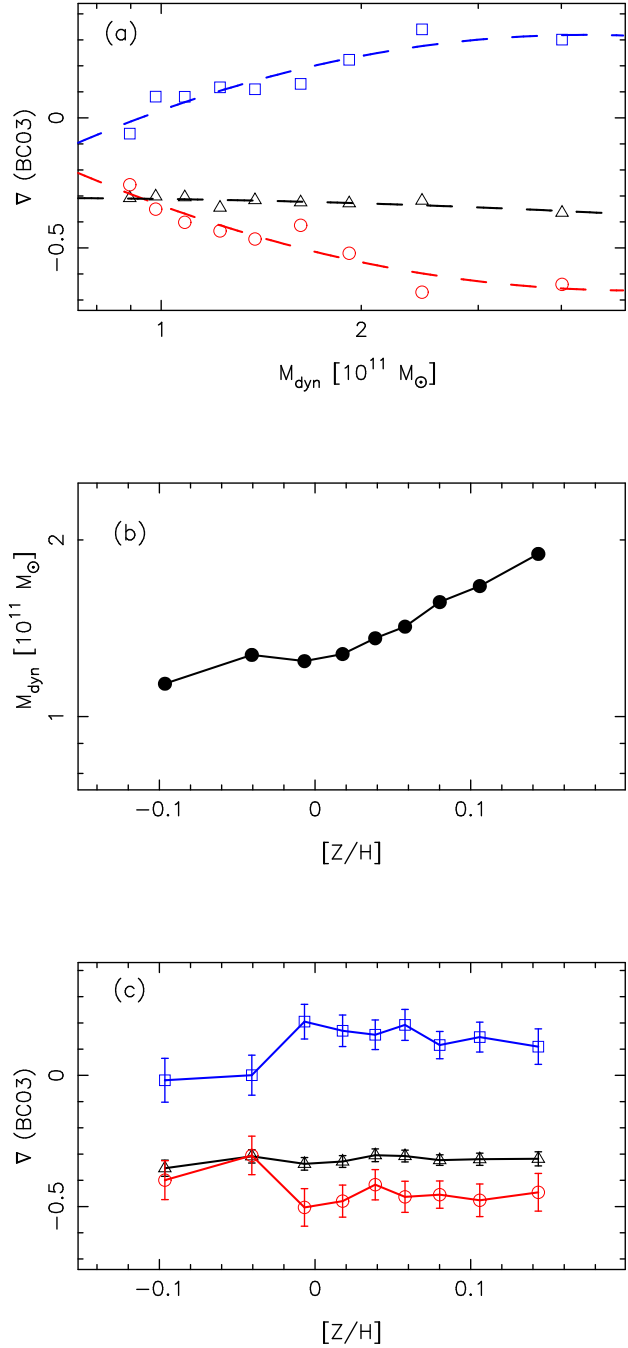}
\caption{Example  of the  correction procedure  adopted to  remove the
  correlation among the $\nabla$'s  and dynamical mass from the trends
  of   \nablas,  \nablaz,   and  \nablat,   $wrt$   galaxy  parameters
  (Figs.\ref{fig:allgrads_pars_low} and~\ref{fig:allgrads_pars_high}).
  Top  panel plots  the  $\nabla$'s  as a  function  of $M_{dyn}$  for
  high-mass galaxies.  Different symbols are  the same as in panel (f)
  of Fig.~\ref{fig:allgrads_pars_high}.  Dashed curves are third-order
  polynomial fits to  the data.  The mid panel  plots the median value
  of $M_dyn$  in different bins of  metallicity, $[Z/H]$.  Metallicity
  bins     as     the     same     as     for     panel     (b)     of
  Fig.~\ref{fig:allgrads_pars_high}.   The   bottom  panel  shows  the
  results of  the correction procedure: for each  metallicity bin, the
  polynomial fits are used to estimate the expected $\nabla$'s in that
  bin.  The expected $\nabla$'s are subtracted from the data-points in
  the top  panel. The resulting values are  re-normalized, by suitable
  additive terms, to match the  median values of the \nablas, \nablaz,
  and \nablat \, distributions.
\label{fig:corr_proc}
}
\end{center}
\end{figure}

\begin{figure}
\begin{center}
\epsscale{1.}
\plotone{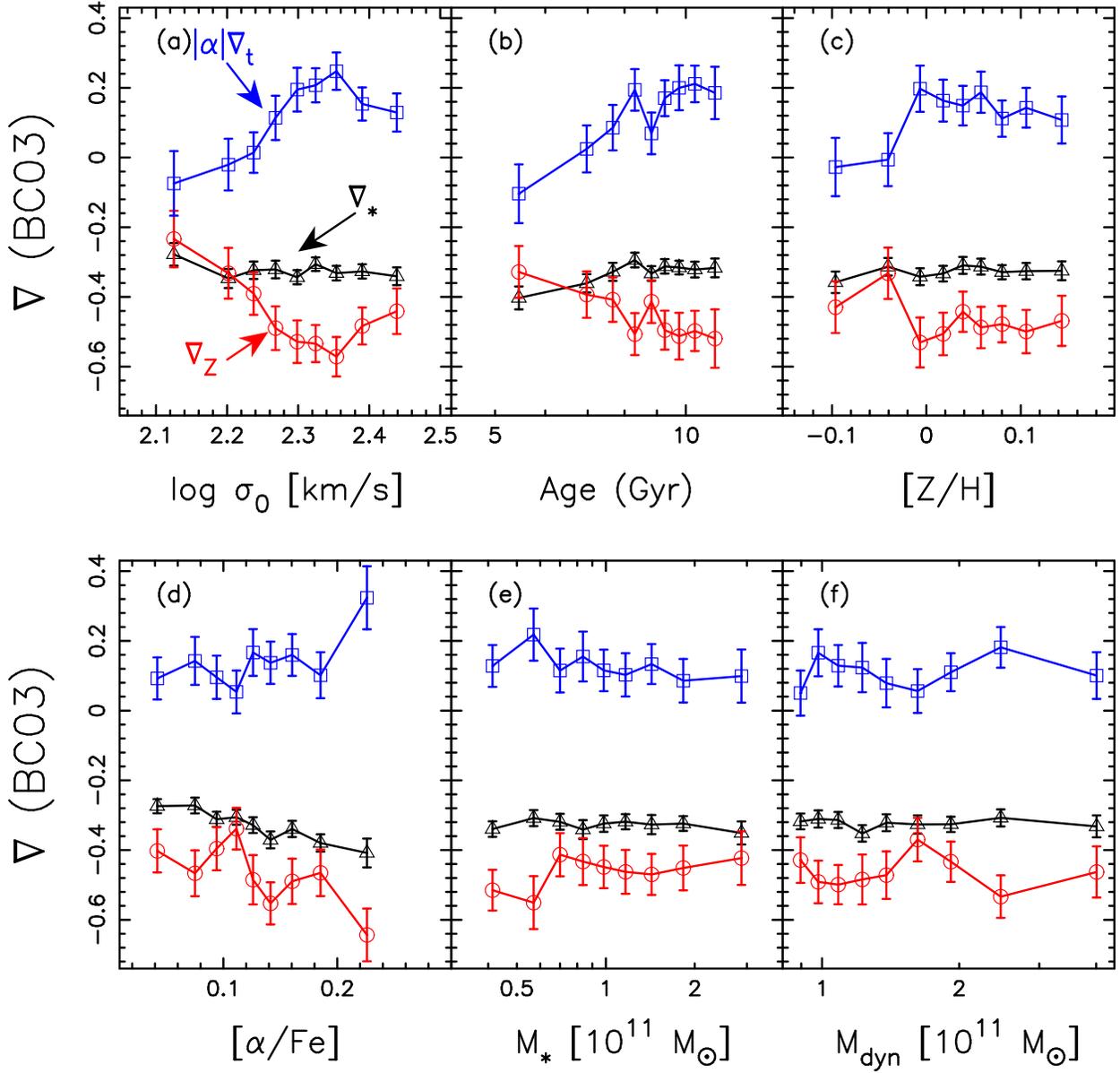}
\caption{The same  as Fig.~\ref{fig:allgrads_pars_high} but correcting
  the $\nabla$ in each panel for the correlations with dynamical mass.
\label{fig:allgrads_pars_highmass_corr_mdyn}
}
\end{center}
\end{figure}

\begin{figure}
\begin{center}
\epsscale{1.}
\plotone{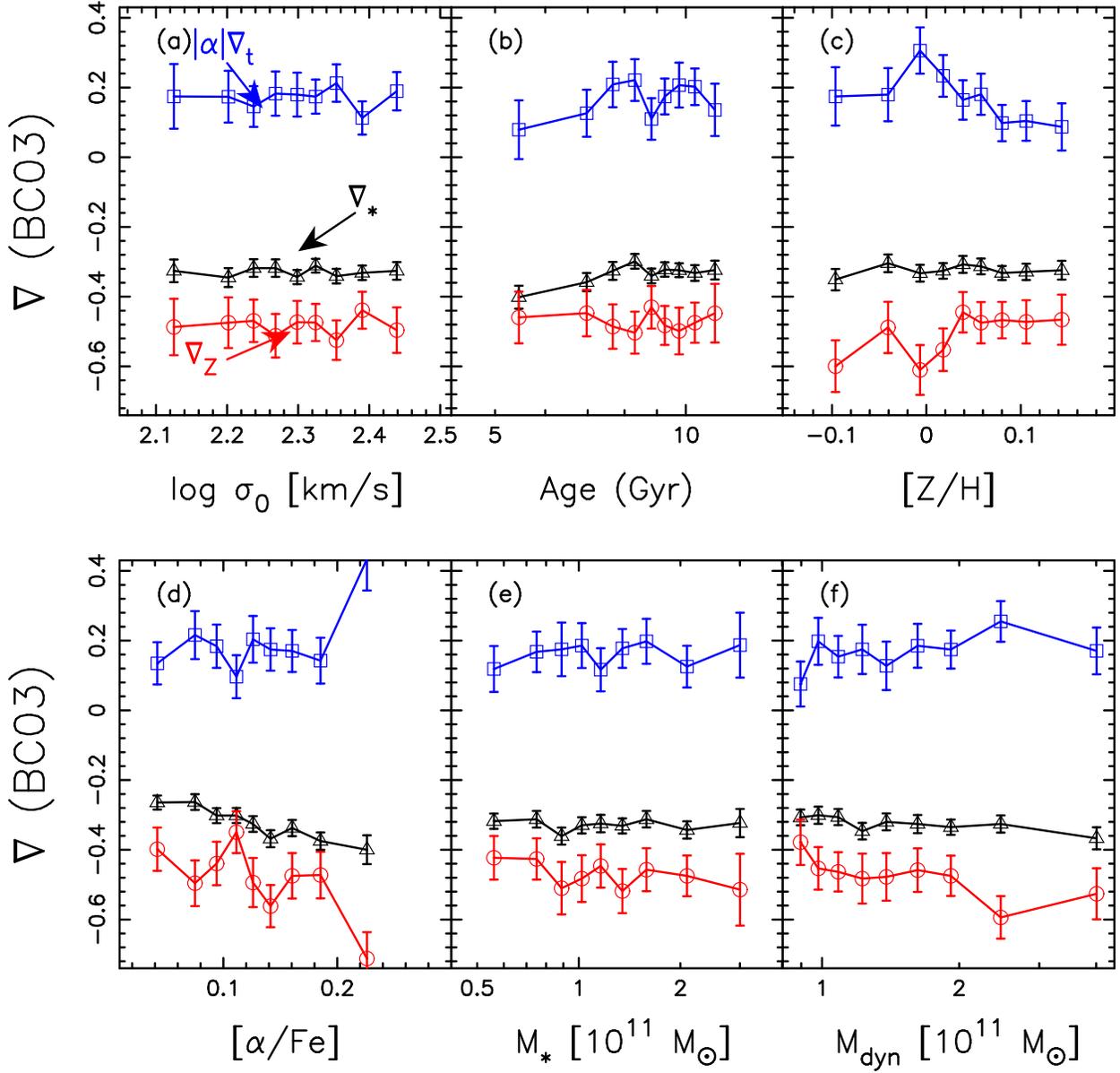}
\caption{The same  as Fig.~\ref{fig:allgrads_pars_high} but correcting
  the  $\nabla$ in each  panel for  the correlations  with $\sigma_0$.
  Note how the correction removes  all the correlations but those with
  the $\alpha$-enhancement.
\label{fig:allgrads_pars_highmass_corr_sig0}
}
\end{center}
\end{figure}

\begin{figure}
\begin{center}
\plotone{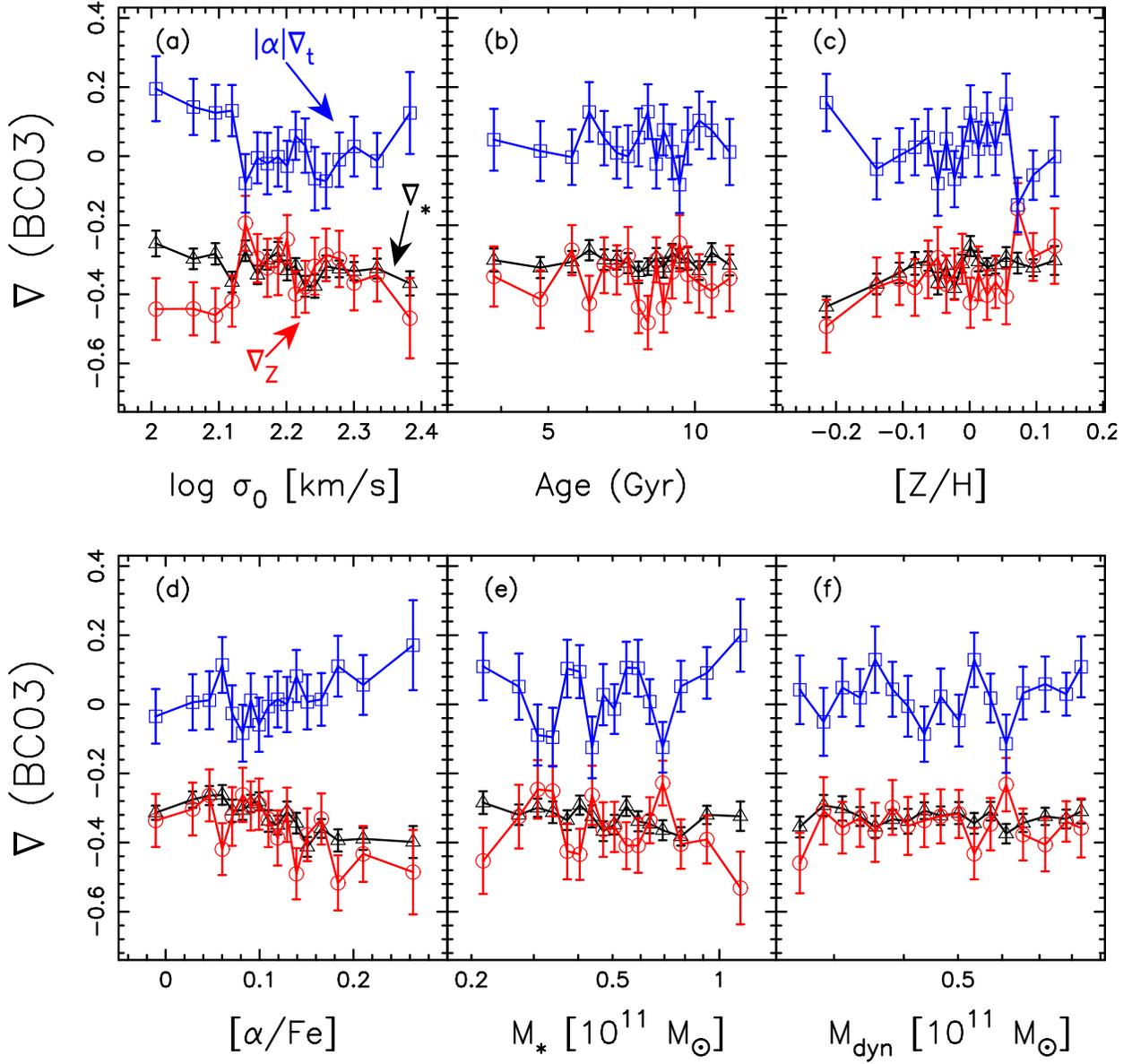}
\caption{The  same as Fig.~\ref{fig:allgrads_pars_low}  but correcting
  the  $\nabla$ in  each panel  for  the correlations  with the  $Age$
  parameter.
\label{fig:allgrads_pars_lowmass_corr_age}
}
\end{center}
\end{figure}

Stellar  population  parameters,   such  as  $Age$,  metallicity,  and
enhancement, are known to correlate  with proxies of galaxy mass, such
as velocity dispersion  and stellar mass (see e.g.~\citealt{Thomas:05,
  Nelan:05,  GALL:06}).  Since for  high-mass  galaxies  we find  that
internal  gradients correlate  with both  mass and  stellar population
properties, a natural question is  if these trends are just because of
the  correlation among  stellar  population properties  and mass.   In
order to address this issue, we apply a correction procedure that {\it
  removes} the  correlations of each $\nabla$ quantity  with mass from
all      the      trends     in      Figs.~\ref{fig:allgrads_pars_low}
and~\ref{fig:allgrads_pars_high}.  The  procedure  is  illustrated  in
Fig.~\ref{fig:corr_proc},  where  we  show   how  the  trends  of  the
$\nabla$'s  $wrt$  to  $[Z/H]$  are  corrected, for  galaxies  in  the
high-mass bin:

\begin{description}
\item[i.] First,  we model the $\nabla$'s-$M_{dyn}$  trends with third
  order polynomials in $M_{dyn}$ (dashed curves in the top panel);
\item[ii.] We compute the median $M_{dyn}$ in each bin of $[Z/H]$ (mid
  panel).    As    expected   from   the    $[Z/H]$-mass   correlation
  (e.g.~~\citealt{Thomas:05}),   the   mass   increases   as   $[Z/H]$
  increases;
\item[iii.]  In each bin  of $[Z/H]$,  using the  corresponding median
  $M_{dyn}$, we calculate the expected values of \nablas, \nablaz, and
  \nablat \, from the best-fit polynomials;
\item[iv.] The expected $\nabla$'s  in each $[Z/H]$ bin are subtracted
  from the  measured $\nabla$ in  that bin; the subtracted  values are
  re-normalized,  by suitable additive  terms, to  the medians  of the
  \nablas,  \nablaz, and  \nablat  \, distributions  (bottom panel  in
  Fig.~\ref{fig:corr_proc}).  Comparing  top   and  bottom  panels  in
  Fig.~\ref{fig:corr_proc}  illustrates how  the  correction procedure
  (partly) removes  the correlations of \nablas,  \nablaz, and \nablat
  \, with $[Z/H]$.
\end{description}
We  apply this  correction scheme  for both  low- and  high-mass ETGs,
modeling   the   $\nabla$'s-$M_{dyn}$    trends   (point   i.   above)
independently for each case. We obtain the following results:
\begin{description}
\item[\it                        --                        High-mass.]
  Fig.~\ref{fig:allgrads_pars_highmass_corr_mdyn}     exhibits     the
  correlations  of  \nablas,  \nablaz,  and  \nablat,  with  different
  quantities,  for  high-mass galaxies  ($M_{dyn}  \ge M_{d2}$),  {\it
    after removing} the  $\nabla$'s-$M_{dyn}$ trends. By construction,
  the  procedure  removes  the  correlations of  the  $\nabla$'s  with
  $M_{dyn}$  (Panel  f).  However,  it  does  not  impact  the  trends
  involving stellar population properties. This is because such trends
  are  due  to  the  $\nabla  - \sigma_0$  correlations  (rather  than
  $\nabla$'s-$M_{dyn}$).         This        is        shown        in
  Fig.~\ref{fig:allgrads_pars_highmass_corr_sig0}, where we repeat the
  entire procedure by  removing the $\nabla$'s-$\sigma_0$ trends.  The
  $\nabla$'s-$\sigma_0$  corrections remove all  the trends,  with the
  remarkable exception of the  correlations among \nablas \, (\nablaz)
  and the enhancement. A  linear fit to the \nablas--$[\alpha/Fe]$ trend
  in Fig.~\ref{fig:allgrads_pars_highmass_corr_sig0}  gives a slope of
  $-0.8 \pm 0.2$, still fully  consistent with that of the uncorrected
  trend ($-0.84 \pm 0.19$, see Tab.~\ref{tab:fit_nablas_high}).
\item[\it --  Low-mass.] In  the mass range  of $M_{d1}$  to $M_{d2}$,
  there  is no  variation of  \nablas \,  with velocity  dispersion or
  dynamical  mass.   Hence,  the  correlations between  gradients  and
  stellar population  parameters can not arise  from age, metallicity,
  and enhancement  varying with mass. However, at  a given $\sigma_0$,
  the stellar  population parameters  are correlated with  each other,
  with   age   being  anti-correlated   with   both  metallicity   and
  enhancement~\citep{Graves:09}. To account for this, we re-derive the
  trends between  the $\nabla$'s and other parameters  by removing the
  correlations  with the  $Age$. The  procedure  is the  same as  that
  adopted to remove the $\nabla$'s-$M_{dyn}$ and $\nabla$'s-$\sigma_0$
  correlations     for     high-mass     galaxies     (see     above).
  Fig.~\ref{fig:allgrads_pars_lowmass_corr_age}  shows the  same plots
  as Fig.~\ref{fig:allgrads_pars_low},  but after the $\nabla$'s-$Age$
  trends  are removed.  We see  that this  correction does  not affect
  significantly the  correlations of the  $\nabla$'s $wrt$ metallicity
  and enhancement.  In particular, after the correction  is applied, a
  linear fit  of \nablas  \, $wrt$  $Z/H$ gives a  slope of  $0.29 \pm
  0.12$, fully  consistent with  that of $0.33  \pm 0.15$  reported in
  Tab.~\ref{tab:fit_nablas_low}.   For \nablas \,  versus $[\alpha/Fe]$,
  the  slope  is $-0.59  \pm  0.15$,  still  consistent with  that  in
  Tab.~\ref{tab:fit_nablas_low} ($-0.84 \pm  0.19$).  It follows that,
  for  low-mass systems,  all the  correlations of  internal gradients
  with stellar  population parameters  are independent of  each other,
  i.e. the  gradients exhibit a  genuine correlation with  each single
  parameter (in particular age and enhancement).
\end{description}

\begin{deluxetable}{l|c|c|c}
\small
\tablecaption{Slopes  of the linear  fits of  median $\nabla_t$
   (col.~2),  $\nabla_\star$ (col.~3), and  $\nabla_Z$ (col.~3),  as a
   function  of different galaxy  parameters. Each  line in  the table
   corresponds to  a different parameter, as reported  in col.~1.  The
   slopes refer to  the sample  of ETGs  in the  {\it low-mass}
   bin.} 
\tabletypesize{\scriptsize}
\tablewidth{0pt}
\tablecolumns{4}
\tablehead{& $\nabla_t$ &  $\nabla_\star$ & $\nabla_Z$ }
\startdata
 $\log \sigma_0            $ & $    0.18 \pm    0.25$ & $   0.050 \pm   0.104$ & $   -0.16 \pm    0.24$  \\ 
 $Age (Gyr)                $ & $    0.59 \pm    0.25$ & $   0.406 \pm   0.091$ & $   -0.10 \pm    0.21$  \\ 
 $[Z/H]                    $ & $   -0.02 \pm    0.30$ & $   0.338 \pm   0.151$ & $    0.33 \pm    0.27$  \\ 
 $[\alpha/Fe]              $ & $    0.16 \pm    0.37$ & $  -0.681 \pm   0.165$ & $   -0.48 \pm    0.38$  \\ 
 $M_\star [10^{11} M_\odot]$ & $   -0.01 \pm    0.17$ & $  -0.111 \pm   0.055$ & $   -0.05 \pm    0.15$  \\ 
 $M_{dyn} [10^{11} M_\odot]$ & $    0.19 \pm    0.16$ & $   0.009 \pm   0.068$ & $   -0.13 \pm    0.16$  \\ 
\enddata
\label{tab:fit_nablas_low}
\end{deluxetable}

\begin{deluxetable}{l|c|c|c}
\small
 \tablecaption{The same as Tab.~\ref{tab:fit_nablas_low} but for
   galaxies in the {\it high-mass} bin.
}
\tabletypesize{\scriptsize}
\tablewidth{0pt}
\tablecolumns{4}
\tablehead{& $\nabla_t$ &  $\nabla_\star$ & $\nabla_Z$ }
\startdata
 $\log \sigma_0            $ & $    1.51 \pm    0.24$ & $  -0.193 \pm   0.129$ & $   -1.53 \pm    0.31$  \\ 
 $Age (Gyr)                $ & $    1.54 \pm    0.33$ & $   0.229 \pm   0.132$ & $   -1.05 \pm    0.30$  \\ 
 $[Z/H]                    $ & $    1.26 \pm    0.42$ & $   0.006 \pm   0.165$ & $   -0.98 \pm    0.37$  \\ 
 $[\alpha/Fe]              $ & $    1.00 \pm    0.54$ & $  -0.756 \pm   0.190$ & $   -1.36 \pm    0.45$  \\ 
 $M_\star [10^{11} M_\odot]$ & $    0.26 \pm    0.13$ & $  -0.020 \pm   0.052$ & $   -0.26 \pm    0.13$  \\ 
 $M_{dyn} [10^{11} M_\odot]$ & $    0.56 \pm    0.15$ & $  -0.082 \pm   0.060$ & $   -0.55 \pm    0.15$  \\ 
\enddata
\label{tab:fit_nablas_high}
\end{deluxetable}

\section{Summary}

In this paper, we examined a sample of 4,546 ETGs with $grizYJHK$ data
available.  {  We consider  an ETG to  be a  bulge-dominated galaxy
  with  a passive  spectrum in  its  central region  (within the  SDSS
  fiber).  As such, the results obtained here do not apply exclusively
  to ellipticals, but  to S0 galaxies and galaxies  with a red central
  bulge (although  the effect of late-type contaminants  is expected to
  be negligible, see Secs.~3.3 and~8.3). However, there is no a priori
  reason  to expect  that  the formation  mechanisms and  evolutionary
  pathways are the same for all sub-classes of what we call ETGs}.

We  measured  structural  parameters  in  all  wavebands  and  stellar
population indicators from the spectra.  We describe a new approach to
constrain age  and metallicity  gradients from the  color information.
Also, we  define an  effective color gradient,  $\nabla_{\star}$, that
uses  all  the  color  terms  provided by  the  photometric  data  and
investigate   its   dependence  on   several   parameters  like   age,
metallicity, mass  and $[\alpha/Fe]$.   { Although the  estimate of
  stellar population  parameters (Sec.~\ref{sec:stellar_pop}) and that
  of stellar  masses and  age/metallicity gradients rely  on different
  stellar population models (BC03 vs.  MILES), we find that all trends
  shown  here remain  unchanged  when using  a  variety of  population
  synthesis  models  to  infer  the  gradients  and  estimate  stellar
  masses.}  In the following we summarize our main findings:

\begin{description}
\item[1)]  We present  a  new scheme  to  quantify stellar  population
  gradients  from color  gradients.  We  introduce an  effective color
  gradient,  $\nabla_{\star}$, which  reflects all  contributions from
  the  optical+NIR colors. $\nabla_{\star}$  correlates well  with all
  color gradients  measured independently  (as expected) and  is model
  independent,  while  $\nabla_{\rm  Z}$  and  $\nabla_{\rm  t}$  {
    (i.e. the metallicity and age gradients)} are not.

\item[2)]  For photometric properties,  we find  that $\nabla_{\star}$
  has a mild correlation with the Sersic index; does not change at all
  with the axis ratio; and strongly correlates with total galaxy color
  - systems   with   bluer  colors   tend   to   have  more   negative
  $\nabla_{\star}$. Using  optical+NIR data  we also find  that larger
  and  more  luminous  ETGs  have  more  negative  $\nabla_{\star}$'s,
  although   if  we   restrict  ourselves   to  only   $K$-band  data,
  $\nabla_{\star}$ shows no dependence on radius or luminosity.

\item[3)] For galaxies more massive than $M_{dyn} > 8.5 \times 10^{10}
  \,  M_\odot$,  the age  gradient  increases,  while the  metallicity
  gradient  decreases  as functions  of  mass,  age, metallicity,  and
  enhancement.  The  trends cancel each  other in the  color gradient,
  with the exception of those  for the enhancement: the color gradient
  decreases  as  a  function  of  $[\alpha/Fe]$.   All  trends  are  a
  consequence  of  the  correlations  of  the  $\nabla$'s  $wrt$  both
  velocity dispersion and $[\alpha/Fe]$. These parameters are the main
  drivers  of the internal  age, metallicity,  and color  gradients in
  massive ETGs.

\item[4)] For less  massive galaxies, $ 2.5 \times  10^{10} \, M_\odot
  \le  M_{dyn}   \le  8.5  \times  10^{10}  \,   M_\odot$,  no  strong
  correlation  of age,  metallicity, and  color gradients  is detected
  $wrt$  mass.   {However}, color  gradients  strongly correlate  with
  stellar   population   parameters,   and  these   correlations   are
  independent of each other.

\item[5)]  In both mass  regimes, there  is a  strong anti-correlation
  between the color gradient and $\alpha$-enhancement, that originates
  from the  metallicity gradient decreasing  with $[\alpha/Fe]$.  This
  could  result from  the  star formation  and metallicity  enrichment
  being regulated  by the interplay  between the input of  energy from
  supernovae, and the temperature and pressure of the hot X-ray gas in
  ETGs.

\item[6)] In both mass  regimes, positive age gradients are associated
  with old galaxy ages  ($>5-7$~Gyr).  For galaxies younger than $\sim
  5$~Gyr,  mostly   at  low-mass,  the   age  gradient  tends   to  be
  anti-correlated  with  the   $Age$  parameter,  with  more  positive
  gradients at younger ages.
\end{description}

{ We have  studied the correlation of color  gradients in ETGs with
  intrinsic  galaxy  properties,   using  an  extensive  sample,  with
  excellent photometric and spectroscopic data over a broad wavelength
  baseline, and state-of-the-art  analysis methods. However, even with
  such tools  and data at our  disposal, it remains  difficult to gain
  substantive  insight  into  the  details of  how  galaxies  actually
  form. The diagnostic value of color gradients will only become fully
  apparent  when  model predictions  become  sufficiently specific  to
  realize the  full potential of  the high quality  observational data
  and methods developed to estimate color gradients.}

\appendix

\section{The $\alpha$-MILES models}
\label{app:MILES}
The  $\alpha$-MILES  SSP  models   are  constructed  as  described  in
Cervantes  et.   al.~(2007),  using  both  empirical  and  theoretical
stellar    libraries.    The    empirical   libraries,    like   MILES
(Sanchez-Blazquez et al.  2006), contain spectra of stars in the solar
neighborhood,  but mostly lack  bulge-stars, with  non-solar abundance
ratios.   Hence, they  do not  allow a  detailed spectral  modeling of
metal-rich,  alpha-enhanced   systems,  such  as   bright  ETGs.   The
$\alpha$-MILES  models  complement  the  empirical libraries,  in  the
non-solar abundance  regime, with the  synthetic library of  Coelho et
al.~(2005),  consisting of  high-resolution synthetic  stellar spectra
covering a wide range  of stellar atmosphere's parameters.  The Coelho
et al.~(2005) library covers both solar and alpha-enhanced mixtures in
a wide baseline, from 3000  \AA \, to 1.4~$\mu$m, superseding previous
versions by Barbuy  et al.  (2003), in the  wavelength range 4600-5600
\AA,  and Zwitter  et  al.(2004),  in the  range  7653-8747 \AA.   The
resulting $\alpha$-enhanced models consist  of SED's covering the same
spectral range  ($3525-7500$ \AA),  with the same  spectral resolution
(2.3 \AA), as the (solar abundance) MILES library.  The $\alpha$-MILES
consists of $1,170$ SSPs, corresponding  to: twenty-six ages from 1 to
18~Gyr; five  $[Z/H]$ from $-1.28$  to $+0.2$, and  nine $[\alpha/Fe]$
abundance ratios from  $-0.2$ to $+0.6$.  From this  set of models, we
run   STARLIGHT   by  extracting   a   sub-set   of  $176$   SSPs~(see
Sec.~\ref{sec:stellar_pop}).

\begin{figure*}
\begin{center}
\plotone{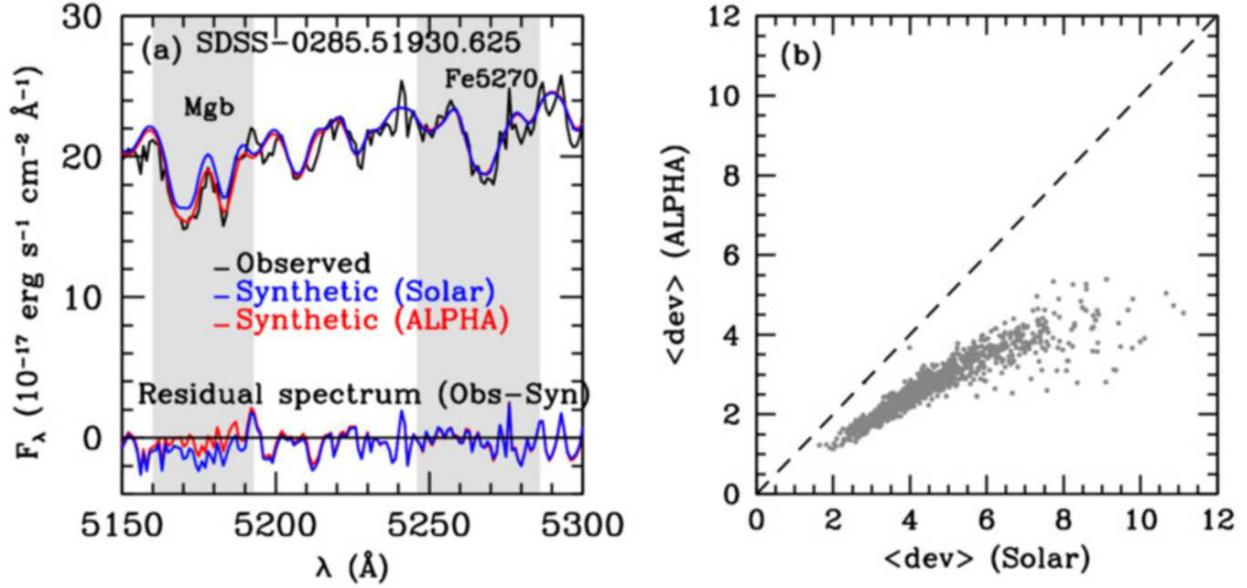}
\caption{Performance of  solar-abundance and $\alpha$-MILES  models to
  model the  spectra of bright  ETGs. (a) The  spectrum of one  ETG is
  shown, as  example, in black  color over a small  spectral interval,
  including  both  $Mgb$ and  $Fe5270$  features  (gray  bands in  the
  plot). The $Mgb$ index is well reproduced only by the alpha-enhanced
  models (red), as  also seen in the residual  spectrum (lower part of
  the plot).  (b)  Comparison of the mean percentage  deviation, $< \!
  dev \! >$, between  the solar-abundance and $\alpha$-MILES synthetic
  spectra produced by STARLIGHT for a subsample of $1,000$ ETGs in the
  SPIDER  (see the  text).  The percentage  deviation  decreases by  a
  factor of $\sim 2$ when using the $\alpha$-enhanced models.
\label{fig:alphaMILES}
}
\end{center}
\end{figure*}

According to  our tests,  these preliminary alpha-enhanced  models are
clearly  more efficient  in fitting  the  ETG's spectra  $wrt$ to  the
solar-scale  counterparts. For  a  subsample of  $1,000$  ETGs in  the
SPIDER sample, we have compared  the quality of the STARLIGHT spectral
fitting when using either solar-abundance or $\alpha$-MILES SSPs.  The
fitting  quality is  measured,  for  each galaxy,  by  using: (1)  the
reduced  $\chi^2$ of  the  fitting, $\chi^2/N_{\lambda_{eff}}$,  where
$N_{\lambda_{eff}}$ is the number of  $\lambda$ values used in the fit
minus the  number of fitting  parameters; and (2) the  percentage mean
deviation, $<  \! dev \!  >$,  over all the fitted  pixels between the
input and  synthetic spectrum.  The  $< \! dev  \!  >$ is  computed by
taking  the mean  of $|Obs_{\lambda}  - Syn_{\lambda}|/Obs_{\lambda}$,
where  $Obs_{\lambda}$  and  $Syn_{\lambda}$  are the  fluxes  of  the
observed and synthetic spectrum  at a given wavelength $\lambda$, over
the  spectral interval  4320-6800  $\AA$. We  found  that the  reduced
$\chi^2$ decreases by $0.1$  when using the alpha-enhanced models. The
improvement     in      fit     quality     is      illustrated     in
Fig.~\ref{fig:alphaMILES}.   Left  panel  shows,  as an  example,  the
synthetic  spectra obtained  for  a  given ETG,  in  a small  spectral
interval including the $Mgb$  and $Fe5270$ features, when using either
solar (blue)  or $\alpha$-MILES (red)  models. Although both  models well
describe the  $Fe$, only the  $\alpha$-MILES synthetic spectrum  (red) is
able   to  reproduce   also   the  $Mg$   region.    Right  panel   of
Fig.~\ref{fig:alphaMILES}  compares the  $<  \! dev  \! >$  parameters
between solar and alpha-enhanced synthetic spectra. The $\alpha$-MILES
models provide a mean  percentage deviation smaller than that obtained
from the solar models by a factor of $\sim 2$.

\begin{figure*}
\begin{center}
\plotone{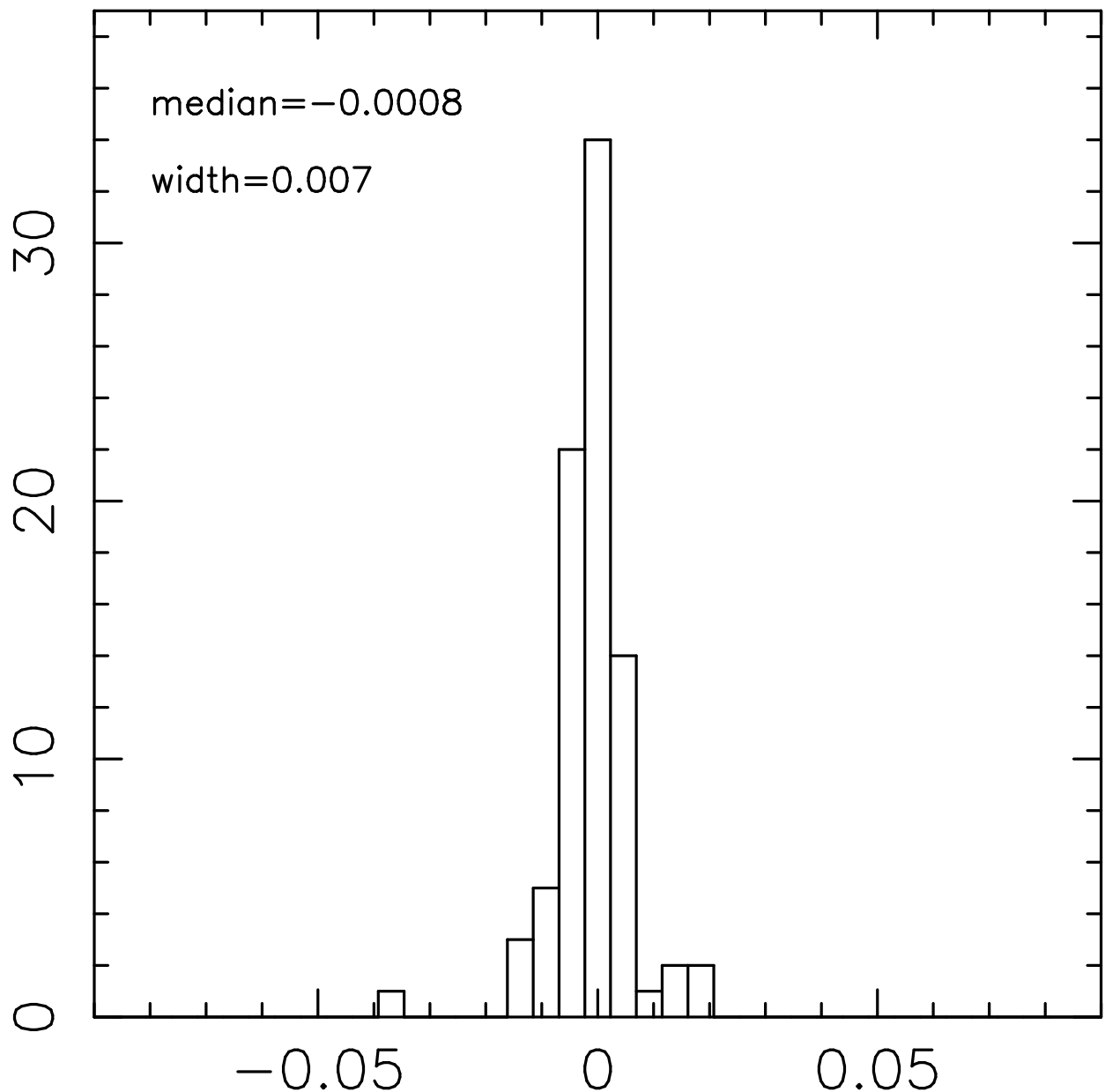}
\caption{{  Comparison of  solar MILES  models  with $\alpha$-MILES
    models having solar  abundance ratio ($[\alpha/Fe]=0$). The Figure
    plots  the   distribution  of  luminosity-weighted  $[\alpha/Fe]$,
    obtained by running  STARLIGHT on SSP models from  the solar MILES
    library, with a variety of  ages and metallicities (see the text).
    The central and  width values of the distribution  are reported in
    the upper--left  corner of the plot.  Notice  how the distribution
    is narrowly  peaked around zero,  proving the consistency  of both
    sets of models for $[\alpha/Fe]=0$.}
\label{fig:MILES_alpha_solar}
}
\end{center}
\end{figure*}

{ As  a further test, we  compared the solar MILES  models with the
  $\alpha$-MILES models having  $[\alpha/Fe]=0$ (i.e., we compared the
  models  which use  the empirical  stellar spectra  with  those using
  synthetic ones, in the regime where both models should give the same
  answer).  Fig.~\ref{fig:MILES_alpha_solar} plots the distribution of
  luminosity-weighted  $[\alpha/Fe]$,  obtained  by running  STARLIGHT
  with a  {\it basis} of $\alpha$-MILES SSPs  on different solar-MILES
  SSP models (see  Sec.~\ref{sec:stellar_pop}).  We considered a total
  of $84$ solar models with a variety of ages (in 21 steps from $1$ to
  $17.8$~Gyr) and metallicities  ($[Z/H]=-0.68, -0.38, 0., 0.2$).  For
  each  solar model,  STARLIGHT  was run  with  a {\it  basis} of  220
  $\alpha$-MILES  SSP  models, covering  the  same  range  in age  and
  metallicity    as    for    the    analysis    of    SDSS    spectra
  (Sec.~\ref{sec:stellar_pop}),                                     and
  $[\alpha/Fe]=-0.2,0,+0.2,+0.4,+0.6$.      The     distribution    of
  luminosity-weighted                 $[\alpha/Fe]$                 in
  Fig.~\ref{fig:MILES_alpha_solar} is sharply peaked around zero, with
  a width less than $\sim 0.01$, proving the consistency of solar- and
  $\alpha$ models for the case of solar abundance ratio.}

\section{Aperture corrections and color gradients.}
\label{sec:apcor}
\begin{figure*}
\begin{center}
\epsscale{0.8}
\plotone{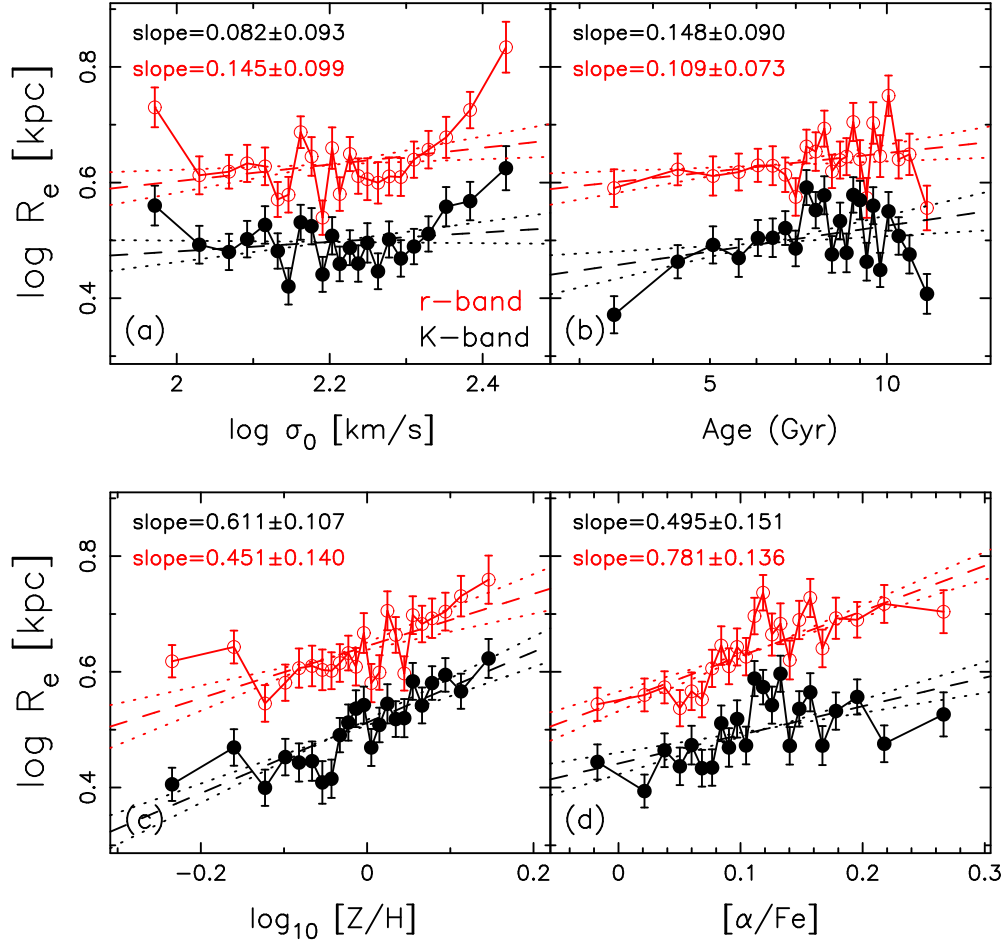}
\caption{Logarithmic median effective radius  as a function of central
  velocity dispersion (Panel a), age (Panel b), metallicity (Panel c),
  and $\alpha$--enhancement  (Panel d). Black and red  colors refer to
  $K$-  and $r$-band  effective  radii. Error  bars denote  1~$\sigma$
  standard error  on the  medians. Each trend  is modeled by  a linear
  fit,  shown as  a dashed  line.  The dotted  lines mark  the $\pm  1
  \sigma$ uncertainties  on the fitted  lines. The slopes of  the fits
  and the  corresponding $1~\sigma$ uncertainties are  reported in the
  left--upper corner of each panel.
\label{fig:re_spar}
}
\end{center}
\end{figure*}

Since the \nablas's correlate with the (optical) $R_e$ (see Panel a of
Fig.~\ref{fig:grads_par}), the fixed size of SDSS spectra might affect
the   correlation   of  the   $\nabla$'s   with   $Age$,  $Z/H$,   and
$[\alpha/Fe]$.     To    address    this    issue,    we    show    in
Fig.~\ref{fig:re_spar} the median galaxy effective radius in different
bins  of   $\sigma_0$,  $Age$,  $Z/H$,  and   $[\alpha/Fe]$,  for  the
optical+NIR sample of ETGs.  Red and black colors correspond to r- and
K-band effective radii.  As  expected, some correlations exist between
the  $R_e$  and  the  spectroscopic parameters.   In  particular,  the
effective  radius increases  in galaxies  with larger  metallicity and
$\alpha$-enhancement.   However,  one can  notice  that  the range  of
variation in  $R_e$ amounts to only $0.1-0.2$~dex,  i.e.  much smaller
than  the range  of  $\sim 1.5$~dex  spanned  by the  optical and  NIR
effective radii (see Panel a of Fig.~\ref{fig:grads_par}). In the case
of  $[\alpha/Fe]$  (Panel d  of  Fig.~\ref{fig:re_spar}), the  optical
$R_e$ changes  from $\sim 0.55$~dex  at $[\alpha/Fe] \sim 0$  to $\sim
0.7$~dex at  $[\alpha/Fe] \sim 0.26$.   From Fig.~\ref{fig:grads_par},
we see  that the corresponding  variation of \nablas \,  is completely
negligible, implying that the aperture  effect is not affecting at all
the correlations exhibited in Fig.~\ref{fig:grads_spar}.

\section{Derivatives of SSP magnitudes with respect to age and metallicity}
\label{sec:polfit}
We describe here how the colors available for the SPIDER sample can be
parametrized  in terms of  age and  metallicity of  stellar population
models. We adopt the ~\citet{BrC03} { (BC03)} models.

\begin{figure*}
\begin{center}
\plotone{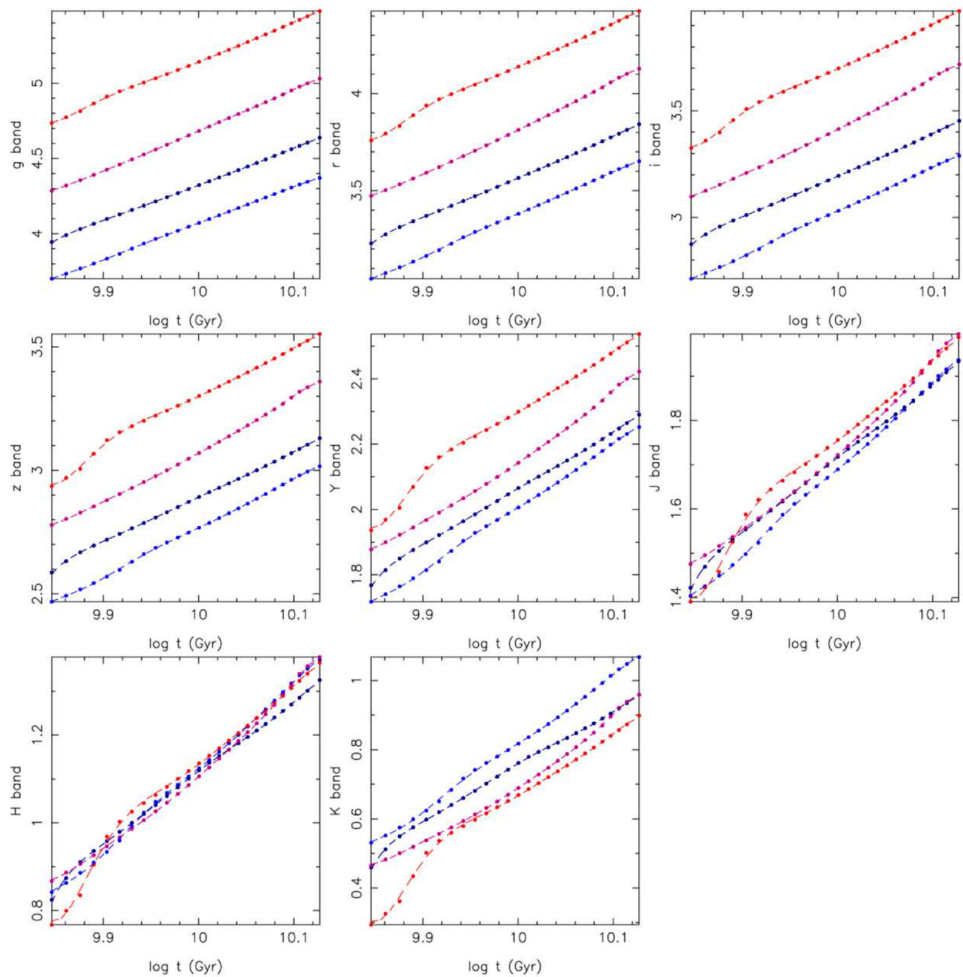}
\caption{{ Magnitudes of simple stellar population models from BC03
    as  a function  of their  formation  epoch, $t$.   The models  are
    normalized to  a total mass  of 1~$M_{\odot}$ (one solar  mass) in
    stars. The  griz and  YJHK magnitudes are  computed in the  AB and
    Vega  systems, i.e. the  photometric systems  of SDSS  and UKIDSS,
    respectively  (see Paper I)}.   Different colors  denote different
  metallicities.  The  color become {  bluer} from higher  to lower
  metallicity. Four  metallicities are plotted: $Z=0.2,  0.4, 1.0, 2.5
  Z_{  \odot  }$.   Dashed  lines  show the  result  of  fitting  {
    magnitudes}   as  a   function   of  age   and  metallicity   with
  two-dimensional polynomials of degree eight.
\label{SP_tau0.1}
}
\end{center}
\end{figure*}

Fig.~\ref{SP_tau0.1} exhibits the magnitudes  of a { simple stellar
  population model,  normalized to a total  mass of one  solar mass in
  stars,} as a function of  its age, $t$, for the $grizYJHK$ wavebands
of the SPIDER dataset. { The  griz and YJHK magnitudes are given in
  the AB  and Vega systems,  respectively.}  In order to  describe the
colors of ETGs, we consider relatively old ages, spanning the range of
$5$ to  $13.5$Gyrs. Different metallicities are also  displayed in the
range  of  $0.2$  to  $2.5$  times solar  ($Z_{\odot}$).   Fluxes  are
computed with the BC03 code, convolving the SDSS and UKIDSS throughput
curves  with the  model  SEDs.   In the  optical  passbands, the  {
  magnitude} is  essentially a linear  function of $\log t$.   This is
consistent with the fact that  the flux of a simple stellar population
is expected to change as  $t^{-\alpha}$, where $\alpha$ depends on the
metallicity  of   the  stellar  population   and  the  slope   of  the
IMF~\citep{Tinsley73}.    In   the   NIR  passbands,   the   power-law
approximation is less accurate, particularly  at the lower age end. In
order  to calculate  the color  derivatives  with respect  to age  and
metallicity,  for  each  waveband  we fitted  the  corresponding  {
  magnitudes} with a two-dimensional  polynomial in $\log t$ and $\log
Z$.  As  shown in Fig.~\ref{SP_tau0.1},  a polynomial of  degree eight
reproduces  model { magnitudes}  very accurately.   The rms  of the
fits are  smaller than $0.01$mag  in all the bands.  Estimating fluxes
using a  composite stellar population with  exponential star formation
rate (SFR)  and an e-folding  time of 1~Gyr  as a function of  age and
metallicity   provides  essentially   similar   results.   Given   the
polynomial coefficients for each  band, the computation of the color's
derivatives is straightforward for any combination of $t$ and $Z$.

\section{Possible sources of bias in the $\nabla$'s.}
\label{sec:age_met_bias}
\begin{figure}
\begin{center}
\epsscale{0.7}
\plotone{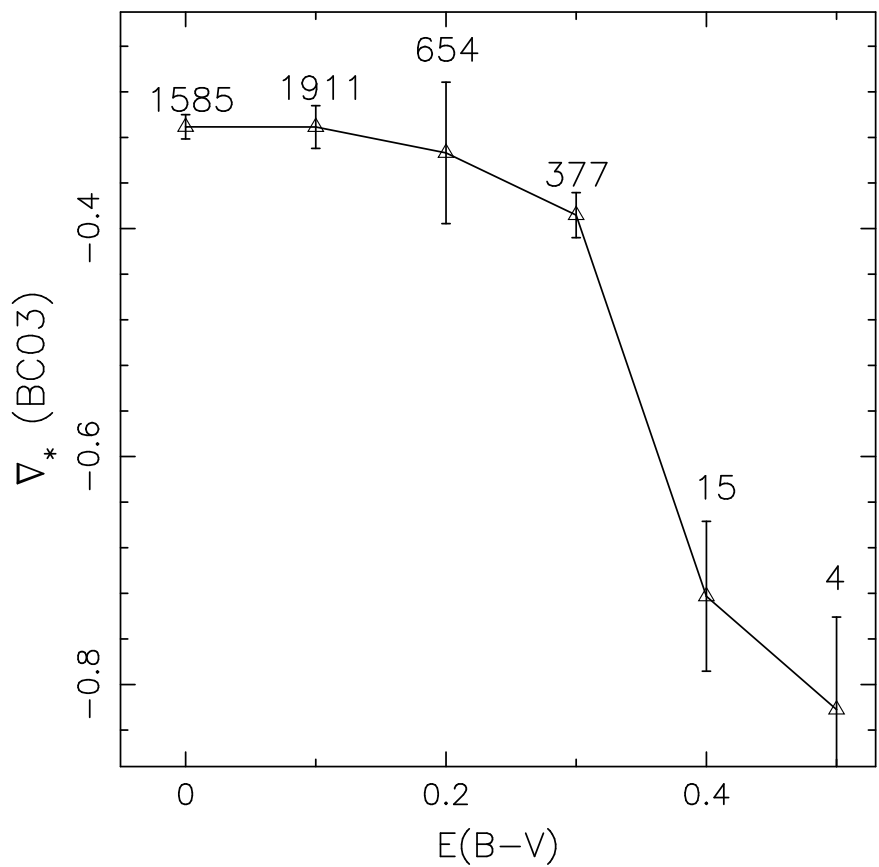}
\caption{Medians  of  the  effective  color gradient,  \nablas,  as  a
  function of the amount of  reddening, $E(B-V)$, in ETGs. The plot is
  obtained  for the  optical+NIR  sample of  ETGs.  Error bars  denote
  1~$\sigma$ uncertainties on median  values.  The numbers of ETGs for
  different $E(B-V)$  are reported above the  corresponding error bars
  in the  plot. Most of the  galaxies ($91\%$) have  $E(B-V) \le 0.2$,
  for which the  variation of \nablas \, with  color excess is smaller
  than $\sim 0.02$.
\label{fig:nablas_ebv}
}
\end{center}
\end{figure}

The  existence of  radial  gradients of  absorption  indices in  ETG's
spectra  (e.g.   $Mgb$  and  $H_\beta$) implies  that  internal  color
gradients  are likely  driven by  a variation  of age  and metallicity
inside    galaxies,     consistent    with    our     assumption    in
Eq.~\ref{eq:grad_gX}.     However,    as    argued    by~\citet{SW:96}
and~\citet{Wise:96}, radial gradients in internal reddening might also
contribute  significantly to  the  observed color  gradients. In  such
case, one  would naturally expect  a correlation to exist  between the
gradient  and the  amount of  dust,  with stronger  gradients in  more
dust-obscured  galaxies.  Although  a  detailed analysis  of the  dust
distribution  in ETGs  is certainly  beyond the  scope of  the present
paper,   we  can   evaluate   if  internal   reddening  might   affect
significantly  our findings  on  the correlation  among gradients  and
stellar  population  properties  (Sec.~\ref{sec:age_met_grads}). As  a
rough estimate of the {\it total} amount of internal reddening, we use
the  color excess,  $E(B-V)$,  obtained  for each  galaxy  by the  SED
fitting             procedure             (Sec.~\ref{sec:cg_galpars}).
Fig.~\ref{fig:nablas_ebv}  plots   the  median  \nablas   \,  for  the
different $E(B-V)$ (see Tab.~1).   A clear correlation is detected, in
the sense that steeper gradients  are found at higher extinctions. The
trend is mostly because of galaxies at high $E(B-V)$ ($\ge 0.3$), with
these  objects amounting  to less  than $10  \%$ of  the  entire ETG's
sample.   We   conclude  that  internal   reddening  might  contribute
significantly  to  color  gradients  only  for  a  minor  fraction  of
galaxies.   Also, we  find that  selecting galaxies  with  $E(B-V) \le
0.2$,  for  which  the  variation  of color  gradients  with  internal
reddening is insignificant, does not change at all the trends shown in
Fig.~\ref{fig:allgrads_pars}.

Another  possible  source of  concern  is  that  color derivatives  in
Eq.~\ref{eq:grad_gX}  are  estimated   by  stellar  population  models
reshifted at  the median redshift,  $z=0.0725$, of the  SPIDER sample,
hence neglecting the effect of k and evolutionary corrections on color
gradients. In order to see  if this approximation impacts our results,
we select only ETGs from the optical+NIR sample in a narrower redshift
range of  $z=0.0625$ to  $0.0825$, leading to  a subsample  of $1,984$
galaxies. For this  subsample, the peaks of the  \nablas, \nablat, and
\nablaz, distributions amount to  about $-0.31$, $0.126$, and $-0.43$,
respectively,  fully  consistent with  that  obtained  for the  entire
sample (see Tab.~\ref{tab:stat_cgrad_optNIR}).  Also, the trends shown
in Fig.~\ref{fig:allgrads_pars} turned out  not to be affected by this
selection.

\begin{acknowledgements}
{  We  would  like  to  thank the  anonymous  referee  for  his/her
  constructive report, which definitely helped improving parts of this
  paper}.  We thank  the staffs in charge of  the cluster computers at
the INPE-LAC  (Sao Jos\'e  dos Campos, Brazil),  H.C.  Velho,  and the
staffs at  INAF-OAC (Naples, Italy), Dr.  A.Grado  and F.I.Getman, for
keep running  the systems  smoothly.  We have  used data from  the 4th
data release  of the  UKIDSS survey, which  is described in  detail in
\citet{War07}.   The  UKIDSS   project  is  defined  in~\citet{Law07}.
UKIDSS uses the  UKIRT Wide Field Camera (WFCAM;  Casali et al, 2007).
The photometric  system is described in  Hewett et al  (2006), and the
calibration  is described  in  Hodgkin et  al.   (2009). The  pipeline
processing and science archive are  described in Irwin et al (2009, in
prep) and Hambly  et al (2008).  UKIDSS data  have been analyzed using
the Beowulf  system at INAF-OAC~\citep{CGP:02}.  Funding  for the SDSS
and SDSS-II has been provided  by the Alfred P.  Sloan Foundation, the
Participating Institutions, the  National Science Foundation, the U.S.
Department   of   Energy,   the   National   Aeronautics   and   Space
Administration, the  Japanese Monbukagakusho, the  Max Planck Society,
and the  Higher Education Funding  Council for England.  The  SDSS Web
Site   is  http://www.sdss.org/.    The   SDSS  is   managed  by   the
Astrophysical Research Consortium  for the Participating Institutions.
The  Participating Institutions  are  the American  Museum of  Natural
History,  Astrophysical   Institute  Potsdam,  University   of  Basel,
University of  Cambridge, Case Western  Reserve University, University
of Chicago,  Drexel University,  Fermilab, the Institute  for Advanced
Study, the  Japan Participation  Group, Johns Hopkins  University, the
Joint  Institute for  Nuclear  Astrophysics, the  Kavli Institute  for
Particle Astrophysics  and Cosmology, the Korean  Scientist Group, the
Chinese Academy of Sciences  (LAMOST), Los Alamos National Laboratory,
the     Max-Planck-Institute     for     Astronomy     (MPIA),     the
Max-Planck-Institute   for  Astrophysics   (MPA),  New   Mexico  State
University,   Ohio  State   University,   University  of   Pittsburgh,
University  of  Portsmouth, Princeton  University,  the United  States
Naval Observatory, and the University of Washington.
\end{acknowledgements}



\end{document}